\documentclass[prb,twocolumn,superscriptaddress,showpacs]{revtex4-1}

\usepackage{amsmath,amsfonts,graphicx,color,amsthm,bm,float}
\usepackage{verbatim}

\def\Id{{\openone}}
\newcommand{\be}{\begin{equation}}
\newcommand{\ee}{\end{equation}}
\newcommand{\bea}{\begin{eqnarray}}
\newcommand{\eea}{\end{eqnarray}}

\newcommand{\mc}[1]{\mathcal{#1}}

\newcommand{\ket}[1]{\vert#1\rangle}
\newcommand{\bra}[1]{\langle#1\vert}
\DeclareMathOperator{\tr}{tr}

\newcommand{\mr}{\mathrm}
\newcommand{\dg}{\dagger}

\renewcommand{\d}{\mathrm{d}}

\begin{document}

\title{Symmetries and boundary theories for chiral Projected Entangled Pair States}

\author{Thorsten B.~Wahl}
\affiliation{Max-Planck Institut f\"ur Quantenoptik, Hans-Kopfermann-Str.~1, D-85748 Garching, Germany}
 \author{Stefan T. Ha{\ss}ler}
\affiliation{JARA Institute for Quantum Information, RWTH Aachen University, D-52056 Aachen, Germany}
\author{Hong-Hao Tu}
\affiliation{Max-Planck Institut f\"ur Quantenoptik, Hans-Kopfermann-Str.~1, D-85748 Garching, Germany}
\author{J.~Ignacio Cirac}
\affiliation{Max-Planck Institut f\"ur Quantenoptik, Hans-Kopfermann-Str.~1, D-85748 Garching, Germany}
\author{Norbert Schuch}
\affiliation{JARA Institute for Quantum Information, RWTH Aachen University, D-52056 Aachen, Germany}


\begin{abstract}
We investigate the topological character of lattice chiral Gaussian fermionic states in two dimensions possessing the simplest descriptions in terms of projected entangled-pair states (PEPS). 
They are ground states of two different kinds of Hamiltonians. The first one, $\mc H_\mr{ff}$, is local, frustration-free, and gapless. It can be interpreted as describing a quantum phase transition between different topological phases. The second one, $\mc H_\mr{fb}$ is gapped, and has hopping terms scaling as $1/r^3$ with the distance $r$. The gap is robust against local perturbations, which allows us to define a Chern number for the PEPS. As for (non-chiral) topological PEPS, the non-trivial topological properties can be traced down to the existence of a symmetry in the virtual modes that are used to build the state. Based on that symmetry, we construct string-like operators acting on the virtual modes that can be continuously deformed without changing the state. On the torus, the symmetry implies that the ground state space of the local parent Hamiltonian is two-fold
degenerate. By adding a string wrapping around the torus one can change one of the ground states into the other. We use the special properties of PEPS to build the boundary theory and show how
the symmetry results in the appearance of chiral modes, and a universal correction to the area law for the zero R\'{e}nyi entropy. 
\end{abstract}

\maketitle

\section{Introduction}

Topological states~\cite{Wen89,Wen90} are quantum states of matter with intriguing properties. They include non-chiral states with topological order, such as the toric code~\cite{Kit03} and string net models~\cite{Lev05}, as well as chiral topological states. The latter have broken time-reversal symmetry, and possess non-vanishing topological invariants. They include celebrated examples like integer and fractional quantum Hall states, as well as Chern insulators~\cite{Hal88} and topological superconductors~\cite{Kit08,Sch08}. They display chiral edge modes which are protected against local perturbations, and cannot be adiabatically connected to states with different values of the topological invariants.

Among others, a remarkable open problem in this field is to classify all topological phases; that is, the equivalence classes of local Hamiltonians that can be connected by a (symmetry preserving) gapped path. For their free fermion versions in arbitrary dimensions, a full classification has been already obtained \cite{Sch08,Kit08}. For interacting spins, this goal has only been achieved in one dimension \cite{Pol09,Che11,Sch11}, based on the fact that ground states of 1D gapped local Hamiltonians are efficiently represented by Matrix Product States (MPS)\cite{Scholl11}. In dimensions higher than one, this problem  remains open. Still, recent developments reveal that there exist deep intrinsic connections between quantum entanglement and topological states. For instance, topological order is reflected in the universal correction to the entanglement area law, also called topological entanglement entropy. A further proposal has been put forward by Li and Haldane \cite{Hal08}, who suggested that the entanglement spectrum, that is, the eigenvalues of the reduced density operator of a subsystem, contains more valuable information than the topological entanglement entropy.

Projected Entangled Pair States (PEPS) \cite{Ver04}, higher dimensional generalizations of MPS, are a natural tool for investigating topological states. By construction, they contain the necessary amount of entanglement required by the entanglement area law. Furthermore, many known topological states, such as the toric code \cite{Kit03}, resonating valence-bond states \cite{And73}, and string nets \cite{Lev05}, possess exact PEPS descriptions \cite{Ver06,Sch10,Bue08,Gu08}. Despite the lack of local order parameters, PEPS nevertheless provide a local description for topological states, with the global topological properties being encoded in a single PEPS tensor. For some of the above examples, the connection of topology and the PEPS tensor has been made precise as originating from a \textit{symmetry} of the PEPS tensor \cite{Sch10} (see also Ref. \onlinecite{Bue13}). This symmetry only affects the virtual particles used to build the PEPS, unlike the physical symmetries of the PEPS. It can be grown to arbitrary regions, and has several intriguing consequences: (i) it leads to the topological entanglement entropy; (ii) it gives rise to a universal part \cite{Sch13} in the boundary Hamiltonian \cite{Cir11} acting on the auxiliary particles at a virtual boundary, whose eigenvalues are related to the entanglement spectrum of the subsystem; (iii) it provides topological protection of the edge modes \cite{Shuo}; (iv) it gives rise to string operators that provide a mapping between the different topological sectors;  (v) it can also be used to build string operators for anyonic excitations and to determine the braiding statistics; (vi) it determines the ground state degeneracy of the parent Hamiltonian.

Chiral topological states are very different from the above mentioned topological states preserving time-reversal symmetry (known as nonchiral topological states), in that they necessarily have chiral gapless edge modes, which cannot be gapped out by weak perturbations due to the lack of a back-scattering channel. There have been doubts that PEPS can describe chiral topological states, until explicit examples with exact PEPS representations have been obtained very recently \cite{Wah13,Dub13}. These chiral PEPS examples are topological insulators and topological superconductors characterized by nonzero Chern numbers, albeit with correlations decaying as an inverse power law. In view of all that, very natural questions arise as whether these chiral topological PEPS fit into the general characterization scheme in terms of a \textit{symmetry} of a single PEPS tensor, and whether useful information characterizing topological order manifests itself in the boundary Hamiltonian.

In this work, we answer these questions in an affirmative way for a family of topological superconductors similar to the one introduced in Ref.~\onlinecite{Wah13}. Those are chiral Gaussian fermionic PEPS (GFPEPS), which are free fermionic tensor network states~\cite{kraus:fPEPS,Cor10}. We give general procedures to build the boundary Hamiltonians and analyze their properties (see also Ref.~\onlinecite{Dub13}). We first show how to determine the boundary Hamiltonian for GFPEPS, and how the Chern number can be obtained by counting the number of chiral modes defined on the boundary Hamiltonian. Then we show that, as in the case of topological (non-chiral) PEPS, there exists a symmetry in the virtual modes that can be grown to any arbitrary region. We connect this symmetry to the chiral modes on the boundary, and show that it also gives rise to a universal correction to the area law in the zero R\'enyi entropy (although not in the von Neumann one), and to the boundary Hamiltonian.

Following Refs.~\onlinecite{Wah13} and~\onlinecite{Dub13}, we also build
two kinds of (parent) Hamiltonian for which the chiral GFPEPS are ground
states, and analyze their properties. The first one, $\mc H_\mr{ff}$, is
Gaussian, local,  frustration free, and gapless. In terms of that
Hamiltonian, our states can be interpreted as being at the quantum phase
transition between different phases characterized by different Chern
numbers. That Hamiltonian is two-fold degenerate on the torus. We use the
symmetry to  build string operators that allow us to characterize the
ground states of $\mc H_\mr{ff}$, and that can be continuously deformed
without changing the state. The second one, $\mc H_\mr{fb}$, is also
Gaussian, although gapped, and has a unique ground state on the torus. It
is not local, possessing hopping terms decaying as $1/|r|^3$ with the
distance $|r|$. We show that it is topologically stable to the addition of
local perturbations. This allows us to consider the state as truly
topological, and to define a Chern number which we find equals -1. We also
compute the momentum polarization~\cite{mom-pol} and show that it has the
expected properties for a topological state. In addition, we provide a
numerical example of a GFPEPS with two Majorana bonds (the number of
Majorana bonds corresponds to twice the logarithm of the bond dimension in
the normal PEPS language) and Chern number 2 having two symmetries of
the above kind.

Given the variety of results obtained in this paper and the different techniques used to derive them, we start with a Section that gives an overview of all of them, and connects them to known properties of topological PEPS. The specific derivations and explicit statements and proofs are given in the following Sections. In Sec. III, a general framework for studying the boundary and edge theories of GFPEPS is developed. Their relation to the Chern number is established. In Sec. IV, we give different examples of GFPEPS, some of them topological and some of them not, in order to provide comparison between the two cases. In Sec. V we completely characterize all PEPS with one Majorana bond with topological character. It turns out that in this case the Chern number can only be $0$ or $\pm 1$; for the latter case we derive necessary and sufficient criteria. In Sec. VI we prove a necessary and sufficient condition on the symmetry that a PEPS tensor has to possess in order to give rise to a chiral edge state for one Majorana bond, and show how those symmetries can be grown to larger regions and to build string-like operators.


\section{Definitions and Results}
\label{DefinitionsResults}

This Section gives an overview of the main results of this paper. It also reviews in a self-contained way the basic ingredients that are required to derive the results, and to interpret them. It is divided in four subsections. The first one contains the definition of GFPEPS, which are the basic objects in our study. It also contains two Hamiltonians for which they are the ground state. The first one is gapped, has power-law hopping terms, and is the one that appears more naturally in the context of topological insulators and superconductors. The second one follows from the PEPS formalism, is gapless, and has a degenerate ground state. In the second subsection we present a simple family of GFPEPS, similar to the one introduced in Ref.~\onlinecite{Wah13}, which we will extensively use to illustrate our findings. The third one contains the construction of boundary and edge theories for GFPEPS, which we explicitly use for the simple family. In the last subsection, we make a connection between the behavior observed for this family, and the one that is known for topological (non-chiral) PEPS (Ref.~\onlinecite{Sch10}). In particular, we show that one can understand it in terms of string operators acting on the so-called virtual particles, which can be moved and deformed without changing the state.

Throughout this Section we will concentrate on the simplest GFPEPS -- those which have the smallest possible bond dimension (which will
correspond to one Majorana bond, see below). This will allow us to
simplify the description and formulas. However, all the constructions given here can be easily generalized to larger bond dimensions, and this will be done in the following Sections. Some of the results, however, explicitly apply to one Majorana bond, so that we will specialize to that case in the following Sections too.

\subsection{Gaussian Fermionic PEPS and parent Hamiltonians}\label{Sec:Intro-GFPEPS}

We consider a square $N_v \times N_h$ lattice of a single fermionic mode per site, with annihilation operators $a_j$, where $j$ is a vector denoting the lattice site. We will consider a state, $\Phi$, of a particular form, and Hamiltonians for which it is the ground state.

\subsubsection{Gaussian Fermionic PEPS}

\begin{figure}[t]
\begin{center}
\includegraphics[width=0.4\textwidth]{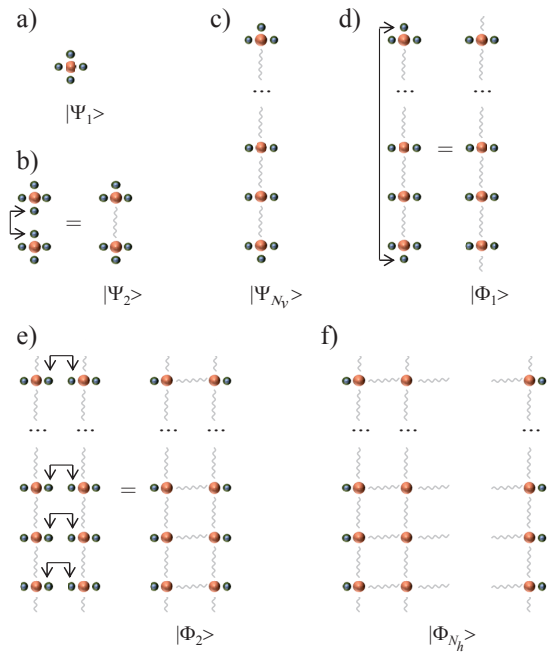}
\end{center}
\caption{Construction of a GFPEPS. (a) We start with a state $\Psi_1$ that is Gaussian and includes one physical fermionic mode (big red ball) and four virtual Majorana fermions (small blue balls) located at site $j$. (b) Two states of this kind in the same column are concatenated by projecting on $\langle \omega_{jn}|$ (see text). (c) Proceeding in the same way one obtains a state $\Psi_{N_v}$ defined on a column with unpaired virtual Majorana modes on the left and right and on the two ends of the column. (d) The remaining up and down modes at the ends are jointly projected out, yielding $\Phi_1$. (e) Afterwards, two columns can be concatenated by pairwise projecting out the left and right virtual Majoranas between them. (f) Continuing in the same way, one obtains a GFPEPS $\Phi_{N_h}$ defined on $N_h$ columns. It can be made completely translationally invariant by pairwise projecting out the remaining left and right virtual Majorana modes, resulting in the final GFPEPS state $\Phi$.\label{Fig:PEPS}}
\end{figure}

We revise here the GFPEPS introduced in Ref.~\onlinecite{kraus:fPEPS}. We will first show how a GFPEPS, $\Phi$, of the fermionic modes is constructed (see Fig.~\ref{Fig:PEPS}).

The basic object in this construction is a fiducial state, $\Psi_1$, of
one fermionic (physical) mode, and four additional (virtual) Majorana
modes~\cite{note-Majorana}, all of them at site $j$
(Fig.\ref{Fig:PEPS}a). The corresponding mode operators,
$c_{j,L},c_{j,R},c_{j,U}$, and $c_{j,D}$ ($L$, $R$, $U$, and $D$, stand
for left, right, up and down, respectively), fulfill standard
anticommutation relations, $\{c_{i,\alpha},c_{j,\beta}\}=2
\delta_{i,j}\delta_{\alpha,\beta}$, are Hermitian and anticommute with the
other fermionic operators. The state $\Psi_1$ is arbitrary, except for the
fact that it must be Gaussian and have a well defined parity. This means
that it can be written as
\be
 |\Psi_1\rangle_j = e^{\mc H_j} |\Omega\rangle, \label{eq:Gaussian-psi1}
 \ee
where $\mc H_j$ is a quadratic operator in all the mode operators, and the
$\Omega$ denotes the vacuum of the virtual and physical modes. One can
easily parametrize $\mc H$, and thus $\Psi_1$, but this will not be
necessary here, since we will make use of the fact that the state is
Gaussian, for which a more appropriate parametrization exists.

The state of the physical fermions, $\Phi$, can be obtained by concatenating all the $\Psi_1$ at different sites in the way we explain now and is illustrated in Fig.~\ref{Fig:PEPS}. First, take two consecutive lattice sites in the same column, $j$ and $n$, and project the up virtual mode of the first and the down of the latter onto a particular state, i.e.
 \be
 |\Psi_2\rangle_{jn} = \omega_{jn} e^{\mc H_j+ \mc H_n}
 |\Omega\rangle,
 \ee
(see Fig.\ref{Fig:PEPS}b). Here $\omega_{jn}= \tfrac{1}{2}(1+ i c_{j,D} c_{n,U})$, which ensures that $c_U$ and $c_D$ are maximally entangled (forming a pure fermionic state)~\cite{note:ME}. Since the modes that we project on are in a well defined state after the projection, we can omit them in the following. In order to simplify the notation, we will denote by $\langle \omega_{jn}|\Psi\rangle$ the state obtained by applying $\omega_{jn}$ and discarding the corresponding modes, and we will say that we have projected onto $\omega_{jn}$. We will also omit the indices representing the lattice sites whenever this does not lead to confusion.

We proceed in the same way, concatenating all the sites corresponding to a column by projecting out the consecutive up and down virtual modes onto the state defined by $\omega_{jn}$. The resulting state is $\Psi_{N_v}$, since we have $N_v$ sites in a column (see Fig.\ref{Fig:PEPS}c). This state contains $N_v$ physical fermionic modes, as well as $2N_v+2$ virtual Majorana modes, $N_v$ on the left, $N_v$ on the right, one up and one down. Since we will consider here periodic boundary conditions along the vertical direction, we also project out the up and down virtual modes, obtaining $\Phi_1$, a state that corresponds to one column (and thus the subindex). Such a state contains $N_v$ physical fermionic modes, as well as $2N_v$ virtual Majorana modes (see Fig.\ref{Fig:PEPS}d). By construction, the state is translationally invariant along the vertical direction.

In order to obtain the state on the lattice, we have to follow a similar procedure in the horizontal direction (see Fig.\ref{Fig:PEPS}e). For that, we take the states of two consecutive columns, and project each of the right virtual modes (at site $j$) of one and the corresponding left virtual mode (at site $n$) of the other onto  $\omega_{jn}'= \tfrac{1}{2}(1+ i c_{j,R} c_{n,L})$. The resulting state, $\Phi_2$, contains $2N_v$ physical fermionic modes, as well as $2N_v$ virtual Majorana modes. We continue adding columns in the same way, until we obtain $\Phi_{N_h}$, containing $N_v\times N_h$ physical fermionic modes and $2N_v$ virtual Majorana ones (see Fig.\ref{Fig:PEPS}f).

In order to obtain a translationally invariant state in the horizontal direction too, we have to project each remaining virtual pair of modes on the left and the right onto the state defined by $\omega'_{jn}$. In this case, we will say that we have a state, $\Phi$, on the torus. Otherwise, we can project the virtual modes on the left and the right onto some other state. If we took a product state (of left and right virtual modes) that is translationally invariant in vertical direction itself, we will still keep that property in the vertical direction and the state $\Phi$ will be defined on a cylinder. A subtle point is that, when we perform this last projection in order to generate the physical state $\Phi$, the result may vanish. This happens, for instance, in some of the examples considered in this paper in the torus case. There, we will have to introduce a string operator in the virtual modes for those particular sizes of our system.

The state $\Phi$ on the torus is fully characterized by the fiducial state $\Psi_1$ (and therefore by $\mc H$), since the construction is carried out by concatenating them with a specific procedure. For the cylinder, $\Phi$ also depends on the states we choose to close the virtual boundaries. From now on we will work on the torus, unless explicitly stated otherwise.

Since the fiducial state $\Psi_1$ is Gaussian and our construction keeps the Gaussian nature, all the states defined above will be Gaussian. For that reason, instead of expressing $\Psi$ and $\Phi$ in the Hilbert space on which the mode operators act, we characterize them in terms of their covariance matrices (CMs). In order to do so, we write each physical fermionic mode operator in terms of two Majorana operators,
 \be
 a_j = (e_{2j-1} - i e_{2j})/2,
 \ee
fulfilling the corresponding anticommutation relations. For a (generally mixed) Gaussian state $\rho$ in a set of Majorana modes, $c_l$, the CM, $\gamma$, is defined through
 \be
 \gamma_{l,m} =\frac{i}{2} \tr(\rho[c_l,c_m]). \label{eq:CM-rho}
 \ee
This is a real antisymmetric matrix, fulfilling $\gamma^\top \gamma \le \Id$, where $\Id$ is the identity matrix. The equality ($\gamma^2=-\Id$) is reached  iff the state $\rho$ is pure. Thus, the original state $\Psi_1$ will have a CM with four blocks,
 \be
 \label{eq:gamma1}
 \gamma_1 = \left(\begin{array}{cc} A & B \\ - B^\top & D \end{array} \right)
 \ee
where $A,D$ are $2\times 2$ and $4\times 4$ antisymmetric matrices, respectively, $B$ is a $2\times 4$ matrix, and they are constrained by $\gamma_1^2=-\Id$ (since the state $\Psi_1$ is pure). Hence, the state $\Phi$ is completely characterized by those matrices. Concatenating states as explained above can be easily done in terms of the CMs (see Ref.~\onlinecite{kraus:fPEPS} and Sec.~\ref{sec:construct-GFPEPS} below).

If we consider the indices $l$ (and $m$) in Eq.~\eqref{eq:CM-rho} as joint indices of the site coordinates $r = (x,y)$ (and $r'$) and the index of the two Majorana modes located at site $r$ ($r'$), the $2 \times 2$ block of 
$\gamma$ of a GFPEPS for given sites $r$ and $r'$ fulfills
\be
\gamma_{r,r'} = \gamma(r-r'), 
\ee
since the construction of the GFPEPS is translationally invariant. Thus, it is convenient to carry out a discrete Fourier transform on $\gamma$. The result is, as outlined in Ref.~\onlinecite{kraus:fPEPS}, a block-diagonal matrix with blocks labelled by the momentum vector $k = (k_x,k_y)$. Due to the purity of the state, they are of the form
\begin{equation}
G(k) = \left(\begin{smallmatrix}
i \hat d_x(k)&\hat d_z(k) + i \hat d_y(k)\\
-\hat d_z(k) + i \hat d_y(k)&- i \hat d_x(k)
\end{smallmatrix}\right) \label{eq:G-param}
\end{equation}
with 
$\hat d_i(k) \in \mathbb{R}$ and $|\hat d(k) | = 1$.  

The above construction can be trivially extended to more general GFPEPS, where there are $4\chi$ virtual Majorana modes and $f$ fermions per site. In Sec.~\ref{sec:Detailed-Analysis} we will show how to carry out such a construction for that general case. The case considered in this Section, $\chi=1$, is much simpler to describe and already possesses all the ingredients to give rise to topological chiral states.

\subsubsection{Parent Hamiltonians}

One can easily construct Hamiltonians for which $\Phi$ is the ground state. For that, we can follow two different approaches. The first one takes advantage of the fact that $\Phi$ is a Gaussian state, whereas the second uses that it is a PEPS.

Our first Hamiltonian is the ``flat band'' Hamiltonian
 \be
 \label{eq:flat}
 {\cal H}_{\rm fb} = -\frac{i}{4} \sum_{l,m} \gamma_{l,m} e_l e_m
 \ee
where $\gamma$ is the CM of the state $\Phi$, and $e$ are the Majorana modes
built out of the physical fermionic modes. Since $\Phi$ is pure,
$\gamma^2=-\Id$ and thus it has eigenvalues $\pm i$. Hence, ${\cal
H}_\mr{fb}$ contains two bands separated by a bandgap of magnitude
$2$, which are flat. As $\gamma$ is antisymmetric,
there exists an orthogonal matrix $O$ such that $O^\top \gamma O$ is block
diagonal. Using this, one can easily convince oneself that $\Phi$ is the
unique ground state of $\mc H_\mr{fb}$. Note also that the Hamiltonian
${\cal H}_\mr{fb}$ will not be local in general, since $\gamma_{l,m}\ne 0$
for all $l,m$. We also remark that for general $\gamma$ the single
particle spectrum of a Hamiltonian of the form~\eqref{eq:flat} is given by
the eigenvalues of $-i \gamma$.

We transform Eq.~\eqref{eq:flat} to reciprocal space
and write it 
in terms of the Fourier transformed Majorana modes 
\be
e_{k,\alpha} = \frac{1}{\sqrt{N_h N_v}} \sum_{r} e_{r,\alpha} e^{i k \cdot r}
\ee
(with $(r,\alpha)$ corresponding to the joint index $l$ above),
so that it takes the form
\be
\mc{H}_\mr{fb} = -\tfrac{i}{4} \sum_k \sum_{\alpha,\beta=1}^2 G_{\alpha, \beta}(k) e_{k,\alpha} e_{k,\beta} \label{eq:H-Gout},
\ee
where $G_{\alpha,\beta}(k)$ is given in Eq.~\eqref{eq:G-param}.

The second Hamiltonian can be constructed by invoking the general theory of PEPS (see, e.g. Ref.~\onlinecite{Sch10}). We can always find a local, positive operator, $h\ge 0$, acting on a sufficiently large plaquette, that annihilates our state, i.e. $h_j|\Phi\rangle=0$. Here $j$ denotes the position of the plaquette. In the case of a GFPEPS, $h_j$ can be chosen to be local. Furthermore, since the state is translationally invariant, we can take
 \be
 \label{eq:HPEPS}
 {\cal H}_{\rm ff} = \sum_{j} h_j.
 \ee
Now, this Hamiltonian is local (i.e., a sum of terms acting on finite regions, the plaquettes), frustration free (thus the subscript), and it is clear that $\Phi$ is a ground state. However, there may still be other ground states, and, additionally, ${\cal H}_{\rm ff}$ may have a gapless continuous spectrum (in the thermodynamic limit).

For the topological states considered later on, we will see that ${\cal H}_{\rm fb}$ is intimately connected to the chiral properties at the edges, as it is well known for topological insulators and superconductors \cite{Qi10, Has10}. The other one, ${\cal H}_{\rm ff}$ will share other topological properties that makes it akin to Kitaev's toric code~\cite{Kit03} and its generalizations.

\subsection{A family of topological superconductors}\label{sec:Intro-Examples}

\subsubsection{Parameterization of the GFPEPS}

Now, we review a family of chiral topological GFPEPS similar to that introduced in Ref.~\onlinecite{Wah13}, which is characterized by a parameter, $\lambda\in[0,1]$. The fiducial state $\Psi_1$ is given by
 \be
 \label{eq:Psi1ex}
 |\Psi_1\rangle=\left(\sqrt{1-\lambda}\Id+\sqrt{\lambda}a^\dagger b^\dagger\right)|\Omega\rangle.
 \ee
Here, $b$ is an annihilation operator acting on the virtual modes as
follows
 \be
 b= \frac{1}{\sqrt{2}}(h+v)
 \ee
where
\be
h= \frac{c_L - i c_R}{2} e^{\frac{i \pi}{4}} \ \mr{and} \ v= \frac{c_U - i c_D}{2}. \label{eq:def-hv}
\ee
The corresponding CM $\gamma_1$ [Eq.~(\ref{eq:gamma1})] is
\begin{align}
A &= \left(\begin{array}{cc}
0&1-2\lambda\\
-1+2\lambda&0
\end{array}\right), \notag \\
B &=  \sqrt{\lambda - \lambda^2}\left(\begin{array}{cccc}
1&-1 &0&-\sqrt{2}\\
-1&-1&-\sqrt{2}&0
\end{array}\right), \notag \\
D &= \left(\begin{array}{cccc}
0&     1-\lambda     &-\frac{\lambda}{\sqrt 2}&-\frac{\lambda}{\sqrt{2}}\\
-1+\lambda&0&\frac{\lambda}{\sqrt 2}&-\frac{\lambda}{\sqrt{2}}\\
\frac{\lambda}{\sqrt 2}&-\frac{\lambda}{\sqrt 2}&0&1-\lambda\\
\frac{\lambda}{\sqrt 2}&\frac{\lambda}{\sqrt 2}&-1+\lambda&0
\end{array}\right) \label{eq:Chern-1}
\end{align}
We have sorted the Majorana mode operators as $e_1,e_2,c_{L},c_{R},c_U,c_D$.

Later on we will consider other states, topological or not, to illustrate the properties of the boundary theories. However, the family of states given here will be a central object of our analysis, since it already possesses all the basic ingredients. As it is evident from the definition, the fiducial state $\Psi_1$ in Eq.~\eqref{eq:Psi1ex} is an entangled state between the physical and one virtual mode, except for $\lambda=0,1$, whereas for $\lambda=1/2$ it is maximally entangled. It has certain symmetries, which will be of utmost importance to understand the topological features of the state $\Phi$ it generates. Explicitly,
 \begin{subequations}
 \label{eq:symmetries}
 \bea
 \left(\sqrt{\lambda} a^\dagger - \sqrt{1-\lambda}b \right) |\Psi_1\rangle &=&0,\\
 \left(\sqrt{1-\lambda} a + \sqrt{\lambda}b^\dagger \right) |\Psi_1\rangle &=&0,\\
 d_1 |\Psi_1\rangle &=& 0 \label{eq:dPsi0}
 \eea
 \end{subequations}
with
 \be
 \label{eq:d}
 d_1 =  \frac{1}{\sqrt{2}}(-h+v).
 \ee
The operators $a,b,$ and $d_1$ define three fermionic modes (one physical, and three virtual). Equations (\ref{eq:symmetries}) just reflect the fact that for a Gaussian state the physical mode can be entangled at most to one virtual mode, since we can always find a basis in which one virtual mode is disentangled. The latter is precisely the one annihilated by $d_1$. In fact, (\ref{eq:symmetries}) completely defines the state $\Psi_1$.

\subsubsection{Algebraic decay of correlations}

\begin{figure}
\begin{center}
\includegraphics[width=0.5\textwidth]{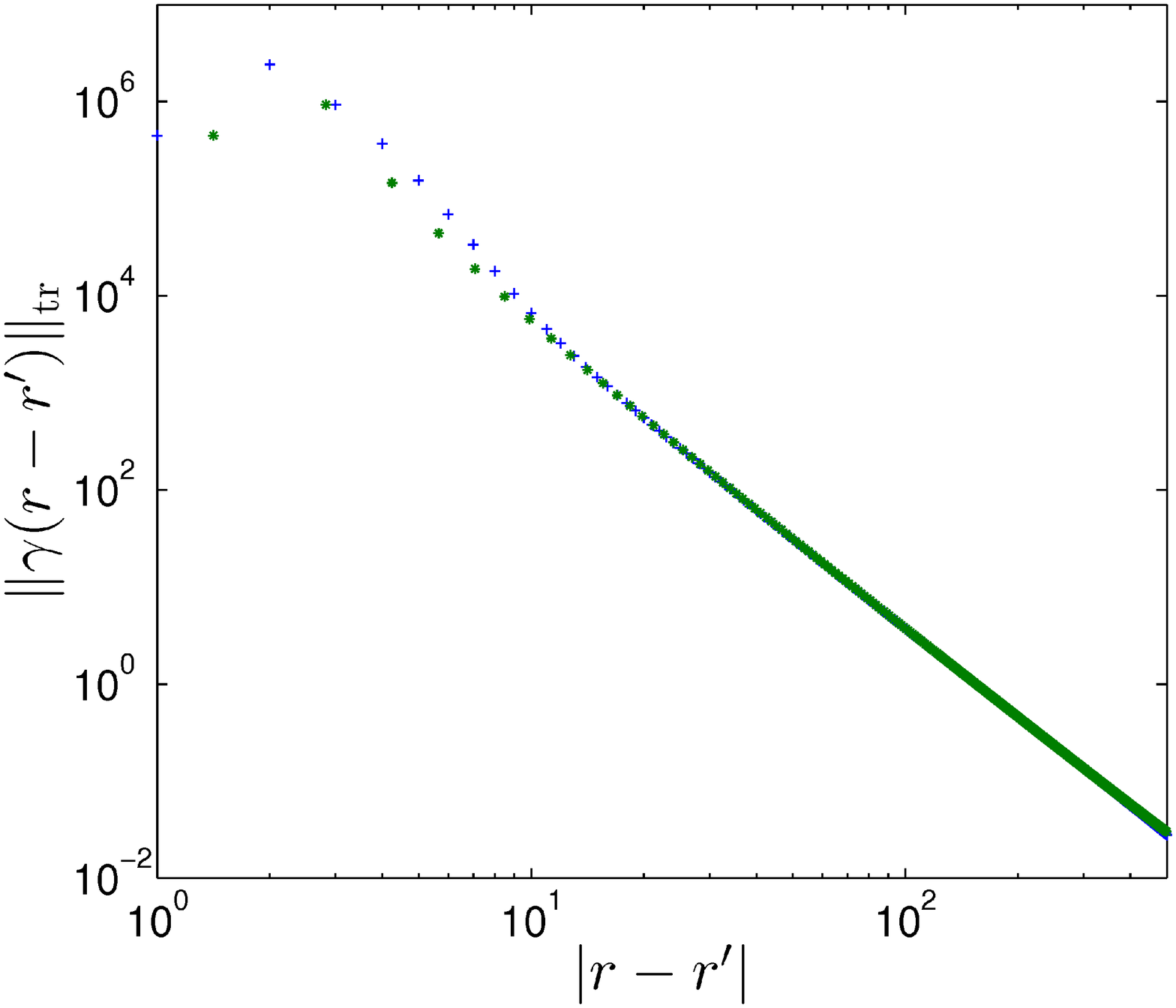}
\end{center}
\caption{Trace norm $\|\gamma(r-r')\|_\mr{tr}$ of the block of the covariance matrix $\gamma_{l,m}$ for $\lambda = 1/2$ in Eq.~\eqref{eq:Psi1ex} corresponding to sites $l$ and $m$ at positions $r$ and $r'$, respectively, as a function of distance $|r - r'|$. Blue crosses correspond to $r-r'$ aligned along the $x$- or $y$-axis (both lie on top of each other), and green stars indicate the case of $r-r'$ aligned along the diagonal of both axes. In this plot, $\gamma$ has been calculated for a $2000 \times 2000$ lattice and it decays as $\tfrac{1}{|r-r'|^{3.05}}$. The exponent converges to $3$ with increasing lattice size.
\label{fig:corr}}
\end{figure}

The correlation functions of the PEPS defined via Eq.~\eqref{eq:Psi1ex} decay algebraically, see Ref.~\onlinecite{Wah13} and Fig.~\ref{fig:corr}. This is most easily understood by considering the Fourier transform~\eqref{eq:G-param}. 
All $\hat d_i(k)$ are continuous for all $k$. 
However, the $\hat d_i(k)$ have a non-analyticity at $k = (0,0)$, where
the first derivatives of $\hat d_x$ and $\hat d_y$ are discontinuous.  For
instance, for $\lambda = 1/2$ in the example of Eq.~\eqref{eq:Psi1ex}, one
obtains
\begin{align}
\hat d_x(k) &= -\frac{2 \sin(k_x) (1 - \cos(k_y))}{3 - 2 \cos(k_x) - 2\cos(k_y) + \cos(k_x) \cos(k_y)}, \label{eq:dx}\\
\hat d_y(k) &= \frac{2 \sin(k_y) (1 - \cos(k_x))}{3 - 2 \cos(k_x) - 2\cos(k_y) + \cos(k_x) \cos(k_y)}, \label{eq:dy} \\
\hat d_z(k) &= \frac{1 - 2 \cos(k_x) - 2 \cos(k_y) + 3 \cos(k_x) \cos(k_y)}{3 - 2 \cos(k_x) - 2\cos(k_y) + \cos(k_x) \cos(k_y)}. \label{eq:dz}
\end{align}
At $k = (0,0)$ both the numerators and the common denominator are zero.
In Appendix~\ref{app:corr}, we show that due to this non-analycity, correlations in real space decay like the inverse of the distance cubed (up to possible logarithmic corrections).

\subsubsection{Frustration free Hamiltonian: fragility}

\begin{figure}[t]
\begin{center}
\includegraphics[width=0.4\textwidth]{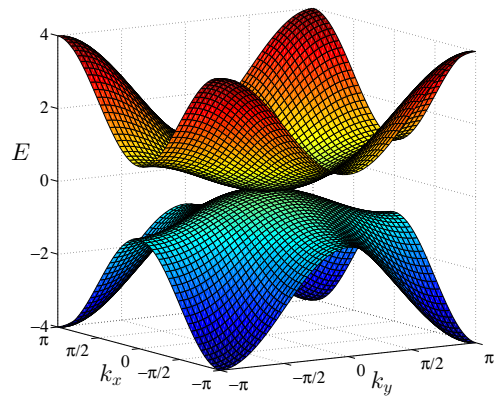}
\end{center}
\caption{Single-particle energy spectrum of the frustration free parent
Hamiltonian $\mc H_\mr{ff}$ of the GFPEPS defined via
Eq.~\eqref{eq:Psi1ex} for $\lambda = 1/2$.  The band-touching point is at
$k = (0,0)$.
\label{fig:parent-Ham}}
\end{figure}

The frustration free parent Hamiltonian for this model is obtained by
explicitly calculating the state $\Psi_{2,2}$ obtained when four $\Psi_1$
on a $2 \times 2$ plaquette are concatenated without closing the
boundaries in horizontal or vertical direction.  Thereafter, one
calculates the fermionic operator $a_{\Box}$, acting only on the physical level, which annihilates $\Psi_{2,2}$, $a_{\Box} |\Psi_{2,2}\rangle = 0$
(it turns out that exactly one such operator exists for any $\lambda \in (0,1)$). This can be done conveniently in the CM formalism. The parent Hamiltonian, $\mc H_\mr{ff}$, can then be obtained by setting 
\be
h_j(\lambda) \propto a_{\Box,j}^\dg(\lambda) \, a_{\Box,j}(\lambda)
\ee 
in Eq.~\eqref{eq:HPEPS}.
For $\lambda=1/2$, for instance, we have 
\begin{align} 
a_{\Box} &= e_{a,1,1} (2 + i) + e_{b,1,1} - e_{a,1,2} (1 + 2i) + i e_{b,1,2} - e_{a,2,1} \notag \\
&+ e_{b,2,1} (-2 + i) + i e_{a,2,2} + e_{b,2,2} (1 - 2i), 
\end{align}
where $e_{a,x,y}$ denotes the first physical Majorana mode located at the site with coordinates $(x,y)$ and $e_{b,x,y}$ the second one. The single-particle spectrum for that case is displayed in Fig.~\ref{fig:parent-Ham}. Note that there is a band-touching point at $k = (0,0)$, and thus this Hamiltonian is gapless and has a continuous many-body spectrum. That is, it is exactly two-fold degenerate for finite systems, and in the thermodynamic 
limit it possesses a continuous spectrum right on top of the ground state.

\begin{figure}
\begin{center}
\includegraphics[width=0.33\textwidth]{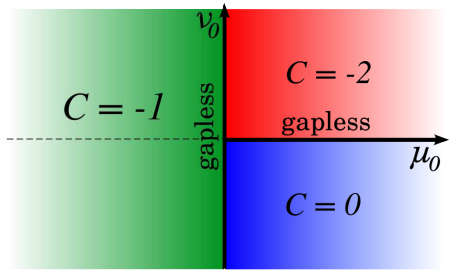}
\end{center}
\caption{Phase diagram of the perturbed Hamiltonian $\tilde{\mc H}_\mr{ff}(\lambda, \mu_0,\nu_0)$ (see text) for $\mu_0$, $\nu_0$ close to zero and $\lambda \in (0,1)$ arbitrary. The vertical gapless line corresponds to a quadratic band touching, whereas the horizontal gapless line ($\mu_0 > 0$) corresponds to four Dirac points. 
All other points in the phase diagram are gapped with the shown Chern numbers.
\label{fig:phases}}
\end{figure}

The frustration free Hamiltonian $\mc H_\mr{ff}$ does not have a protected chiral edge mode, as it is gapless in the bulk: Let us add a translationally invariant perturbation [with variable GFPEPS parameter $\lambda \in (0,1)$],
\begin{align}
&\tilde{\mc H}_\mr{ff}(\lambda,\mu_0,\nu_0) \notag \\
&= H_\mr{ff}(\lambda) - \tfrac{i}{4}\sum_{x,y} [\mu_0 e_{a,x,y} e_{b,x,y} + \nu_0 (e_{a,x+1,y} e_{b,x,y} \notag
\\ &- e_{a,x,y} e_{b,x+1,y} + e_{a,x,y+1} e_{b,x,y} - e_{a,x,y} e_{2,x,y+1})]
\end{align}
where $\mu_0, \nu_0 \in \mathbb{R}$. Note that only $\mu_0 = \nu_0 = 0$ corresponds to a GFPEPS ground state. After carrying out a Fourier transform, the Hamiltonian can be brought into the form 
\be 
\tilde{\mc{H}}_\mr{ff}(\lambda,\mu_0,\nu_0) = \sum_{i=x,y,z} \sum_k d'_i(k) (a_k^\dg, a_{-k}) \sigma_i \left(\begin{smallmatrix}
a_{k} \\
a_{-k}^\dg
\end{smallmatrix}\right)
\ee 
with $\sigma_i$ the Pauli matrices, the Chern number can be calculated via~\cite{Qi06}
\be 
C = \frac{1}{4 \pi} \int_{-\pi}^\pi \int_{-\pi}^\pi \hat d'(k) \cdot (\tfrac{\partial \hat d'(k)}{\partial k_x} \times \tfrac{\partial \hat d'(k)}{\partial k_y}) \, \mr{d} k_x \mr{d} k_y
\label{eq:Chern}
\ee
with $\hat d'(k) = \tfrac{d'(k)}{| d'(k)|}$.
Depending on the signs of the parameters $\mu_0$ and $\nu_0$, the Hamiltonian can be driven by infinitesimally small perturbations to gapped phases with Chern number $C = 0$ (trivial), $C = -1$ or $C = -2$ as shown in Fig.~\ref{fig:phases}. This phase diagram does not depend on the parameter $\lambda$ as long as $|\mu_0|$ and $|\nu_0|$ are sufficiently small. 
Hence, with respect to the frustration free Hamiltonian, the states defined by Eq.~\eqref{eq:Psi1ex} describe
critical points in the transition between different topological phases with Chern numbers $C = -2$ and $C = -1$ and a topologically trivial phase ($C = 0$). 

We conclude that the frustration free Hamiltonian is gapless and thus not topologically protected. Instead, it is at the critical point between free fermionic topological phases with different Chern numbers. 

\subsubsection{Flat band Hamiltonian: robustness}

Let us now consider the stablity of the flat band Hamiltonian $\mc
H_{\mr{fb}}$ against perturbations.  First, we will show analytically that
the Hamiltonian is robust even against long-ranged translationally
invariant perturbations; and second, we will demonstrate numerically the
stability against local disorder.  This shows that the Hamiltonian is
topologically protected and its Chern number is therefore a meaningful
quantity.

Let us first consider translational invariant perturbations where we
assume that the perturbation decays faster than $1/|r|^3$ in real space (with
$|r|$ the distance). Then, it can be shown (see, e.g., Ref.~\onlinecite{grafakos},
Proposition 3.2.12) that the perturbation $\mathcal H$ is differentiable
in Fourier space, and thus, the perturbed flat band Hamiltonian 
$\tilde{\mc{H}}_\mr{fb} = \mc{H}_\mr{fb} + \epsilon \mc H$ 
is differentiable as well.  Moreover, since the Fourier components of
$\mathcal H$ are uniformly bounded,
the gap of $\tilde{\mathcal H}_\mathrm{fb}$ stays open for sufficiently small
$\epsilon$.  Thus, the bands of $\tilde{\mathcal H}_\mathrm{fb}$ are a
smooth function of $\epsilon$, and thus, the Chern number cannot change
under sufficiently small perturbations.

\begin{figure}[t]
\begin{center}
\includegraphics[width=0.5\textwidth]{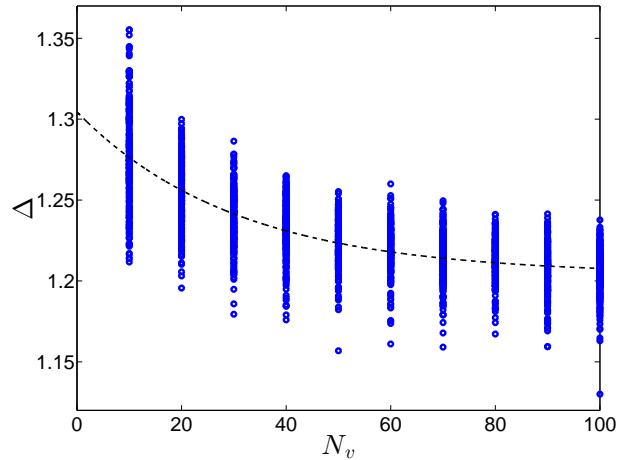}
\end{center}
\caption{Energy gap $\Delta$ of the flat band Hamiltonian $\mc H_\mr{fb}$ after the addition of disorder terms (see text). The Hamiltonian is defined on a torus of size $N_v \times N_v$ ($N_h = N_v$) and for each system size 225 random samples have been considered. The dashed line represents a fit with the function $f(N_v) = a \exp(-b N_v) + c$, which gives $a = 0.101\pm 0.005$, $b = 0.033\pm 0.004$ and $c = 1.204 \pm 0.003$ ($95 \%$ confidence intervals), i.e., the gap saturates at a value that is roughly $60 \%$ of the unperturbed gap. \label{fig:disorder}}
\end{figure}

Let us now turn towards the stability of $\mathcal H_\mathrm{fb}$ against
random disorder, which we have verified numerically. To this end, we randomly
added local disorder terms $\sum_{j} \mu_j a_j^\dg a_j$ ($\mu_j \in
[-1,1]$) to the flat band Hamiltonian for $\lambda = {1}/{2}$ defined on
an $N_v \times N_v$ torus ($N_h = N_v$) as a function of its length $N_v$.
In Fig.~\ref{fig:disorder} we plot the energy gap obtained for 225 random realizations
 for each system size $N_v$. As can be
gathered from the figure, its gap stays non-vanishing in the thermodynamic
limit, indicating that it is topologically protected against disorder.

To summarize, the gap of the flat band Hamiltonian $\mc H_\mr{fb}$ is topologically protected against the addition of on-site disorder and (small) translationally invariant perturbations whose hoppings decay faster than the inverse of the distance cubed. Its Chern number is $-1$.

\subsection{Boundary and Edge Theories}\label{sec:boundary-edge}

In Ref.~\onlinecite{cirac:peps-boundaries} a formalism was introduced for spin PEPS to map the state in some region ${\cal R}$ to its boundary. This bulk-boundary correspondence associates to each PEPS a boundary Hamiltonian, ${\cal H}^{\rm b}$, that acts on the virtual particles. The Hamiltonian faithfully reflects the properties of the original PEPS. In particular, for the toric code \cite{Kit03}, or the resonating valence-bond states \cite{And73}, that boundary Hamiltonian features their topological character \cite{schuch:topo-top}. In this Section we review that theory for GFPEPS and show how one can determine ${\cal H}^{\rm b}$ for GFPEPS.

Chiral topological insulators and superconductors, on the other hand, are characterized by the presence of chiral edge modes, featuring  robustness against certain bulk perturbations. Here, we also analyze how those features are reflected in ${\cal H}^{\rm b}$, as well as the relation of that Hamiltonian with that found for the toric code.

\subsubsection{Boundary Theories}

Given the GFPEPS $\Phi$, let us take a region ${\mc R}$ of the lattice, trace all the degrees of freedom of the complementary region, $\bar{\mc R}$, and denote by $\rho_{\mc R}$ the resulting mixed state. As it was shown in Ref.~\onlinecite{cirac:peps-boundaries}, $\rho_\mc{R}$ can be isometrically mapped onto a state of the virtual particles (or modes) that are at the boundary of the region ${\mc R}$. That is, there exists an isometry $\mc V_{\mc R}$, such that $\rho_\mc{R} = \mc V_{\cal R} \sigma_\mc{R} \mc V_{\cal R}^\dagger$, where $\sigma_\mc{R}$ is a mixed state defined on those virtual modes.

\begin{figure}[t]
\begin{center}
\includegraphics[width=0.22\textwidth]{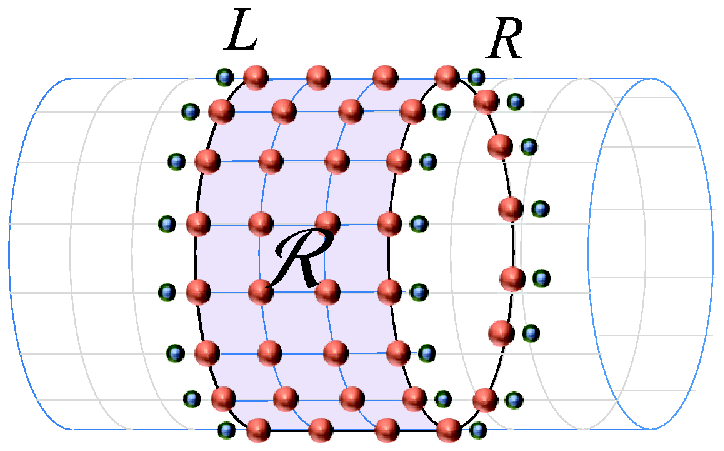}
\includegraphics[width=0.22\textwidth]{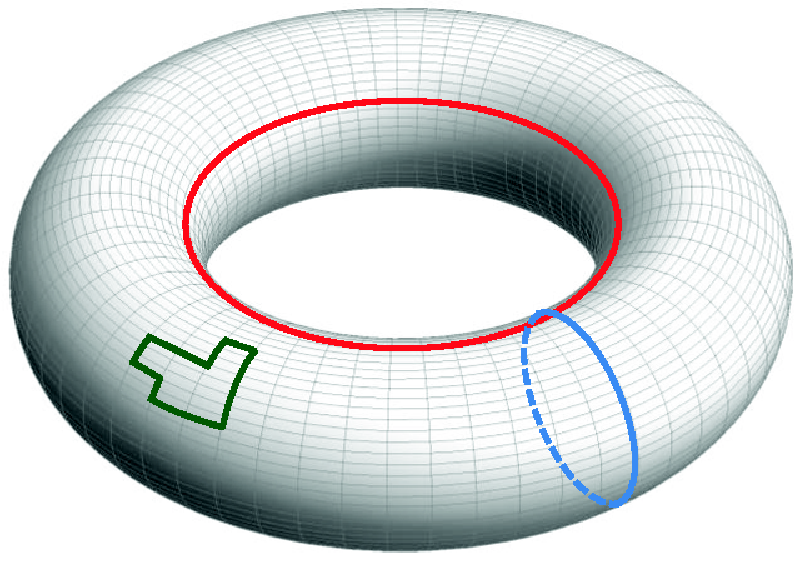}
\end{center}
\caption{Left: The state obtained after cutting out $N$ columns (region $\mc R$) from a translationally invariant GFPEPS is still translationally invariant in the vertical direction. Hence, it can be understood as being defined on a cylinder. The Majorana modes on the left and the right boundary (small blue balls) remain unpaired.
Right: Illustration of string operators. Those are defined as operators acting on the virtual Majorana modes lying on a closed string (such as the blue, red and green examples). The projection onto the final physical state $\Phi$ is carried out after applying one or more of those string operators.}
\label{fig:cylinder}
\end{figure}

Here we will take as region ${\mc R}$ a cylinder with $N$ columns, see Fig.~\ref{fig:cylinder}. There we have drawn the (red) physical fermions, as well as the (blue) virtual Majorana modes, as they appear in the construction explained above (Fig. \ref{Fig:PEPS}).

The state $\sigma_\mc{R}$ is Gaussian and is thus also characterized by a CM, which we will denote by $\Sigma_N$. In Sec.~\ref{sec:Detailed-Analysis} we will show how to determine it in terms of $\gamma_1$. Here, we just quote the results. We can write
 \be
 \sigma_\mc{R} = \frac{1}{Z_N} e^{-{\cal H}^{\rm b}_{N}},
 \ee
where
 \be
 {\cal H}_N^{\rm b} = -\frac{i}{4} \sum_{j,k} (H_N^{\rm b})_{j,k} c_j c_k, \label{eq:boundary-Ham}
 \ee
is the so-called boundary Hamiltonian, with $c_j$ the Majorana operators acting on the left and right boundaries, and $H_N^{\rm b}$ a $2N_v\times 2N_v$ antisymmetric matrix, given by
 \be
 \label{eq:HN}
 H_N^{\rm b}= 2 \arctan ( \Sigma_N).
 \ee
The spectrum of ${\cal H}^{\rm b}_N$ coincides with the so-called entanglement spectrum~\cite{Hal08}. Here we will be interested in the corresponding single-particle spectrum, i.e. that of $H^\mr{b}_N$.

Since ${\cal H}^{\rm b}_N$ is translationally invariant in the vertical direction, we can easily diagonalize it by using Fourier transformed Majorana modes. It is convenient to define
 \be
 \hat c_{k_y}= \frac{1}{\sqrt{N_v}} \sum_{y=1}^{N_v} e^{i k_y y} c_y,
 \ee
separately for the left and right virtual modes, so that ${\cal H}_N^{\rm b}$ displays a simple form in their terms. Here, the quasi-momentum is $k_y=2\pi n/N_v$, with $n=-{N_v}/{2}+1,\ldots,{N_v}/{2}$. Up to a factor of two, the operators $\hat c_{k_y}^\dagger = \hat c_{-k_y}$ fulfill canonical commutation relations for fermionic operators, $\{\hat c_{k_y},\hat c_{k_y'}^\dagger\}=2\delta_{k_y,k_y'}$, for $k_y\ne 0,\pi$. For $k_y=0,\pi$, they are Majorana operators (i.e., $\hat c_0^\dagger=\hat c_0$, and $\hat c_\pi^\dagger=\hat c_\pi$). This latter fact is crucial to understand the topological properties of the original state $\Phi$, as we will discuss in Sec.~\ref{sec:full-chi1}.

The single-particle spectrum (dispersion relation, since we have translational invariance) will be labeled by $k_y$. For the GFPEPS determined by Eq.~\eqref{eq:Psi1ex} for $\lambda \in (0,1)$ we will show that in the limit $N\to\infty$ one can write
 \be
 H_\infty^{\rm b} = \bigoplus_{k_y\ne 0,\pi} \left(\hat H_\infty^L(k_y) \oplus \hat H_\infty^R(k_y)\right) \oplus \hat H^{LR}_\infty(0) \oplus \hat H^{LR}_\infty(\pi) \label{eq:decouple-H}
 \ee
where $\hat H_\infty^{L}(k_y)$ and $\hat H_\infty^{R}(k_y)$ correspond to virtual fermionic modes on the left and right, respectively, which are decorrelated from each other. For $k_y=0$ and $k_y=\pi$, however, there is a single unpaired Majorana mode in each boundary. For the above family of chiral GFPEPS, the $k_y = 0$ Majorana modes
pair up, giving rise to an entangled state between the left and the right boundaries, which is why we obtain the structure of Eq.~\eqref{eq:decouple-H} for the single-particle boundary Hamiltonian.

The Chern number, $C$ (up to a sign), is given by the number of right-movers minus the number of left-movers on one of the boundaries. For the simple case considered in this Section, with one Majorana bond, $|C|=0,1$. For GFPEPS with more Majorana bonds, one can build the boundary Hamiltonian in the same fashion, as we will show in the next Section. In that case, the Chern number is determined ditto, but it may be larger than one.

\begin{figure}
\centering
\includegraphics[width=0.5\textwidth]{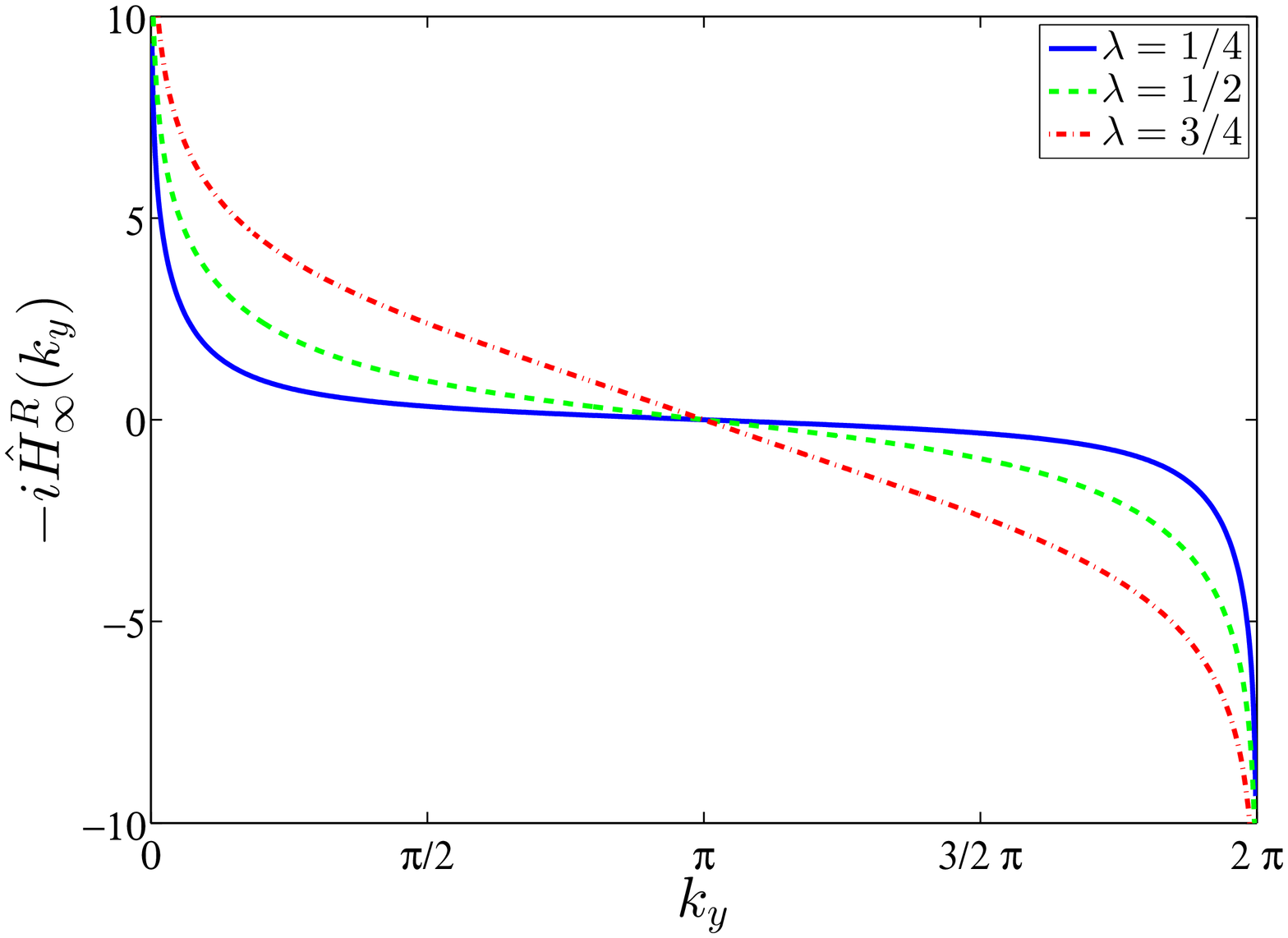}
\caption{
Dispersion relation corresponding to the right boundary Hamiltonian
for the chiral state defined via Eq.~\eqref{eq:Psi1ex}. 
We plot $-i \hat H^R_\infty(k_y)$ (which is a $1 \times 1$ matrix), for $\lambda = {1}/{4}$ (blue solid line), $\lambda = {1}/{2}$ (green dashed line) and $\lambda = {3}/{4}$ (red dash-dotted line) and $N \rightarrow \infty$. For convenience, we have plotted it for $k_y\in[0,2\pi)$. Note the divergence at $k_y = 0$, where there is a maximally entangled virtual Majorana pair between the left and the right boundary. The lines cross
the Fermi level from above at $k_y = \pm \pi$, thus $C = -1$.
\label{Gamma_R1}}
\end{figure}

In Fig.~\ref{Gamma_R1} we plot the single-particle dispersion relation of
the right boundary as a function of $k_y$, for the state generated by
(\ref{eq:Psi1ex}) for different values of $\lambda$ and $N\to\infty$ (we
will provide an analytical formula for that limit in
Sec.~\ref{sec:full-chi1}). It displays chirality, and the Chern number is
$-1$. The mode at $k_y=\pi$ has zero ``energy", indicating that the state
of the left and right Majorana modes with such a momentum are in a
completely mixed state. If we construct a fermionic operator using those
two modes, the boundary state $\sigma_\mc{R}$ at momentum $\pi$ has
infinite temperature, and thus is an equal mixture of zero and one
occupation. If we do the same with the modes at $k_y=0$, the opposite is
true, namely they are in a pure state (the vacuum mode of the fermion mode
built out of the two Majorana modes from the left and the right). Thus, as
anticipated, the left and right boundaries are in an entangled state,
which reflects the topological properties of the state. In
Sec.~\ref{sec:Detailed-Analysis} we will show that all the features
displayed by this example are intimately related.

\begin{figure}
\centering
\includegraphics[width=0.5\textwidth]{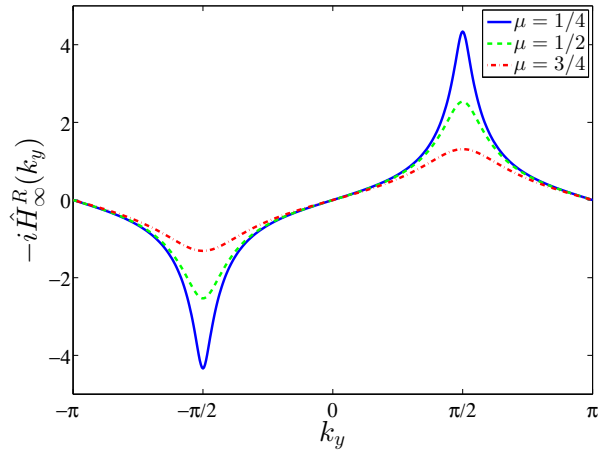}
\caption{Dispersion relation at the right boundary for the non-chiral
state defined via Eq.~\eqref{eq:D-chi0}. We plot $-i \hat H^R_\infty(k_y)$
(which is a $1 \times 1$ matrix), for $\mu = {1}/{4}$ (blue solid line),
$\mu =
{1}/{2}$ (green dashed line) and $\mu = {3}/{4}$ (red dash-dotted line)
and $N \rightarrow \infty$. It crosses the Fermi level twice with slopes
of different signs, hence $C = 0$.
\label{fig:chi0}}
\end{figure}

As a second example, we take a state that does not display any topological features. Its explicit form is given in Sec.~\ref{sec:C0-nontrivial}. The dispersion relation for the right boundary is shown in Fig.~\ref{fig:chi0}. Since the energy band of the boundary Hamiltonian does not connect the valence and conduction band for any $\mu$, the Chern number is zero. Furthermore, both at $k_y=0,\pi$ the "energy" vanishes, showing that the right and left boundaries are unentangled.

In Sec.~\ref{sec:examples}, we present further examples: We give an example of a GFPEPS displaying $C=2$. We also investigate the Chern insulator presented in Ref.~\onlinecite{Wah13}, provide a topologically trivial GFPEPS as well as the non-chiral state introduced in Ref.~\onlinecite{kraus:fPEPS}.

\subsubsection{Edge theories}

The definition of the boundary theory used above may look a bit artificial; the Hamiltonian ${\cal H}_N^{\rm b}$ does not generate any dynamics, but is just the logarithm of the density operator, and thus comes from the interpretation of the boundary operator as a Gibbs state. However, it is well known \cite{fidkowski:freeferm-bulk-boundary} that for free fermionic (i.e. Gaussian) states, its spectrum is intimately related to the one of another Hamiltonian that indeed generates the dynamics at the physical edges of the system in question. In the PEPS representation, there is a way of constructing such an edge Hamiltonian \cite{Shuo}, which we review here and we explicitly illustrate such a relation.

Let us consider the flat band Hamiltonian (\ref{eq:flat}), but in the case of a cylinder with open boundary conditions. For that, we restrict the sum in Eq.~\eqref{eq:flat} to the modes that correspond to region ${\cal R}$ (the cylinder in Fig.~\ref{fig:cylinder}), and denote by ${\cal H}_{\cal R}$ the corresponding Hamiltonian. The state $\Phi_N$ (see Fig. \ref{Fig:PEPS}f) has extra (virtual) modes, which we can project onto an arbitrary state, say $\phi_v$. The energy (in absolute value) of the resulting state will typically be much smaller than the gap of the system on the torus. Thus, there is a subspace spanned by all the states resulting from this construction with a low energy. By choosing a set of linearly independent vectors $\phi_v$, and orthonormalizing the resulting state, we can project ${\cal H}_{\cal R}$ onto that subspace. This is precisely the procedure given in Ref.~\onlinecite{Shuo}, and the resulting Hamiltonian, which has as many degrees of freedom as there are virtual Majorana modes, is the edge Hamiltonian, ${\cal H}_N^{\rm e}$. We now write
 \be
 \label{eq:Hedge}
 {\cal H}_N^{\rm e} = -\frac{i}{4} \sum_{l,m} (H_N^{\rm e})_{l,m} c_l c_m,
 \ee
and in Sec.~\ref{sec:edge} we show that one obtains that $H_N^{\rm e}=\Sigma_N$. Thus, up to a scale transformation (cf. Eq.~\eqref{eq:HN}), we see that the edge Hamiltonian is nothing but the boundary Hamiltonian, whenever we take the flat-band Hamiltonian as the parent Hamiltonian of our GFPEPS. This agrees with the statement of Ref.~\onlinecite{fidkowski:freeferm-bulk-boundary}, and indicates that our results on the boundary Hamiltonian can be translated to the edge Hamiltonian constructed in the outlined way.

\subsection{Symmetries, degeneracy, and Topological Entropy}\label{sec:symm-top}

Here, we will first briefly review how the topological properties of PEPS in spin systems are reflected in the symmetries of the corresponding fiducial state $\Psi_1$. Then we will show that for the GFPEPS considered in previous subsections, a similar behavior is present.

\subsubsection{Spins}

For PEPS in spin systems, all the properties are encoded in the single tensor which is used to build the state. In the language used in this paper, this tensor is equivalent to $\Psi_1$, since it is given by its coefficients in a basis. In particular, for topological states like the double models \cite{schuch:topo-top}, there exist operators $U_g$, where $g$ is an element of a group $\mc G$ and $U_g$ a unitary representation of it, acting on the virtual particles which leave $\Psi_1$ invariant. Those operators can be concatenated to string operators defined on the virtual modes on the boundary, so that for any state appearing during the construction of the PEPS $\Phi$, there exist other operators fulfilling the same property. Those operators can be built starting out from $U_g$ in a systematic way. This implies that for any region ${\cal R}$, there exists operators $U_g$ acting on the virtual particles at the boundary, such that
 \be
 \label{Eq:Ug}
U_{g} \sigma_\mc{R} = \sigma_\mc{R} U_g = \sigma_\mc{R}.
 \ee
For double models the operators $U_g$ can be written as products of operators acting on each of the virtual particles of the boundary.

From Eq. (\ref{Eq:Ug}) it follows that $\sigma_{\cal R}$ is supported on a proper subspace of the virtual system, that corresponding to the eigenvalue 1 of all $U_g$, i.e.,
\be
 \label{Eq:sigmasym}
 \sigma_\mc{R} = \frac{1}{Z_N} P e^{-{\cal H}^{\rm b}_N} P.
 \ee
Here $P$ is a non-local operator which projects onto that subspace. This fact has two consequences: (i) the zero R\'enyi entropy (which is the logarithm of the dimension of that subspace) does not coincide with the logarithm of the dimension of the Hilbert space of the virtual particles on the boundary of ${\cal R}$; (ii) there is a non-local constraint on the boundary and edge Hamiltonian. Those two features are thus related to the topological character of the PEPS. Note that (i) may also imply in some cases that there is a correction to the area law, what is usually called the topological entropy. That is, the von Neumann entropy of $\sigma_\mc{R}$ scales like the number of virtual particles on the boundary of ${\cal R}$ minus a universal constant, which is directly related to the topological properties of the model under study. The property (ii) acts as a superselection rule in the boundary and edge theories, since any perturbation in the bulk will not change that subspace. Additionally, in the spin lattices studied in Ref.~\onlinecite{schuch:topo-top}, ${\cal H}_N^{\rm b}$ is local (contains hoppings that decay exponentially with the distance) whenever the frustration free parent Hamiltonian of the state $\Phi$ is gapped.

Another consequence of (\ref{Eq:Ug}) is apparent if we take a PEPS defined on the torus. Then, we can attach different string operators $U_g$ and $U_{g'}$ around the two different cuts of the torus (see Fig.~\ref{fig:cylinder}, right). This means that during the construction of the PEPS, we apply those operators to the virtual particles at the position where the strings appear before applying the projections $\omega$ and $\omega'$. Because of the symmetry, those string operators can be moved without changing the state. However, they cannot be discarded given the topology of the torus. The states for each pair of $U_g$ and $U_{g'}$ are ground states of the parent, frustration free Hamiltonian of the PEPS as well, and for some particular $g,g'$ they are linearly independent. Thus, that Hamiltonian is degenerate and in fact all its ground states can be generated by applying the string operators on circles around the torus. Furthermore, anyonic excitations can be understood as the extreme points of open strings, and the braiding properties related to the group $\mc G$.

\subsubsection{Fermionic systems}

Now we show that an analogous phenomenon is present in our chiral topological models. That is, as PEPS, they also possess a symmetry in $\Psi_1$ which is inherited for larger regions, and that gives rise to properties (i) and (ii).
Besides that, the parent Hamiltonian ${\cal H}_{\rm ff}$ is degenerate on the torus, and the different ground states can be obtained by attaching to the virtual modes string operators around the torus. The strings can be deformed, without changing the state. However, there are some differences, too. First of all, the von Neumann entropy of $\sigma_\mc{R}$ does not display a universal correction, which we attribute to the long-range properties of the parent Hamiltonian $\mc H_\mr{fb}$ of the state $\Phi$ (see Refs.~\onlinecite{Wah13} and~\onlinecite{Dub13}). For the same reason, the hoppings in ${\cal H}^\mr{b}_N$ decay according to a power law. Furthermore, the ground-state subspace of the parent Hamiltonian, ${\cal H}_{\rm ff}$, is doubly degenerate on the torus, and some topologically inequivalent string configurations give rise to the same state.

Let us consider any region ${\cal R}$, and denote by $\Psi_{\cal R}$  the state obtained by projecting all the virtual modes within region ${\cal R}$ onto the state generated by $\omega_{jn}$ or $\omega'_{jn}$, as they appear in the PEPS construction. We arrive at a state of the physical modes in ${\cal R}$ and the virtual ones sitting at the boundary of $\mc R$. For instance, if we take as ${\cal R}$ a cylinder with $N$ columns, the state $\Psi_{\cal R} =\Phi_N$ (see Fig.~\ref{fig:cylinder}). We can write
 \be
 |\Phi\rangle = \langle \omega_{\partial \mc R, \partial \bar{\mc R}} |\Psi_{\cal R},\Psi_{\bar{\cal R}}\rangle,
 \ee
where $\omega_{\partial \mc R, \partial \bar{\mc R}}$ projects out all the virtual modes at the boundaries of ${\cal R}$ and its complement $\bar{\cal R}$.

If a contour ${\cal C}$ encloses a connected region ${\cal R}$, for chiral GFPEPS with one Majorana bond, there is a fermionic operator $d_\mc{C}$ such that
 \be
 \label{eq:dC}
 d_{\cal C} |\Psi_{\cal R}\rangle =0.
 \ee
For any contour, we will say that the state
 \be
 |\Phi_{\cal C}\rangle = \langle \omega_{\partial \mc R, \partial \bar{\mc R}} | d_{\cal C}| \Psi_{\cal R},\Psi_{\bar{\cal R}}\rangle,
 \ee
is a GFPEPS with a string along the contour ${\cal C}$. In Sec.~\ref{sec:symmetry-GS} we will show how this string operator can be deformed continuously for a chiral GFPEPS without changing the state we are building. However, if a contour wraps up around one of the sections of the torus, we cannot get rid of it by continuous deformations.

Let us denote by ${\cal C}_{h,v}$ contours wrapping the torus horizontally and vertically, respectively.
We show in Sec.~\ref{sec:symmetry-GS}
that if we build the family of chiral GFPEPS starting out from $\Psi_1$ according to Eq.~\eqref{eq:Psi1ex}, we obtain $\Phi=0$ after the last projection. However, the states obtained if we add a certain string along any of those contours coincide, $\Phi_{{\cal C}_h} \propto \Phi_{{\cal C}_v}$, and in the following that is the state that we will consider. We also show that if we insert string operators along the two contours $\mc{C}_h$ and $\mc{C}_v$,
the state $\Phi_{{\cal C}_h,{\cal C}_v}$ we obtain is orthogonal to the previous one, but it is also a ground state of ${\mc H}_{\rm ff}$.

The frustration free Hamiltonian has certainly very interesting properties, although we cannot determine them unambiguously given our results. It is
not only at a quantum phase transition point between free fermionic (gapped) phases with Chern numbers $C = 0, -1$ and $-2$, but it furthermore carries features of states described by PEPS with long-range topological order: Its ground state manifold is obtained by inserting strings along the non-trivial loops of the torus. Hence, our results also allow to interpret the local parent Hamiltonian as being at the edge of a topologically ordered interacting phase.

The existence of the operators $d_{\cal C}$ in Eq.~\eqref{eq:dC} for any
simply connected region $\mc R$ has another important consequence. It
follows that we can build a unitary operator $U= \Id-2 d_{\mc C}^\dagger
d_{\mc C}$ such that Eq.~\eqref{Eq:Ug} is fulfilled for the boundary operator.
As a consequence, we also have Eq.~\eqref{Eq:sigmasym} with $P = \Id - d_{\mc
C}^\dagger d_{\mc C}$. Note that in our case $\mc G = \mathbb{Z}_2$ is
represented by $\{\Id, U\}$. Thus, we conclude that the properties of the
previous paragraph (i) (topological correction to zero R\'e{}nyi entropy)
and (ii) (non-local constraint on boundary and edge Hamiltonian) are
fulfilled as in the standard PEPS case. Note that if $\mc R$ lies on a
cylinder as in
Fig.~\ref{fig:cylinder}, we can also give the interpretation that, as in
the case of a Majorana chain, there are two Majorana modes at the
boundaries building a fermionic mode in the (pure) vacuum state. As a
consequence, we can write $\sigma_\mc{R}$ for the cylinder as in Eq.~\eqref{Eq:sigmasym},
where $P$ projects onto the subspace where that mode
is in the vacuum.

\begin{figure}
\centering
\includegraphics[width=0.5\textwidth]{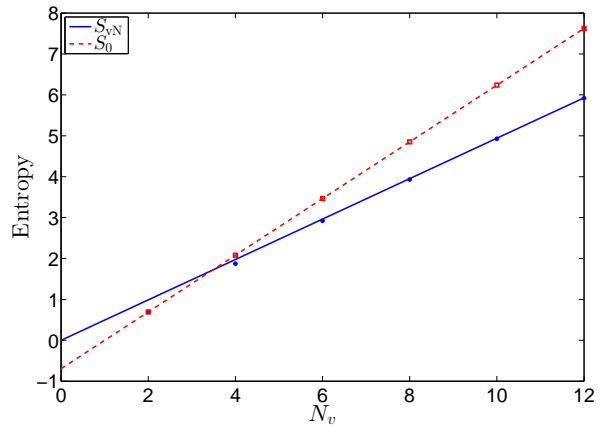}
\caption{Von Neumann entropy $S_\mr{vN}$ (blue circles) and zero R\'enyi entropy $S_0$ (red squares) versus length of the cylinder in vertical direction, $N_v$, for the example given in Eq.~\eqref{eq:Psi1ex} for $\lambda = 1/2$. The lines indicate linear fits, which have been done for 31 data points distributed equally between $N_v = 4000$ and $N_v = 4600$. These yield $S_\mr{vN} = 0.49401 N_v - 2.0 \cdot 10^{-7}$ (the constant converges to zero for intervals containing increasing $N_v$'s)  and $S_0 = \mathrm{ln}(2) N_v - \mathrm{ln}(2)$.}
\label{fig:entropy}
\end{figure}

In addition to the zero R\'e{}nyi entropy $S_0(N_v)$, we have also
numerically computed the von Neumann entropy $S_{\mathrm{vN}}(N_v)$ for
the example given in Eq.~\eqref{eq:Psi1ex} for $\lambda=1/2$. Both are shown in Fig.~\ref{fig:entropy} as
a function of $N_v$: While the zero R\'e{}nyi entropy clearly shows a
topological correction of $\ln(2)$, similar to the toric code model, the
von Neumann entropy does not exhibit such a correction. As we prove in
Appendix~\ref{app:euler-maclaurin}, this follows from the fact
that $S_{\mathrm{vN}}(N_v)$ forms a discrete approximation to the integral
over the modewise entropy, which is sufficiently smooth in $k_y$ to ensure
fast convergence. The same happens for all R\'enyi entropies $S_\alpha$ except for $\alpha = 0$. This is 
consistent with the result of, e.g., Ref.~\onlinecite{Fla09} (where, however, only non-chiral topological
states have been considered).

In order to further investigate the topological properties of our model,
we have also computed the so-called \textit{momentum
polarization}~\cite{mom-pol} (see also Refs.~\onlinecite{Zha12}, \onlinecite{Cin13} and \onlinecite{Zal13}), 
which measures the topological spin and
chiral central charge of an edge~\cite{note-Zaletel}.  For a state
$\ket\varphi$ on a cylinder, it is defined as $\mu(N_v)=\bra\varphi
T_L\ket\varphi$, where $T_L$ is the translation operator on the left half
of the cylinder.  It can thus be rephrased in terms of the (many-body)
entanglement spectrum $\lambda_\ell$ of the left half, which implies that
in the framework of PEPS, it can be naturally evaluated on the virtual
boundary between the two parts of the system. In particular, for GFPEPS it
can be expressed as a function of the (single-particle) spectrum of the
boundary Hamiltonian $H^{\mr b}_N$, as shown in Fig.~\ref{Gamma_R1}.
In Ref.~\onlinecite{mom-pol}, it has been shown that (for systems with CFT edges)
$\mu(N_v)=\exp(-\alpha N_v-2\pi i \tau/N_v+\dots)$, with a non-universal
$\alpha$, and a universal $\tau$ which carries information about the
topological properties of the system.  In
Appendix~\ref{app:euler-maclaurin}, we prove that for GFPEPS, $\mu(N_v)$
exactly follows the above behavior, and $\tau$ is indeed universal:
Remarkably, it only depends on whether the boundary Hamiltonian exhibits a
divergence, but not at all on its exact form. In particular, for our
example, we analytically obtain a $\tau$ which corresponds to a chiral
central charge of $c=1/2$, independently of $\lambda$, in accordance with
expectations.

\begin{figure}
\centering
\includegraphics[width=0.5\textwidth]{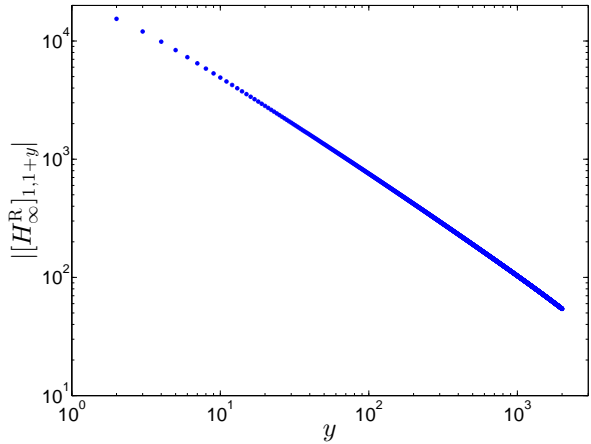}
\caption{Hopping amplitudes $|[H_\infty^{R}]_{1,1+y}|$ of the boundary Hamiltonian of the example given in Eq.~\eqref{eq:Psi1ex} for $\lambda = 1/2$ versus $y$. For large $y$ the curve has an inclination of $-1$ (on the log-log scale) indicating a decay as $1/y$, consistent with the fact that the logarithmic correction gets less important. The plot has been generated for $N \rightarrow \infty$ and $N_v = 2 \cdot 10^4$ sites in vertical direction.}
\label{fig:Ham}
\end{figure}

Finally, an interesting behavior is also observed for the boundary Hamiltonian,
Eq.~\eqref{eq:boundary-Ham}, for $N \rightarrow \infty$. On the right
boundary we perform the Fourier transform to position space
$[H_\infty^R]_{n,m}$. Then, for $y \gg 1$, $|[H_\infty^{R}]_{n,n+y}|
\propto \log(y)/y + \mathcal{O}(1/y)$, see Appendix~\ref{app:poly-hopping}. Thus, the decay is not exponential
as it is the case for gapped phases in spins, but follows a power law. We
plot the hopping amplitudes $|[H_\infty^{R}]_{1,1+y}|$ of the above chiral
family for $\lambda = 1/2$ in Fig.~\ref{fig:Ham}.

\section{Detailed Analysis}\label{sec:Detailed-Analysis}

In this Section, we provide a detailed derivation of the boundary and edge
theories for GFPEPS. We start in Sec.~\ref{sec:Detailed-Analysis}~A by
formally introducing GFPEPS, and then provide the derivation of boundary
theories (\ref{sec:Detailed-Analysis}~B) and edge theories
(\ref{sec:Detailed-Analysis}~C) for GFPEPS.

\subsection{GFPEPS}\label{sec:construct-GFPEPS}

The construction of GFPEPS given in Sec.~\ref{Sec:Intro-GFPEPS} can be
defined more generally for $f$ physical fermionic modes per site and
$\chi$ Majorana bonds between them. We again start with an $N_h \times
N_v$ lattice, now with $\chi$ left, right, up and down Majorana modes per
site, $c_{j,L,\kappa}, c_{j,R,\kappa}, c_{j,U,\kappa}$ and
$c_{j,D,\kappa}$, respectively, where $\kappa = 1, \ldots, \chi$ is the
index of the Majorana bonds. At each site $j$ they are jointly with the
physical modes in a Gaussian state as in Eq.~\eqref{eq:Gaussian-psi1}. The
procedure to construct the GFPEPS is the same, except that there are
now $\chi$ virtual bonds between any two neighboring sites, i.e., here we have
to set 
\begin{subequations}
 \label{eq:omegas}
 \bea
 \omega_{jn} &=& \frac{1}{2^\chi} \prod_{\kappa=1}^\chi (1 + i c_{j,D,\kappa} c_{n,U,\kappa}),\\
 \omega'_{jn} &=& \frac{1}{2^\chi} \prod_{\kappa=1}^\chi (1 + i c_{j,R,\kappa} c_{n,L,\kappa}),
 \eea
 \end{subequations}
for the vertical and horizontal bonds, respectively. We will again denote
by $\langle \omega_{jn}|$ ($\langle \omega_{jn}'|$) the map which applies
$\omega_{jn}$ ($\omega_{jn}'$) and discards the corresponding virtual
modes.  For simplicity, in the following we will call the states generated
by the operators~\eqref{eq:omegas} out of the vacuum {\em maximally
entangled states}.  The remaining procedure of how to concatenate them is
the same as in Sec. \ref{Sec:Intro-GFPEPS}, cf.~also Fig.~\ref{Fig:PEPS}.

In this scenario the CM is likewise given by Eq.~\eqref{eq:gamma1}, just that
the blocks $A$, $B$, and $D$ now have sizes $2 f \times 2 f$, $2f
\times 4 \chi$, and $4 \chi \times 4 \chi$, respectively. We are
interested in how to determine the CMs of the different states $\Psi_{N_v}$,
$\Phi_N$, and $\Phi$ involved in the construction of the GFPEPS.
It is based on two operations (see Fig.~\ref{Fig:PEPS}): \emph{(i)} building
the state of $l+m$ modes out of two states of $l$ and $m$ modes, respectively, i.e.,
taking tensor products; \emph{(ii)} projecting some of the modes onto some
state (given by $\omega$ or/and $\omega'$). Apart from that, we
will also extensively use in other parts of this paper: \emph{(iii)} tracing out some modes.

In terms of the CM, those operations are performed as follows
\cite{Bra05}. \emph{(i)---joining two systems}: the resulting CM is
a $2\times2$ block diagonal matrix, where the two diagonal  
blocks are given by the CM of the state of the $l$ and $m$ modes,
respectively. 
The operation
\emph{(ii)---projecting out some of the modes}, is slightly more elaborate.
Let us consider an arbitrary state (pure or mixed) with CM $\gamma_1$ with
blocks $A,B,D$ [as in Eq.~\eqref{eq:gamma1}], and we want to project
the last modes (corresponding to matrix $D$) onto some other state
of CM $\omega$. The resulting CM is given by~\cite{Bra05,kraus:fPEPS}
\be
\gamma_1' = A + B(D+\omega^{-1})^{-1}B^\top. \label{eq:proj-ME}
\ee
Typically, we will have to project onto the states generated by (\ref{eq:omegas}). Their CM is very simple,
 \begin{equation}
    \label{eq:max-ent-def}
    \omega = \left(\begin{matrix}0&-\openone\\
    \openone&0\end{matrix}\right).
 \end{equation}
Finally, in the case of operation \emph{(iii)---tracing out some of the
modes}, one simply has to take the corresponding subblock of the CM. This
block is the CM of the reduced state. For instance, if one traces out the
physical degrees of freedom of the state described by the CM
\eqref{eq:gamma1}, one obtains a (generally mixed) state defined on the
virtual degrees of freedom with CM $D$.  Conversely, one can also build
the CM of a purification of a mixed state $D$, as
\be
 \left(\begin{array}{cc}
 -D&\sqrt{\Id + D^2}\\
  -\sqrt{\Id + D^2}&D
 \end{array}\right) \ .
 \ee
Operations \textit{(i)} and \textit{(ii)} can be used to build the CM of
the state $\Phi$ out of that of $\Psi_1$. In this Section we will
extensively use all presented operations to construct the boundary and
edge states and Hamiltonians.

\subsection{Boundary Theories}

\subsubsection{Boundary Theories in GFPEPS%
\label{sec:bnd-theories-in-gfpeps}}

\begin{figure}
\includegraphics[width=0.48\textwidth]{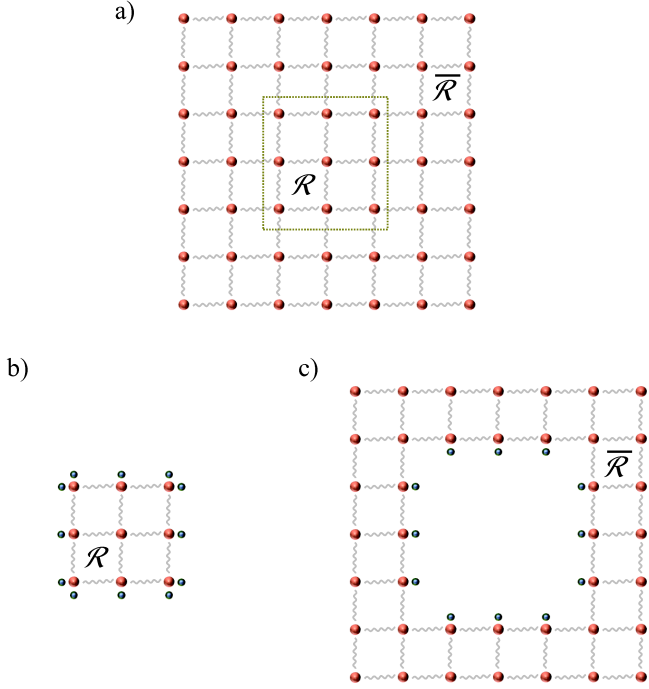}
\caption{(a) Partition of a lattice with a GFPEPS defined on it into a
region $\mc R$ and its complement $\bar{\mc R}$. 
The blue balls represent 
the virtual Majorana modes and the big red balls (connected
by wavy lines indicating the prior projection on maximally entangled pairs of virtual Majorana modes) the physical fermions.
 (b) After cutting the bonds as indicated in (a), the
region $\mc R$ has unpaired virtual indices at its boundary which are
collectively denoted by $\partial \mc R$. (c) The same is true for the
(inner) boundary of region $\bar{\mc R}$, whose virtual degrees of freedom
are denoted by $\partial \overline {\mc R}$.\label{fig:peps-bipartition}}
\end{figure}

We will now show how to derive boundary theories in the
framework of fermionic Gaussian states, by only using their description in
terms of CMs rather than the full state.  We consider a bipartition of the
PEPS $\Phi$ into two regions $\mc R$ and $\bar{\mc R}$
(Fig.~\ref{fig:peps-bipartition}) and are interested in the reduced state
$\rho_{\cal R}={\rm tr}_{\bar{\cal R}} (|\Phi\rangle\langle\Phi|)$. We
proceed as follows. First, we consider the states where all virtual
bonds \textit{within} those regions have been projected out, leaving only
virtual particles at the boundaries of those regions (which are denoted by $\partial
\mc R$ and $\partial \bar{\mc R}$, respectively) unpaired. Hence, we are
left with two states, which are defined on the physical degrees of freedom
of these regions plus the virtual degrees of freedom of the respective
boundaries (see Fig.~\ref{fig:peps-bipartition}b,c). We define their CMs as
\begin{equation} \label{eq:Gamma-R-Rbar}
\Gamma=\left(\begin{matrix}
L&F\\-F^\top & G\end{matrix}\right) \mbox{\ \ and\ \ }
\bar\Gamma=\left(\begin{matrix} \bar L&\bar F\\-\bar F^\top & \bar
G\end{matrix}\right)\ ,
\end{equation}
respectively, where the first (second) block corresponds to the physical
(virtual) degrees of freedom. The whole \mbox{GFPEPS} $\Phi$ could be obtained by
pairwise projecting their virtual degrees of freedom on maximally
entangled states, and thus, according to Eq.~\eqref{eq:proj-ME}, its CM is
 \be
 \gamma =
 \left(\begin{array}{cc}
 L&0\\
 0&\bar L
 \end{array}\right) +
 \left(\begin{array}{cc}
 F&0\\
 0&\bar F
 \end{array}\right)
 \left(\begin{array}{cc}
 G&\Id\\
 -\Id&\bar G
 \end{array}\right)^{-1}
 \left(\begin{array}{cc}
 F&0\\
 0&\bar F
 \end{array}\right)^\top. \label{eq:CM-GFPEPS}
 \ee
The CM of $\rho_\mc{R}$ is given by the (1,1) block of
Eq.~\eqref{eq:CM-GFPEPS}, that is
\be
\gamma_\mc{R} = L + F(G + \bar G^{-1})^{-1} F^\top. \label{eq:CM-R}
\ee

As explained in Sec.~\ref{sec:boundary-edge}, we are interested in a state
$\sigma_{\partial \mc{R}}$ defined on the virtual degrees of freedom located on
$\partial {\cal R}$, which is isometric to $\rho_\mc{R}$. Naively, one could think
that its CM is given by the (2,2) block of $\Gamma$, i.e., $G$, which
corresponds to a reduced state acting on that boundary. However, this is
not the case in general, since the state described by the CM $G$ is usually not isometric
to $\rho_\mc{R}$. As outlined in Ref.~\onlinecite{cirac:peps-boundaries},
$\sigma_{\partial \mc{R}}$ is given by a symmetrized version which takes into account
$\partial {\cal R}$ and $\partial \bar{\cal R}$. In fact, we can construct $\sigma_{\partial \mc{R}}$
by first finding
the appropriate purification of $\rho_{\mc{R}}$, and then tracing the
physical modes. We will carry out that task in two steps. First, we will
conveniently rotate the basis of the physical Majorana modes in
region $\mc R$ and afterwards truncate the redundant degrees of freedom
(projection). Both taken together correspond to the application of an isometry on $\rho_\mc{R}$.

We start with an orthogonal basis change in the basis of physical Majorana
modes $\{e_l\}$ in region $\mc R$. The new ones are given by an orthogonal
matrix $M$,
\begin{equation}
\label{eq:Q-trafo}
e_m' = \sum_{l} M_{m,l} e_l\ .
\end{equation}
This obviously does not change
the spectrum of $\rho_\mc{R}$.  By performing this basis change, the CM
$\Gamma$ gets modified to
 \[
 \Gamma' =
 \left(\begin{array}{cc}
 M&0\\
 0&\Id
 \end{array}\right)
 \left(\begin{array}{cc}
 L&F\\
 -F^\top&G
 \end{array}\right)
 \left(\begin{array}{cc}
 M^\top&0\\
 0&\Id
 \end{array}\right).
 \]
Note that this CM corresponds to a pure state, as $\Gamma$ does. We choose
$M$ in such a way that $\Gamma'$ decouples into a purification of the
virtual state and a trivial part on the remaining physical level. This is
always possible if the region $\mc R$ contains more degrees of freedom
than $\partial \mc R$ and can be done practically by using a singular value
decomposition of $F$. Then,
\be
 \Gamma' = \left(\begin{array}{ccc}
 Z&0&0\\
 0&-G&\sqrt{\Id + G^2}\\
 0&-\sqrt{\Id + G^2}&G
 \end{array}\right), \label{eq:P-CM}
 \ee
where $Z$ is the CM of a pure state defined on the physical level and the
remaining non-trivial part of $\Gamma'$ corresponds to a purification of
$G$ (note that the first and second block correspond to the physical
degrees of freedom and only the third block to the virtual ones). We
discard the decoupled physical part and project the virtual degrees of
freedom (together with those of region $\bar{\mathcal R}$, given by $\bar
G$) on the maximally entangled state. This yields the relevant part of
Eq.~\eqref{eq:CM-R}, which is the CM of $\sigma_\mc{R}$
\begin{equation}
    \label{eq:symmetrized-es}
 \Sigma_N = -G + \sqrt{\openone + G^2} (G+\bar G^{-1})^{-1}
    \sqrt{\openone + G^2}
 \ ,
\end{equation}
which is defined on the modes at the boundary. (We denote it by
$\Sigma_N$, since $\mc R$ will be typically taken to lie on a cylinder,
cf.~Fig.~\ref{fig:cylinder}, with $N$ columns. However,
Eq.~\eqref{eq:symmetrized-es} is true for any bipartition $\mathcal R$,
$\bar{\mathcal R}$.)

In order to obtain the boundary Hamiltonian $\mc H_N^{\rm b} =
-\tfrac{i}{4}\sum_{l,m} [H_N^{\rm b}]_{l,m} c_l c_m$, which reproduces the
entanglement spectrum, we can then use the relation
$H_N^\mathrm{b} = 2\arctan(\Sigma_N)$, Eq.~(\ref{eq:HN}). Note that for
$G=\bar G$, Eq.~(\ref{eq:symmetrized-es}) yields a trivial entanglement
spectrum, $\Sigma_N=0$, while for $G=-\bar G$, one finds $\Sigma_N = - 2 G
(\openone-G^2)^{-1}$, which gives a factor
of $\tfrac12$ in the entanglement temperature (i.e., the effective
strength of $H_N^{\rm b}$) with respect to $G$, $H_N^{\rm b} =
4\arctan(G)$, corresponding to the case $\sigma_L=\sigma_R^\top$
in Ref.~\onlinecite{cirac:peps-boundaries}.

A crucial point to observe in the result for the boundary theory is that
$\Sigma_N$ only depends on the CMs $G$ and $\bar G$, which characterizes
the reduced state of the virtual degrees of freedom at the boundaries of
$\mathcal R$ and $\bar{\mathcal R}$. We can therefore trace the physical
degrees of freedom from the beginning and only ever need to consider $G$ and
$\bar G$. While this observation is also true for general PEPS, it is
particularly useful when working with GFPEPS in terms of CMs, as it allows
us to completely neglect the physical part of the CM right from the
beginning. 

Let us finally briefly comment on the relation of the boundary theory as
given by $\Sigma_N$ to the construction of the boundary theory for general
PEPS derived in Ref.~\onlinecite{cirac:peps-boundaries}. There, the part of the
PEPS which describes $\mathcal R$ (corresponding to the CM $\Gamma$) is
interpreted as a linear map $\mathcal X_\mc{R}$ from the boundary to the bulk
degrees of freedom, which is then decomposed as $\mathcal X_\mc{R}=\mathcal
V_\mc{R}\,\mathcal P_\mc{R}$, with $\mathcal V_\mc{R}$ an isometry and $\mathcal
P_\mc{R}=\sqrt{\tau_{\mc{R}}^\top}$, where $\tau_{\mathcal R}$ is the reduced
density matrix of $\mathcal R$ on the virtual system (corresponding to
$G$). This is exactly identical to the decomposition (\ref{eq:P-CM}); in
particular, $M$ describes the isometry $\mathcal V_\mc{R}$, and the (2+3,2+3)
block of $\Gamma'$ describes the map $\nu\rightarrow \sqrt{\tau_{\mathcal
R}^\top} \nu \sqrt{\tau_{\mathcal{R}}^\top}$ (realized by projecting the (3,3)
part onto $\nu$).  Finally, $\bar G$ describes the analogous state
$\tau_{\bar{\mathcal R}}$ obtained from the part $\bar{\mathcal R}$, and
thus, $\Sigma_N$ is exactly identical to the boundary theory
$\sqrt{\tau_{\mathcal R}^\top}\tau_{\bar{\mathcal{R}}}
\sqrt{\tau_{\mathcal R}^\top}$ derived in
Ref.~\onlinecite{cirac:peps-boundaries}.

\subsubsection{Boundary Theories on the
torus\label{sec:bnd-theories-on-torus}}

We will focus now on the situation where the GFPEPS is placed on a square
lattice on a long torus, where we take the length of the torus to
infinity. The two regions $\mc R$ and $\bar{\mc R}$ are then obtained
by cutting the torus into two halves, and are thus given by (identical)
long cylinders with diameter $N_v$ and length $N \rightarrow \infty$,
cf.~Fig.~\ref{fig:cylinder}. As we have seen, the central object in the
description is the CM $G$ at the boundary of region $\mc R$, $\partial \mc{R}$, obtained after
tracing out the physical system (and correspondingly for $\bar{\mc R}$).
In the case of a cylinder, $\mc R$ is given by the left
and right boundary of the cylinder together.  In the following, we will
show how to determine $G$ given the CM $\gamma_1$ defining the GFPEPS,
without having to construct the CM of the whole state $\Phi_N$.

As we have seen in the preceding Subsection, the boundary theory is
entirely determined by the CM of the virtual part of the initial state
$\Psi_1$.  We thus start by decomposing the CM of the virtual system of
$\Psi_1$ into
 \begin{equation}
    \label{eq:CM-singlesite-virtonly-def}
 D=\left(\begin{matrix} H & K \\ -K^\top & V\end{matrix}\right)\ .
 \end{equation}
Here, $V$ corresponds to the vertical and $H$ to the horizontal Majorana
modes, respectively.  We now concatenate one column of tensors, closing
its vertical boundary, leaving us with a CM which describes the left and
right virtual indices of the column (cf.~Fig.~\ref{Fig:PEPS}b-d). This is done by employing
Eq.~\eqref{eq:proj-ME} for the corresponding subblocks of $V$ of each pair
of (cyclically)
consecutive states $\Psi_{1,j}$ and $\Psi_{1,k}$. Due to translational
invariance, this is conveniently expressed in the Fourier basis (with
$k_y$ the quasi-momentum in $y$-direction): In this basis the $D$'s of one
column form a block-diagonal matrix, while
 \[
 \hat \omega(k_y) = \left(\begin{matrix}
    0 & e^{ik_y} \openone_\chi \\ -e^{-ik_y} \openone_\chi & 0
 \end{matrix}\right)
 \]
($\Id_\chi$ denoting the $\chi \times \chi$ identity matrix) since the
$\omega$'s of one column form a circulant matrix with the two blocks
coupling the ``up'' and ``down'' indices of adjacent $V$'s.  In
Fourier space, the CM describing the left and right virtual modes of one
column is thus
 \begin{equation}
 \label{eq:hat-D1-fromHVK}
 \hat D_1 = H + K \,(V+\hat \omega^{-1})^{-1}\, K^\top \ .
 \end{equation}
(We use the hat to denote dependence on $k_y$ in the following; the
subscript $N$ of $\hat D_N$ indicates the number of columns.)
Taking advantage of the fact that the matrix inverse can be written in terms of determinants, one immediately finds that each entry of $\hat D_1$ is a complex ratio of trigonometric polynomials (i.e., polynomials in $e^{\pm i k_y})$ with a degree bounded by the dimension of $\hat\omega$, i.e., $2\chi$.

The matrix $\hat D_1$ consists itself of four blocks,
 \be
 \hat D_1=\left(\begin{matrix} \hat R_1 & \hat S_1
 \\ -\hat S_1^\dagger & \hat T_1 \end{matrix}\right)\ , \label{eq:defD1}
 \ee
corresponding to the left and right indices, respectively.  Let us now see
what happens if we contract two columns. We will consider the general case
where the two columns can be different -- for instance, each of them could
have been derived by contracting some number of single columns $\Phi_1$;
this will allow us to easily derive recursion relations. We thus have two
columns described by
\[
 \hat D=\left(\begin{matrix} \hat R & \hat S \\
 -\hat S^\dagger & \hat T \end{matrix}\right)
 \mbox{\ \ and \ \ }
 \hat D'=\left(\begin{matrix} \hat R' & \hat S'
 \\ -(\hat S')^\dagger & \hat T' \end{matrix}\right)\ ,
 \]
with a column of maximally entangled states connecting them: The CM of both blocks concatenated is then according to Eq.~\eqref{eq:proj-ME}
 \begin{equation}
 \label{eq:general-D-recursion}
 \hat D'' =
 \left(\begin{matrix} \hat R & 0 \\ 0 & \hat T' \end{matrix}\right)
 +
 \left(\begin{matrix} -\hat S & 0 \\ 0 & (\hat S')^\dagger \end{matrix}\right)
 \left(\begin{matrix} \hat T & \Id \\
 -\Id & \hat R' \end{matrix}\right)^{-1}
 \left(\begin{matrix} -\hat S^\dagger & 0 \\ 0 & \hat S' \end{matrix}\right)\ .
 \end{equation}
Using the Schur complement formula for the matrix inverse in the middle,
this gives a recursion relation for the blocks $\hat R$, $\hat S$, and
$\hat T$, which serves several purposes. In particular, by choosing $\hat
D=\hat D'$, we can obtain an iteration formula for $\hat D_{2^\ell}$
describing $2^\ell$ columns, which quickly converges towards the infinite
cylinder limit $\hat D_\infty$, thus being very useful for numerical study.
Moreover, as we will see in Sec.~\ref{sec:full-chi1}, in certain cases it
can also be used to analyze the convergence of the transfer operator, or, by
choosing $\hat D'=\hat D''$
and $\hat D=\hat D_1$, to determine the explicit form of the fixed point
$\hat D_\infty$.

Finally, given the fixed point $\hat D_\infty$, as well as $\hat{\bar
D}_\infty$ corresponding to the boundary  $\partial \bar{\mc R}$, it is
now straightforward to determine the boundary Hamiltonian using
eqs.~(\ref{eq:symmetrized-es}) and \eqref{eq:HN} for $N \rightarrow
\infty$. Note that in the particular case of a torus which we consider,
$\hat{\bar D}_\infty$ can be obtained from $\hat D_\infty$ by exchanging
the blocks corresponding to the left and right boundary.

\subsection{Edge theories}\label{sec:edge}

\subsubsection{Derivation of edge theory}

We will now turn our attention towards the edge Hamiltonian, which
describes the effective low-energy physics obtained at an edge of the
system.

As explained in Sec.~\ref{sec:boundary-edge}, the GFPEPS $\Phi$ is the
ground state of the flat band Hamiltonian $\mc H_\mr{fb}=
-\tfrac{i}{4}\sum_{l,m} \gamma_{l,m} e_l e_m$, Eq.~(\ref{eq:flat}), where
$\gamma$ is the CM of the whole state $\Phi$, Eq.~(\ref{eq:CM-GFPEPS}).
The restriction of $\mathcal H_\mathrm{fb}$ to a region $\mathcal R$ of
the system is then given by
\be
\mc H_\mc{R}= -\tfrac{i}{4} \sum_{l,m} [\gamma_\mathcal{R}]_{l,m} e_l e_m,
\label{eq:Ham-truncated}
\ee
where the sum now only runs over modes in $\mathcal R$, and
$\gamma_{\mathcal R}$ is determined by Eq.~(\ref{eq:CM-R}).

Let us now perform the basis transformation $M$, Eq.~(\ref{eq:Q-trafo}):
Following Eq.~(\ref{eq:CM-R}), the CM of $\mathcal R$,
$\gamma_\mathcal{R}$, is then transformed to
\[
\gamma_\mathcal{R}' = \left(\begin{matrix}
    Z & 0 \\ 0 & \Sigma_N\end{matrix}\right)\ ,
\]
with $\Sigma_N$ given by Eq.~(\ref{eq:symmetrized-es}), and at the same time,
$\mathcal H_\mathcal{R}$ is transformed into an isomorphic Hamiltonian
$\mc H_\mc{R}'= -\tfrac{i}{4} \sum_{l,m} [\gamma'_\mathcal{R}]_{l,m} e_l
e_m$.  We thus see that the spectrum of $\mathcal H'_\mathcal R$ (and thus
of $\mathcal H_\mathcal R$) consists of two parts: First, the (1,1)-block of
$\gamma_\mathcal{R}'$ corresponds to bulk modes at energy $\pm 1$.
Second, the (2,2) bock $\Sigma_N$ corresponds to modes at generally
smaller energy, which are thus related to restricting $\mathcal
H_{\mathrm{fb}}$ to region $\mathcal R$; those modes are related to the
boundary degrees of freedom via the purification in the (2+3,2+3) block of
$\Gamma'$, Eq.~\eqref{eq:P-CM}.  We thus find that the edge Hamiltonian,
i.e., the low-energy part of the truncated flat band Hamiltonian, is given
by
\begin{equation}
\label{eq:Hedge-eq-Sigma}
 H_N^\mr{e} = \Sigma_N \ ,
\end{equation}
with $\mathcal H_N^\mathrm{e}=-\tfrac{i}{4}\sum_{l,m} [H_N^\mathrm{e}]_{l,m} c_l
c_m$, Eq.~\eqref{eq:Hedge}.  Except for additional bulk modes with energy
$\pm 1$, $H_N^\mr{e}$ in fact exactly reproduces the spectrum of
the truncated flat band Hamiltonian.  The relation
(\ref{eq:Hedge-eq-Sigma}) allows us to transfer the results on the
boundary theory $\Sigma_N$ of GFPEPS one-to-one to their edge Hamiltonian
$H_N^\mr{e}$.  Note that the resulting relation between entanglement
spectrum and edge Hamiltonian, $H_N^\mr{b} = 2 \arctan(H_N^\mr{e})$,
corresponds to the one derived by Fidkowski~\cite{fidkowski:freeferm-bulk-boundary}.

The derivation of the edge Hamiltonian in this section is again identical
to the edge Hamiltonian introduced for general PEPS in Ref.~\onlinecite{Shuo}.
Using the same notation as in the last paragraph of
Sec.~\ref{sec:bnd-theories-in-gfpeps}, the edge Hamiltonian for general
PEPS is obtained by projecting the physical Hamiltonian onto the boundary
using the isometry $\mathcal V_\mc{R}$. This projection is exactly accomplished
by rotating with $M$ and subsequently considering only the (2,2) block of
$\gamma_{\mathcal R}'$, and thus, the edge Hamiltonian obtained here is
identical to the one of Ref.~\onlinecite{Shuo}, with the bulk Hamiltonian
taken to be the flat band Hamiltonian.

\subsubsection{Localization of edge modes}

In the case of a cylinder, on which we focus, the edge Hamiltonian
$H^{\mathrm{e}}_N$ is supported on the auxiliary modes both on the left
and the right edge (cf.~Fig.~\ref{Fig:PEPS}f). However, as we will show in
the following, the edge Hamiltonian (as well as the boundary theory) on
the two edges decouples for almost all $k_y$, and moreover, the
corresponding physical edge modes are localized at the same edge as the
virtual modes.  An important consequence of that is that we can use the
\emph{virtual} edge Hamiltonian to compute the Chern number of the system,
as it is known that the Chern number corresponds to the winding number of
the edge modes localized at one of the edges of the system~\cite{Hat93,
kitaev:honeycomb-model}.

In order to answer both of these questions, we will first need to
demonstrate some properties of the CM $\Gamma\equiv \Gamma_{N}$,
Eq.~\eqref{eq:Gamma-R-Rbar}, which describes the GFPEPS $\Phi_{N}$
(Fig.~\ref{Fig:PEPS}f) on a cylinder of length $N\gg1$. Since the system
is translational invariant in vertical direction, we can equally well
carry out our analysis in Fourier space, and we will do so in the
following. By combining Eqs.~\eqref{eq:gamma1},
(\ref{eq:CM-singlesite-virtonly-def}), and \eqref{eq:hat-D1-fromHVK} we
immediately find that $\Phi_1$ is described by a CM of the form
\[
\hat\Gamma_1 = \left(\begin{array}{c|cc}
    \hat A_1 & \hat B_{1,R} & \hat B_{1,T} \\\hline
    -\hat B_{1,R}^\dagger & \hat R_1 & \hat S_1 \\
    -\hat B_{1,T}^\dagger & -\hat S_1^\dagger & \hat T_1
\end{array}\right)\ ,
\]
with $\hat R_1$, $\hat S_1$, and $\hat T_1$ defined in Eq.~\eqref{eq:defD1}. The concatenation of $N$ columns is then given by the Schur complement
\begin{equation}
\label{eq:PhiN-bnd-coupling}
\hat \Gamma_{N} =  \hat P_{N} +
    \hat Q_{N} \hat V_{N}^{-1} \hat Q_{N}^\dagger\ ,
\end{equation}
with
\begingroup
\allowdisplaybreaks
\begin{align*}
\hat P_{N} & = \left(\begin{array}{c|cccc|c}
    \hat R_1 &  -\hat B_{1,R}^\dagger & 0 & \cdots &  \\\hline
    \hat B_{1,R} & \hat A_1  & 0 & & \\
      0		 & 0 & \hat A_1 & & \\
      \vdots & & \ddots & \ddots& \\
           &  &   & 0 & \hat A_1 & -\hat B_{1,T}^\dg \\\hline
		 & &  &   &  \hat B_{1,T}  & \hat T_1
\end{array}\right)\ ,\\[1ex]
\hat Q_{N} & = \left(\begin{matrix}
    \hat S_1 \\\hline
    \hat B_{1,T} \\
    & \hat B_{1,R} & \hat B_{1,T} \\
    & & & \cdots \\
    & & &  & \hat B_{1,R} & \hat B_{1,T} \\
    & & & & & & \hat B_{1,R} \\\hline
    & & & & & & -\hat S_1^\dagger
\end{matrix}\right)\ ,\\[1ex]
\hat V_{N} &= \left(\begin{matrix}
    \hat T_1 & \openone \\
    -\openone & \hat R_1 & \hat S_1 \\
    0 & -\hat S_1^\dagger & \hat T_1 & \openone \\
    & & -\openone & \ddots & \ddots  \\
    & & & \ddots & \ddots & \openone \\
    & & & & -\openone & \hat R_1 & \hat S_1 \\
    & & & & & -\hat S_1^\dagger & \hat T_1 & \openone \\
    & & & & & & -\openone & \hat R_1
\end{matrix}\right)\ ,
\end{align*}
\endgroup
where we have moved the virtual modes on the left (right) boundary to the
left (right) corner of the CM, as indicated by the lines above.

Let us now first show that the two virtual edges are decoupled.  To this
end, we consider the reduced state of the virtual system of
$\Phi_{N}$, which is given by the CM $G$ in Eq.~\eqref{eq:Gamma-R-Rbar};
evidently, vanishing off-diagonal blocks in $G$ (and $\bar G$) imply that
any coupling between the two boundaries in $\Sigma_{N}$,
Eq.~\eqref{eq:symmetrized-es}, vanishes as well.
$G$ is given by the two outer blocks of $\hat\Gamma_{N}$. Obviously, the
only way in which these two blocks can couple is via $\hat V_{N}^{-1}$.

We now invoke a result on the inverse of banded
matrices~\cite{Dem84}: Given a banded matrix $A_b$, it holds that
$|(A_b^{-1})_{ij}|\le \mr{const.} \times \beta^{|i-j|}$, where $\beta <1$ depends on the
ratio of the largest and smallest eigenvalue of $A_b A_b^\dagger$ (and
$\beta\rightarrow1$ if the ratio diverges). Using this result, we find
that the coupling between the two edges in $G$ is exponentially suppressed
in the
length $N$ of the cylinder, as desired, as long as the ratio of the
eigenvalues of $\hat V_N \hat V_N^\dagger$ does not diverge.  Its largest
eigenvalue is clearly bounded by $4$,
since $\hat V_{N}$ is the sum of two CMs.  To lower bound the smallest eigenvalue,
observe that $\hat V_{N}\hat V_{N}^\dagger$ is again a banded Toeplitz
matrix, which we can regard as a subblock of a larger circulant matrix.
This
circulant matrix can in turn be diagonalized using a Fourier transform,
and we find that it is of the form $(\hat D_1 +
\hat\omega^{-1}(k_x,k_y))(\hat D_1+\hat\omega^{-1}(k_x,k_y))^\dagger$.  On
the other hand, $\mathrm{det}\,\big[\hat D_1+\hat \omega^{-1}(k_x,k_y)\big]$
is exactly the energy spectrum of the local parent Hamiltonian as
constructed in Ref.~\onlinecite{kraus:fPEPS}, and thus, $\hat
V_{N}^{-1}\equiv \hat V_{N}^{-1}(k_y)$ has exponentially decaying entries
if and only if the parent Hamiltonian is gapped for the given value of
$k_y$ (which is the case for almost all $k_y$).

As we have seen, (almost) all virtual edge modes on the left and right of the
cylinder decouple. In the following, we will show that also the physical
modes corresponding to these edge modes are exponentially localized around
the corresponding boundary.  To this end, we fix $N\gg1$ and consider the CM
$\Gamma'$, Eq.~(\ref{eq:P-CM}), which is obtained by an orthogonal
transformation from the original CM $\Gamma\equiv \Gamma_{N}$.
In $\Gamma'$, the edge modes are supported on the $(2,2)$ block, and we
need to figure out how the inverse of the orthogonal transformation $M$,
Eq.~(\ref{eq:Q-trafo}), maps these back to the physical
modes.

To this end, note that in order to prepare an arbitrary state in
the $(2,2)$ block of $\Gamma'$, Eq.~\eqref{eq:P-CM}, we just need to project the $(3,3)$
block on an (unphysical) CM $X$ via the Schur complement formula
Eq.~\eqref{eq:proj-ME}. In particular, we can use this to occupy or
deplete a specific mode. (We assume $\chi$ to be even; otherwise, one can simply
group pairs of modes.) Consequently, by projecting the original CM
$\Gamma_{N}$ onto the very same $X$, we will exactly occupy or deplete the
corresponding physical mode. Now, we can make use of
Eq.~(\ref{eq:PhiN-bnd-coupling}), together with the aforementioned result
on inverses of banded matrices: Given $X$ and $X'$ such that projecting
onto $X$ ($X'$) occupies (depletes) a certain mode at one boundary, and
denoting by $\gamma(X)$ [$\gamma(X')$] the corresponding CMs after the
projection, we have that
\[
\gamma(X) - \gamma(X') = \hat Y
    \left[
	(\hat Z+X^{-1})^{-1}
	-(\hat Z+X'^{-1})^{-1}
    \right]
\hat Y^\dagger\ ,
\]
where $\hat Y$ and $\hat Z$ denote the corresponding submatrices of
$\hat\Gamma_N$. Importantly, $\hat Y$ decays exponentially in distance as
it is a column of $\hat\Gamma_{N}$. Since we also have that
\[
[\gamma(X)-\gamma(X')]_{l,m} = 2i( v_l v^*_m - w_l w^*_m)
\]
with $\sum_l v_l c_l$ and $\sum_l w_l c_l$ the creation/annihilation operator
for the corresponding physical mode, it follows that $v_l$ and $w_l$ decay
exponentially with the distance from the corresponding boundary, i.e., the
physical edge mode corresponding to a given virtual edge mode is localized
around that edge.

\section{Further Examples}\label{sec:examples}

In this Section, we will present further examples for both chiral and
non-chiral GFPEPS, and discuss their respective boundary theories.
In Subsection A, we discuss a Chern insulator with $C=-1$; in Subsection
B, we discuss a model with $C=2$ which has entangled edge modes at
incommensurate values of $k_y$; and in Subsections C and D, we discuss two
non-chiral models.

\subsection{GFPEPS describing a Chern insulator with $C = -1$}\label{sec:Chern}

In the following, we study the family of chiral GFPEPS presented in
Ref.~\onlinecite{Wah13}, which are particle number conserving and describe a
Chern insulator with $C = -1$.  They can be decoupled into two
copies of a topological superconductor which closely related to the family
of Eq.~\eqref{eq:Psi1ex}~\cite{note-differ}.
This family has $f = 2$ physical fermionic modes per site, $\chi
= 2$ Majorana bonds, and $\gamma_1$ [Eq.~\eqref{eq:gamma1}] is defined via
\begin{align}
\begin{split}
A &= (-1 + 2 \eta) \left(\begin{array}{cc}
W&0\\
0&-W
\end{array}\right) \\[0.5ex]
B &= \sqrt{\frac{\eta - \eta^2}{2}} \left(\begin{matrix}
\Id - W&\Id + W&-\sqrt{2}\,W&\sqrt{2}\,\Id\\
\Id - W&-\Id - W&\sqrt{2}\,\Id&-\sqrt{2}\,W
\end{matrix}\right) \\[0.5ex]
D &= \left(\begin{matrix}
0&(-1+\eta)\,\Id&-\frac{\eta}{\sqrt{2}}\,\Id&\frac{\eta}{\sqrt{2}}\,\Id\\\
(1-\eta)\,\Id&0&-\frac{\eta}{\sqrt{2}}\,\Id&-\frac{\eta}{\sqrt{2}}\,\Id\\
\frac{\eta}{\sqrt{2}}\,\Id&\frac{\eta}{\sqrt{2}}\, \Id&0&(-1+\eta)\,\Id\\
-\frac{\eta}{\sqrt{2}}\,\Id&\frac{\eta}{\sqrt{2}}\,\Id&(1-\eta)\,\Id&0
\end{matrix}\right)
\end{split}
\label{eq:Chern-ins}
\end{align}
where $\Id= \left(\begin{smallmatrix} 1&0\\  0&1
\end{smallmatrix}\right)$,
$W = \left(\begin{smallmatrix} 0&1\\ -1&0
\end{smallmatrix}\right)$ and $\eta \in (0,1)$. The ordering of the physical Majorana modes is
$(c_{1\uparrow},c_{2\uparrow},c_{1\downarrow},c_{2\downarrow})$, and
the blocks of $D$ are ordered according to left, right, up, and down virtual modes.

\begin{figure}[b]
\centering
\includegraphics[width=0.95\columnwidth]{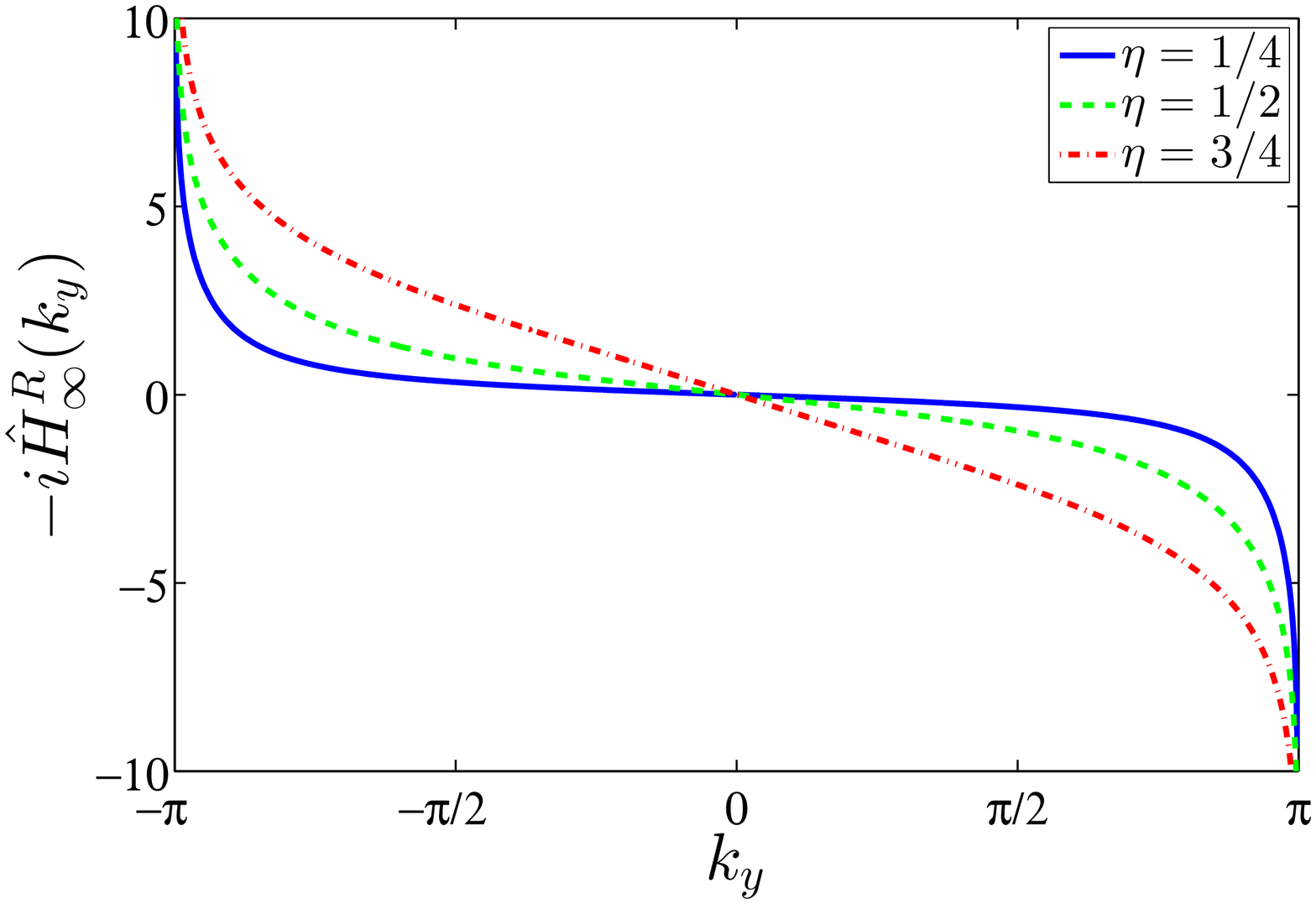}
\includegraphics[width=0.95\columnwidth]{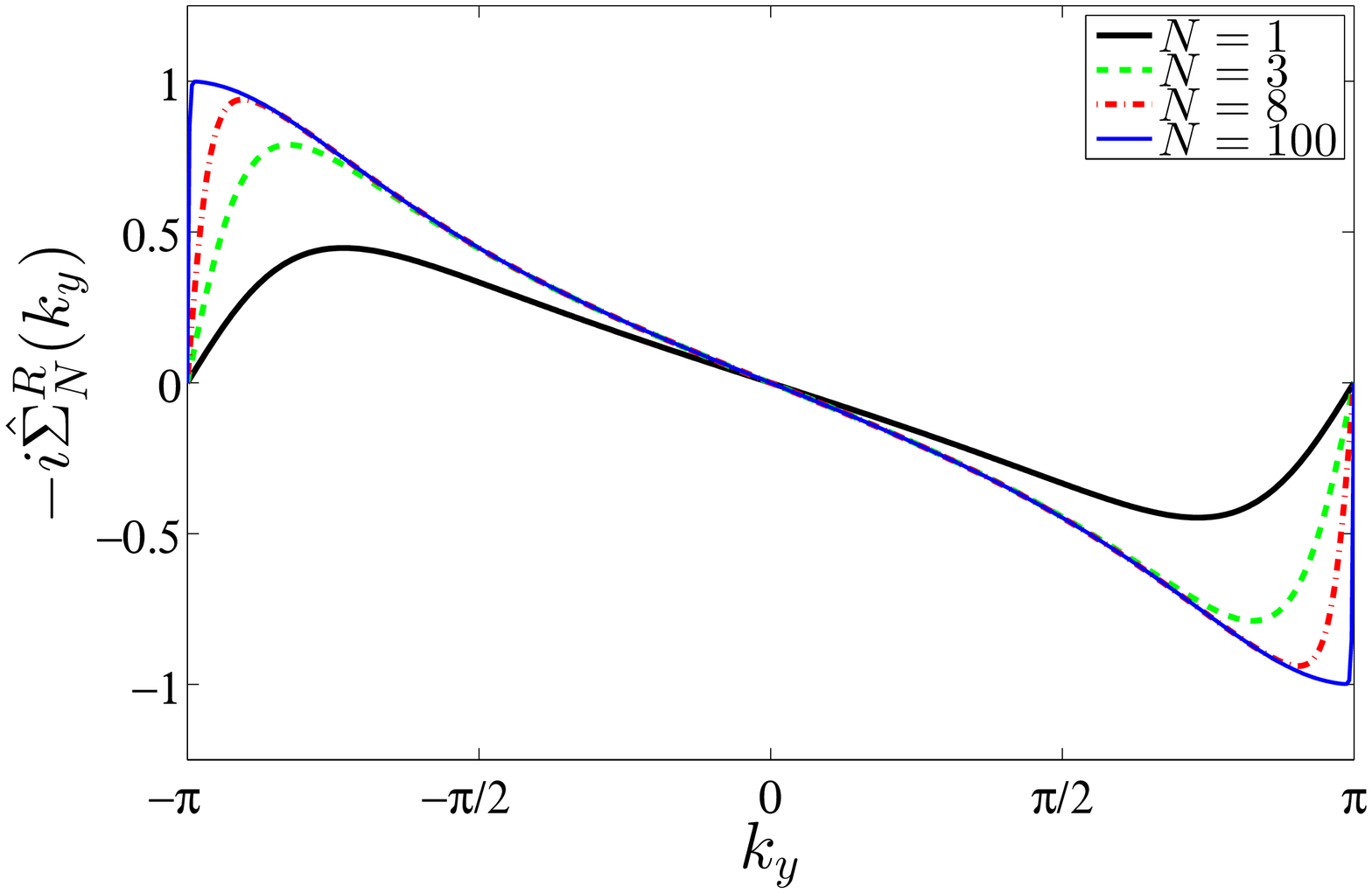}
\includegraphics[width=0.95\columnwidth]{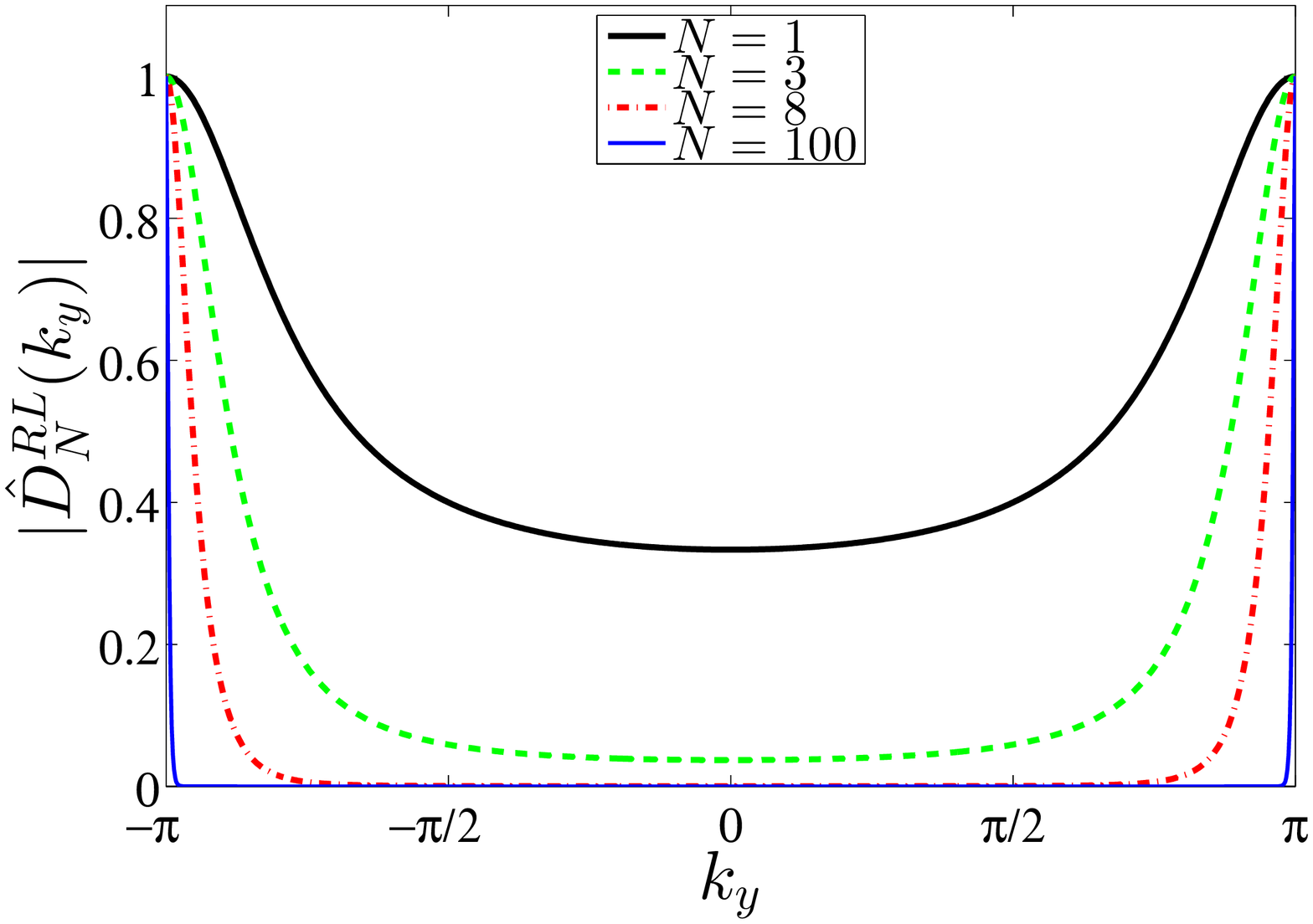}
\caption{
\label{fig:Convergence_R1}
Analysis of the boundary and edge Hamiltonian of the model of
Sec.~\ref{sec:Chern}, using the form of Eq.~\eqref{eq:sigma-chernins}.
Top: Boundary Hamiltonian $-i\hat H_\infty^{R}(k_y)$ for different values
of $\eta$. Middle: Convergence of the right edge spectrum $\hat
\Sigma^{R}_N(k_y)$ (i.e., the block of $\hat\Sigma_{N}(k_y)$ corresponding
to the right edge) for $\eta = 1/2$ with increasing cylinder length
$N=1,3,8,100$.  For $N = 100$, the spectrum is already well 
converged.  Bottom: Magnitude of the corresponding off-diagonal element of
$\hat D_{N} (k_y)$ which describes the coupling of the two boundaries
(cf.~Sec.~\ref{sec:bnd-theories-on-torus}), for cylinder lengths
$N=1,3,8,100$, illustrating the exponential decoupling in $N$.}
\end{figure}

The boundary $\hat \Sigma_\infty(k_y)$ can be computed using the
results of Sec.~\ref{sec:full-chi1}, and we find it to be of the form
\begin{equation}
\label{eq:sigma-chernins}
\Sigma_\infty = \left(\bigoplus_{k_y \neq \pi} \left(\hat
\Sigma^L_\infty(k_y) \oplus \hat \Sigma^R_\infty(k_y) \right)\oplus \hat
\Sigma^{LR}_\infty(\pi)\right) \otimes \Id
\end{equation}
with $\hat \Sigma^{LR}_N(\pi) = \left(\begin{smallmatrix} 0&\pm 1 \\ \mp
1& 0 \end{smallmatrix}\right)$, the sign depending on whether the
horizontal length $N$ of the cylinder is even or odd.
In Fig.~\ref{fig:Convergence_R1}, we show the spectrum of the boundary
Hamiltonian of the above model (top panel). Moreover, we illustrate how
for $N\rightarrow\infty$ the edge Hamiltonian for a single edge converges
(middle panel) and how the coupling between the two edges vanishes (bottom
panel).

The Chern number can now be determined by counting the number of times the
bands of $-i \hat \Sigma_\infty^R(k_y) \otimes \Id$ (or, alternatively, of
$-i \hat H_\infty^R(k_y) \otimes \Id$) cross the Fermi level.  Obviously,
the spectrum of the boundary and edge Hamiltonian consists of two bands
lying on top of each other. In the language of topological
superconductors, this would give rise to a Chern number of $-2$. However,
since we assume particle number conservation (as we deal with a Chern
insulator), the Chern number is given by the number of \textit{fermionic}
chiral modes of the edge or boundary Hamiltonian, respectively. There is
only one such fermionic chiral mode (annihilation operator $\hat
a_{k_y}$), which is obtained by combining the two chiral Majorana modes on
the right edge, $\hat c_{1,k_y}$ and $\hat c_{2,k_y}$, with equal dispersion to
$\hat a_{k_y} = \frac{1}{2} (\hat c_{1,k_y} - i \hat c_{2,k_y})$. In this
case, combining the Majorana modes does not make the system topologically
trivial, since both of them have the same chirality.  Therefore, the
(particle number conserving) Chern number is $C = -1$.

\subsection{GFPEPS with Chern number $C = 2$}\label{sec:Chern2}

In the following, we provide an example of a topological superconductor
with $\chi=2$ and Chern number $C=2$. The model has been constructed
numerically such that it exhibits discontinuities in $\hat\Sigma_\infty(k_y)$
and thus pure fermionic 
modes between the edges maximally entangled modes between the edges at $k_y=\pm1$;
it thus demonstrates that for $\chi>1$, there is no constraint (in terms
of simple fractions of $\pi$) on the possible values of $k_y$.  The CM $D$
of the example is given by
\[
D \!\approx\!\! \left(\begin{smallmatrix}
    0     &  -0.326 &  -0.250 &   0.510  &  0.295  &  0.071 &  -0.434 &  -0.030  \\
    0.326 &        0&   0.044 & -0.074   & -0.513  &  0.032 &  -0.051 &   0.577  \\
    0.250 &  -0.044 &   0     &  -0.467  &  0.036  &  0.603 &  -0.423 &  -0.125 \\
   -0.510 &   0.074 &   0.467 &       0  &  0.148  &  0.156 &   0.169 &   0.216 \\
   -0.295 &   0.513 &  -0.036 &  -0.148  &  0      & -0.161 &  -0.296 &   0.237 \\
   -0.071 &  -0.032 &  -0.603 &  -0.156  &  0.161  &  0     &   0.042 &   0.521 \\
    0.434 &   0.051 &   0.423 &  -0.169  &  0.296  & -0.042 &   0     &   0.047 \\
    0.030 &  -0.577 &   0.125 &  -0.216  & -0.237  & -0.521 &  -0.047 &   0
\end{smallmatrix}\right).
\]
It has been obtained by numerically optimizing $D$ such that one of the
eigenvalues of $\hat\Sigma_N^R(k_y)$ (where $N=2^{29}$) jumps from $\pm i$
to $\mp i$ for some $k_y\in[0.999,1.001]$, while restricting half of the
eigenvalues of $D$ to be between $\pm0.6i$  such as to prevent $D$ from
converging to a pure state. As
$\hat\Sigma_N^R(-k_y)=[\hat\Sigma_N^R(k_y)]^*$ (with $^*$ indicating the complex
conjugate), this automatically yields another identical discontinuity at
$k_y=-1$.
Note that $D$ can be purified to a state with $f = 2$ physical fermions.

\begin{figure}[t]
\centering
\includegraphics[width=0.52\textwidth]{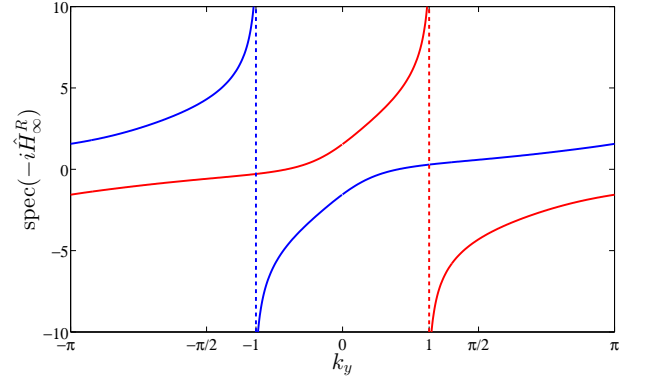}
\caption{Eigenvalue spectrum of $- i \hat H^{R}_\infty(k_y)$ for the
example of Sec.~\ref{sec:Chern2}. Since $\chi = 2$, there are two bands.
They diverge at $k_y = \pm 1$, respectively, where only one mode of $\hat
H^{R}_\infty(k_y)$ is defined. Since the Fermi level at $E = 0$ is crossed
two times from below, $C = 2$. The dispersionless bulk bands of
the (truncated) flat band Hamiltonian $\mc H_\mr{R}$ correspond to an
energy of $\pm \infty$.} \label{Gamma_L2}
\end{figure}

The spectrum of $-i \hat H^R_\infty(k_y)$ is plotted in
Fig.~\ref{Gamma_L2}. Due to the discontinuities at $k_y\pm1$, it crosses
the Fermi energy twice from below, thus describing a topological
superconductor with Chern number $C = 2$.  At $k_y = \pm 1$, one of
the eigenvalues of $-i \hat H_\infty^R(k_y)$ diverges, and thus $\hat
\Sigma^{LR}_\infty(k_y)$ is non-trivial, coupling one of the two virtual
Majorana modes between the left and the right end of the cylinder.

\subsection{GFPEPS with Chern number $C = 0$}\label{sec:C0-nontrivial}

The following example provides a family of non-topological GFPEPS with Chern
number $C=0$.  It has one parameter $\mu$, and its matrix $D$ is given by
\be
D = \left(\begin{array}{cccc}
0&0&-\frac{\mu}{2}&f(\mu)\\
0&0&f(\mu)&-\mu\\
\frac{\mu}{2}&-f(\mu)&0&0\\
-f(\mu)&\mu&0&0
\end{array}\right) \label{eq:D-chi0}
\ee
with $f(\mu) = \sqrt{1-\frac{3 \mu}{2}+\frac{\mu^2}{2}}$ and $\mu \in
(0,1)$. ($A$ and $B$ can be obtained by choosing an arbitrary
purification.)

We find that the left and right boundary, Eq.~\eqref{eq:decouple-H},
decouple for all $k_y$. The dispersion relation for the right boundary is
shown in Fig.~\ref{fig:chi0}. Since the energy band of the boundary
Hamiltonian crosses the Fermi energy once with positive and once with
negative slope for all $\mu \in (0,1)$, the Chern number is always zero.

\subsection{GFPEPS with flat entanglement spectrum and $C = 0$}\label{sec:ex:nochern-flatspec}

The last example we consider is taken from Ref.~\onlinecite{kraus:fPEPS}; it
does not display any topological features. It is given by
\be
\label{eq:ex:nochern-flatspec}
\gamma_1 = \left(\begin{array}{cccccc}
0&0&\frac{1}{\sqrt 2}&-\frac{1}{\sqrt 2}&0&0\\
0&0&0&0&\frac{1}{\sqrt 2}&-\frac{1}{\sqrt 2}\\
-\frac{1}{\sqrt 2}&0&0&0&\frac{1}{2}&\frac{1}{2}\\
\frac{1}{\sqrt 2}&0&0&0&\frac{1}{2}&\frac{1}{2}\\
0&-\frac{1}{\sqrt{2}}&-\frac{1}{2}&-\frac{1}{2}&0&0\\
0&\frac{1}{\sqrt 2}&-\frac{1}{2}&-\frac{1}{2}&0&0
\end{array}\right)\ .
 \ee
Since $\hat D_N=\hat{\bar D}_N$, and thus $G=\bar G$, the entanglement
spectrum and edge Hamiltonian of this model are totally flat, i.e.,
$\Sigma_N=0$ according to Eq.~\eqref{eq:symmetrized-es}, and the Chern
number is zero.


\section{Full solution for $\chi=1$}\label{sec:full-chi1}

In this Section, we will use the recursion relation
(\ref{eq:general-D-recursion}) to explicitly
derive the boundary and edge theories for GFPEPS with one Majorana mode
per bond, $\chi=1$. We will then use this result to show that the presence
of chiral edge modes is related to the occurrence of Majorana modes maximally
correlated between the two edges, i.e., a fermionic mode in a pure state
shared between the two edges.

We start by deriving a closed expression for the boundary and edge
Hamiltonian for $\chi=1$.  In this case,
\be
\hat D = \left(\begin{matrix} i \hat r & i\hat s \\
    i\hat s^* & i\hat t\end{matrix}\right)\ , \label{eq:D-column}
\ee
with scalar functions $\hat r\equiv \hat r(k_y)\in \mathbb R$, $\hat
t\equiv \hat t(k_y)\in \mathbb R$, and $\hat s\equiv \hat s(k_y)$.  Note
that for given $k_y$, the eigenvalues need not to come in complex conjugate
pairs. However, they are still bounded by one, which implies that for
$\hat r\,\hat t\ge0$,
\begin{equation}
\label{eq:sqrtacb-eq}
1-\sqrt{\hat r\,\hat t} \ge |\hat s|
\mbox{\ with equality iff $|\hat s|=1$\ ,}
\end{equation}
which in turn implies that for all $\hat r$ and $\hat t$,
\begin{equation}
\label{eq:acb-eq}
1-\hat r\,\hat t \ge |\hat s|
\mbox{\ with equality iff $|\hat s|=1$\ .}
\end{equation}
(For $\hat r\ge0$ and $\hat t\ge0$, Eq.~\eqref{eq:sqrtacb-eq} follows from
$2\ge -i\,(1,\tfrac{\hat s}{|\hat s|})\,\hat D_1 \,(1,\tfrac{\hat s}{|\hat
s|})^\dagger = \hat r + \hat t +2 |\hat s| \ge
2\sqrt{\hat r\hat t} +2|\hat s|$,  and similarly for
$\hat r\le0$ or $\hat t\le0$.)

Let us now study what happens when we concatenate cylinders; for the CM of
$N$ columns, we will write $\hat D_N$ and $\hat r_N$, $\hat
s_N$, and $\hat t_N$.  The iteration relation
(\ref{eq:general-D-recursion})
yields the following iteration relations for the matrix elements:
\begin{subequations}
\label{eq:onemode-recursion}
\begin{align}
\hat r''  & = \hat r + \hat r'\, \frac{|\hat s|^2}{1-\hat r'\hat t}
    \label{eq:onemode-recursion-a}\\
\hat t''  & = \hat t' + \hat t\, \frac{|\hat s'|^2}{1-\hat r'\hat t}
    \label{eq:onemode-recursion-c}\\
\hat s''  & = \frac{\hat s\hat s'}{1-\hat r'\hat t}\quad .
    \label{eq:onemode-recursion-b}
\end{align}
\end{subequations}
For $\hat z= \hat z'= \hat z_N$ ($\hat z= \hat r, \hat s, \hat t$), and $\hat z''= \hat z_{2N}$, with $N$ a power of
$2$ (i.e., doubling the number of columns in each step), we obtain
\begin{subequations}
\begin{align}
\hat r_{2N}  & = \hat r_N\, (1+\hat \xi_N)\\
\hat t_{2N}  & = \hat t_N\, (1+\hat \xi_N)\\
\hat s_{2N}  & = \frac{\hat s_N}{\hat s_N^*} \,\hat \xi_N\quad,
\end{align}
\end{subequations}
with
\begin{equation}
\hat \xi_N = \frac{|\hat s_N|^2}{1-\hat r_N \hat t_N}\ .
\end{equation}
Assume for now $|\hat s_N|<1$: Then, \eqref{eq:acb-eq}
$\Rightarrow\;\hat \xi_N<1\;\Rightarrow \;  |\hat s_{2N}|<1$, and thus
$|\hat s_1|<1$ implies  $|\hat s_N|<1$. Moreover, Eq.~\eqref{eq:acb-eq}
implies $|\hat s_{2N}|<|\hat s_N|$, which in turn implies that $|\hat
s_N|$ converges; similarly, since $\hat \xi_N\ge0$, $\hat
r_N$ and $\hat t_N$ monotonously move away from zero and thus
converge. We therefore find that for $|\hat s_N|<1$, all matrix elements
converge.

On the other hand, $|\hat s_1|=1$ implies that $\hat r_1=\hat t_1=0$ (as
$\hat D_1$ must have spectral radius $\le 1$), and thus $\hat r_\infty=\hat
t_\infty=0$, while $|\hat s_\infty|=1$.  An explicit analysis of the
possible $\hat D_1$ for $\chi=1$, using Eq.~\eqref{eq:hat-D1-fromHVK},
shows that $\hat r_1(k_y)=\hat t_1(k_y)=0$ can only be the case for
$k_y=0$ or $k_y=\pi$, unless both vanish identically (in which case the
fixed point and the GFPEPS are trivial).

In order to determine the fixed point for $|\hat s_1| < 1$ ($N\rightarrow\infty$),
 we now consider the scenario where
$\hat z=\hat z''= \hat z_\infty$ and $\hat z'= \hat z_1$, i.e., where we append a single column to an
infinite cylinder. From (\ref{eq:onemode-recursion-a}), we find that
\[
\hat r_\infty = \hat r_\infty + \
    \hat r_1 \frac{|\hat s_\infty|^2}{1-\hat r_1\hat t_\infty}
\]
and thus $\hat s_\infty=0$ for $\hat r_1\ne0$ (and similarly if $\hat
c_1\ne 0$); if $\hat r_1 = \hat t_1 =0$,
(\ref{eq:onemode-recursion-b}) yields $\hat s_\infty=\hat s_\infty \hat
s_1$ which as well implies $\hat s_\infty=0$ as long as $|\hat s_1|<1$. On the
other hand, Eq. (\ref{eq:onemode-recursion-c})
yields a quadratic equation for $\hat t_\infty$,
\be
    \label{eq:cinfty-quad-eq}
\hat r_1\hat t_\infty^2 -(1+\hat r_1\hat t_1-|\hat s_1|^2)\hat t_\infty
+\hat t_1=0,
\ee
and similarly for $\hat r_\infty$ by exchanging $\hat r$ and $\hat t$.
 Of the two solutions
\[
\hat t_\infty^{\pm} = \frac{\hat\Delta_1 \pm
\sqrt{\hat\Delta_1^2-4\hat r_1\hat t_1}}{2\hat r_1}
\]
[where $\hat\Delta_1 =
1+\hat r_1\hat t_1 -|\hat s_1|^2 = \det(i\hat D_1)+1\ge 0$], the fixed
point is always given by $\hat t_\infty^-$.  This is seen by noting that
$-1\le\hat t_\infty^+\le1$ implies that $\pm 2\hat r_1-\hat
\Delta_1\ge\sqrt{\hat\Delta_1^2-4\hat r_1\hat t_1}$ (with $\pm$ the sign
of $\hat r_1$), squaring which yields $0\ge(\hat r_1\mp1)(\hat
t_1\mp1)-|\hat s_1|^2 = \det(i\hat D_1\mp\openone)$, and thus $t_\infty^+$
can only be physical if $i\hat D_1$ has an eigenvalue $\pm1$; and these
remaining cases can be easily analyzed by hand.
We thus find that the fixed point CM is of the form
\[
\hat D_\infty = \left(\begin{matrix} i \hat r^-_\infty
& 0 \\ 0 & i\hat t^-_\infty\end{matrix}\right)\ ,
\]
except when $|\hat s_1|=1$, which we found can only happen at $k_y=0,\pi$
(in which case $\hat s_1(k_y)$ is real).

In order to obtain the boundary theory, we need to combine the expression
for $\Sigma_N$, Eq.~(\ref{eq:symmetrized-es}), with the fact
that $\hat G$ and $\hat{\bar G}$ are given by $\hat G=i\hat r_\infty\oplus i \hat t_\infty$
and $\hat{\bar G} = i \hat t_\infty\oplus i \hat r_\infty$, with the exception of the
singular points in $k_y$-space where $|\hat s_\infty|=1$. In particular, the
two boundaries can be described independently almost everywhere, and we
obtain for the edge theory of the right edge ($\hat t_\infty = \hat t_\infty^-$, $\hat r_\infty = \hat r_\infty^-$)
\begin{align}
\hat \Sigma^{R}_\infty(k_y) & = -i\hat t_\infty-i(1-\hat t_\infty^2)
    (\hat t_\infty - \hat r_\infty^{-1})^{-1} \nonumber \\
\label{eq:sigma_es_chi1}
& = i\frac{\hat r_1-\hat t_1}{\sqrt{ {\hat\Delta}_1^2-4\hat r_1 \hat t_1}}\ ,
\end{align}
with the boundary Hamiltonian given by $\hat H^R_\infty(k_y)=
2\arctan(\hat \Sigma_\infty^R(k_y))$;
for the
opposite edge, $\hat r$ and $\hat t$ need to be interchanged.
For the points with $|\hat s_1|=|\hat s_\infty|=1$, on the other hand,
the two boundaries are in a maximally entangled state of the
Majorana modes with the corresponding $k_y$.

Clearly, $\hat \Sigma_{\infty}^R(k_y)$ [Eq.~(\ref{eq:sigma_es_chi1})] is
continuous unless the denominator becomes zero.  For the latter to happen,
one first needs that $\hat r_1\hat t_1\ge 0$, and with this, $\hat
\Delta_1^2-4\hat r_1\hat t_1=0$ is equivalent to $1-\sqrt{\hat r_1\hat
t_1}=|\hat s_1|$, which using Eq.~(\ref{eq:sqrtacb-eq}) implies that
$|\hat s_1|=1$, which can only be the case for $k_y=k_y^0=0,\pi$. In order
to analyze how $\hat \Sigma^{R}_\infty(k_y)$
behaves around such a point, we expand to first order in $\delta
k_y=k_y-k_y^0$:
Then, $\hat r_1=\hat r_1'\,\delta k_y+O(\delta k_y^2)$, $\hat t_1=\hat
t_1'\,\delta k_y+O(\delta k_y^2)$, and $|\hat s_1|=1+O(\delta k_y^2)$
(since $|\hat s_1|\le 1$). One immediately finds that
\begin{align*}
\hat \Sigma^{R}_\infty(k_y) &= i\frac{(\hat r_1'-\hat t_1')\delta k_y + O(\delta k_y^2)}{
    \sqrt{-4\hat r_1'\hat t_1'\,\delta k_y^2 + \mc O(\delta k_y^4)}}
\\
& =i\,\mathrm{sign}(\delta k_y)\,
    \frac{\hat r_1'-\hat t_1'}{\sqrt{-4\hat r_1'\hat t_1'}}+ \mc O(\delta k_y)\ ,
\end{align*}
this is, $\hat \Sigma^{R}_\infty(k_y)$ exhibits a discontinuity unless $\hat
r_1'=\hat t_1'$. In order to relate $\hat r'_1$ and $\hat t'_1$, we
observe that the eigenvalues of $\hat D_1$ around $k_y^0$ are $i(\pm
1+\tfrac12(\hat r_1'+\hat t_1') \delta k_y +O(\delta k_y^2))$, and thus $\hat r_1'+\hat
t_1'=0$, which implies that
\[
\hat \Sigma^{R}_\infty(k_y) = i\,\mathrm{sign}(\delta k_y) \,
    \mathrm{sign}(\hat r_1'(k_y^0))\ ;
\]
this is, the edge Hamiltonian exhibits a jump between $\pm1$, and the
boundary Hamiltonian derived from the entanglement spectrum diverges, as
we have seen in the examples.  The case of vanishing
first order terms, $\hat r_1'= \hat t_1'=0$, can be dealt with using the
explicit form of $\hat D_1$ for $\chi=1$, which yields that $\hat r_1=\hat
t_1=0$ vanish identically for all $k_y$, making the fixed point trivial; 
if $\hat r'_1$ changes its sign, this corresponds to a transition point
between $C=+1$ and $C=-1$.  Note that according to
Eq.~\eqref{eq:hat-D1-fromHVK},  $\hat r_1=\hat t_1=0$ happens if and only
if $K$ is either diagonal or off-diagonal (as the other terms are
antihermitian $2\times 2$ matrices). This means that the virtual CM $D$
does not couple the left with the down Majorana mode and the right 
with the up Majorana mode (or the other way round).

We thus find that $|\hat s_1(k_y^0)|=1$ at $k_y^0=0$ or $k_y^0=\pi$ is
equivalent to having a discontinuity in the edge Hamiltonian, which jumps
between $\pm 1$. Since $H^\mathrm{e}_N$ is otherwise continuous, and we
will see that for $\chi=1$, $|\hat s_1(k_y^0)|=1$ can occur for at most one
$k_y$ (see Sec.~\ref{sec:necessity}),  it follows that $|\hat
s_1(k_{y}^0)|=1$, i.e., the existence of a maximally entangled mode
between the left and right edge of the cylinder $\hat D_1$
at $k_y = k_y^0$ is equivalent to having a chiral mode at the edge.


\section{Symmetry and chirality}

As we have seen in the preceding section, the existence of a
chiral edge mode is equivalent to the existence of a maximally entangled
Majorana mode
between the left and right edge of the cylinder at $k_y^0=0$
or $k_y^0=\pi$. In the following, we will show that this mode can be
understood as arising from a \emph{local} symmetry of the state $\Psi_1$
which defines the GFPEPS (Eq.~\eqref{eq:Gaussian-psi1}).

Concretely, in part A we will demonstrate that a certain symmetry of
$\Psi_1$ leads to a maximally entangled Majorana pair between the left and
right edge and thus a chiral edge state. In part B we will show the
opposite -- that a maximally entangled Majorana pair between the left and
the right implies $\Psi_1$ having a certain symmetry.  In part C we
uncover these kinds of symmetries in the examples presented in the
previous sections. In part D we consider again the example given by
Eq.~\eqref{eq:Psi1ex} and outline how strings of symmetry operators can be
used to construct all ground states of its frustration free parent
Hamiltonian $\mc H_\mr{ff}$.

We will generally restrict the discussion in this Section to the case of
$\chi=1$ Majorana mode per bond, though some of the results (in particular
in Subsection A) directly generalize to larger $\chi$.

\subsection{Sufficiency of local symmetry}\label{sec:sufficiency-symmetry}

\begin{figure}[b]
\begin{center}
\includegraphics[width=0.4\textwidth]{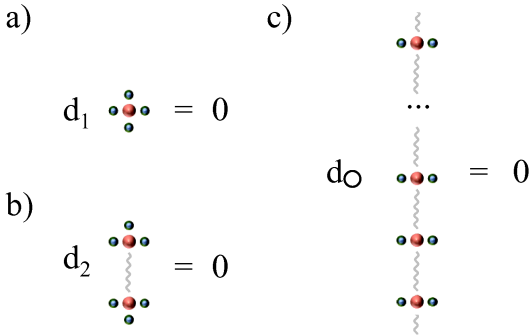}
\caption{Concatenation of a symmetry. (a) Symmetry $d_1$ annihilating the state $\Psi_1$ of virtual and physical Majorana fermions on one site. For a chiral state with $\chi = 1$ it can be concatenated as described in the text to a symmetry $d_{2}$ annihilating the state $\Psi_2$ defined on two sites (b). Proceeding in the same manner and closing the vertical boundary, one obtains a symmetry $d_\bigcirc$ annihilating one column $\Phi_1$ (c). \label{fig:symmetry}}
\end{center}
\end{figure}

We start by showing how a symmetry in $\Psi_1$ induces a symmetry on a
whole column, $\Phi_1$, and how this subsequently gives rise to a maximally
correlated mode between the two edges of a cylinder.
Since $\Psi_1$ is a pure Gaussian state
where four virtual Majorana modes are entangled with one physical
fermionic mode, there must be a virtual fermionic mode which is in the vacuum, i.e.,
\begin{equation}
\label{eq:d-sym-def}
d_1=\alpha_L c_L +\alpha_R c_R + \alpha_U c_U + \alpha_D c_D
\end{equation}
on the virtual system which annihilates $\Psi_1$,
\begin{equation}
\label{eq:P-annih-d}
d_1\,\ket{\Psi_1} = 0\ ,
\end{equation}
as already discussed in Sec.~\ref{sec:symm-top}.
[$d_1$ corresponds to the eigenvector of $D$, Eq.~\eqref{eq:gamma1},
with eigenvalue $-i$, and describes a fermionic mode].  We will refer
to $d_1$ as a \textit{symmetry}, since it corresponds to a $\mathbb{Z}_2$
symmetry of $\Psi_1$ with $U_1 = \Id - 2 d_1^\dg d_1$.
On the other hand, for the virtual fermionic modes
$\omega_{12}$ (the indices denoting the vertical positions),
Eq.~\eqref{eq:omegas}, it
holds that $\bra{\omega_{12}}(1-i c_{1,D} c_{2,U})=0$ and thus
\begin{equation}
\label{eq:annih-omega}
\bra{\omega_{12}}(c_{1,D}+i c_{2,U})=0.
\end{equation}
By combining Eqs.~\eqref{eq:P-annih-d} and eq.~\eqref{eq:annih-omega}, we can now
study how the symmetry (\ref{eq:d-sym-def}) behaves when we concatenate
two or more sites by projecting onto $\bra{\omega_{12}}$
(we assume $\alpha_U\ne0$ for now, and define
$\theta:=i\alpha_D/\alpha_U$):
\begin{align*}
0   &=   \bra{\omega_{12}}\Big[
	(\alpha_L c_{1,L} + \alpha_R c_{1,R} + \alpha_U c_{1,U} + \alpha_D c_{1,D})|\Psi_1,\Psi_1\rangle_{1,2} \\
    &\qquad+ \theta
    (\alpha_L c_{2,L} + \alpha_R c_{2,R} + \alpha_U c_{2,U} + \alpha_D c_{2,D})|\Psi_1,\Psi_1\rangle_{1,2} \Big] 
\\
&=  d_2 \langle\omega_{12} \ket{\Psi_1, \Psi_1}_{1,2} \ \equiv \
d_2\ket{\Psi_2}\ ,
\end{align*}
with
\[
d_2=\alpha_L (c_{1,L} + \theta c_{2,L})
         + \alpha_R (c_{1,R} + \theta c_{2,R})
	 + \alpha_U c_{1,U}
	 +\theta \alpha_D c_{2,D}
\]
the symmetry of the concatenated state $\Psi_2$,
Fig.~\ref{Fig:PEPS}b and Fig.~\ref{fig:symmetry}b.  The argument can be
easily iterated, and we find that
\[
d_{N_v} \ket{\Psi_{N_v}} = 0\ ,
\]
with
\begin{align*}
d_{N_v} &=
    \alpha_L \sum_{y=1}^{N_v}\theta^{y-1} c_{y,L}
    +\alpha_R \sum_{y=1}^{N_v}\theta^{y-1} c_{y,R}
	 + \alpha_U c_{1,U} \\
	 &+\theta^{N_v-1} \alpha_D c_{N_v,D}\ .
\end{align*}
Let us now see what happens when we close the boundary between sites $N_v$
and $1$, which yields $\ket{\Phi_1}\equiv
\langle\omega_{N_v,1}\ket{\Psi_{N_v}}$, Fig.~\ref{Fig:PEPS}d: Since
$\bra{\omega_{N_v,1}}(c_{N_v,D}+i c_{1,U})=0$, we find that
\be
\d_{\bigcirc}\ket{\Phi_1} = 0\ , \label{eq:d-column}
\ee
with
\begin{equation}
\label{eq:d-column-2}
d_{\bigcirc}=
    \alpha_L \sum_{y=1}^{N_v}\theta^{y-1} c_{y,L}
    +\alpha_R \sum_{y=1}^{N_v}\theta^{y-1} c_{y,R}
\end{equation}
(Fig.~\ref{fig:symmetry}c) whenever $\theta^{N_v}=1$.  This leads to two
requirements for the existence of $d_\bigcirc$ fulfilling
Eq.~\eqref{eq:d-column}: First, $|\alpha_U|=|\alpha_D|$, and
second, the momentum $k_y^0$ of $d_\bigcirc$ (defined via $e^{ik_y^0} =
\theta$) must be commensurate with the lattice size.  Whenever these
requirements are fulfilled, we thus find that the local symmetry $d_1$,
Eq.~\eqref{eq:d-sym-def}, gives rise to a symmetry $d_{\bigcirc} \propto
\alpha_L \hat c_{L,k_y} + \alpha_R  \hat c_{R,k_y}$,
Eq.~\eqref{eq:d-column}, on the whole column (i.e., on $\hat D_1$), at
momentum $e^{ik_y^0} = i \alpha_D/\alpha_U$.  Note that we only need to
assume that either $\alpha_U$ or $\alpha_D$ is non-zero; if both are zero,
the condition \eqref{eq:P-annih-d} implies that the horizontal virtual modes
entirely decouple from the physical system, and the GFPEPS describes a
product of one-dimensional vertical chains.

We have thus found that a certain local symmetry induces a symmetry on a
column $\Phi_1$, which forces the Majorana modes with a specific momentum on
both ends of the column to be correlated.  This is equivalent to demanding
that for this $k_y = k_y^0$, $\hat D_1(k_y^0)$ has an eigenvalue $-i$. For
$k_y^0=0,\pi$, this implies that $|\hat s_1(k_y^0)|=1$, as the diagonal
elements of $\hat D_1(k_y^0)$ are zero due to $\hat D_1(-k_y) = \hat
D_1^*(k_y)$.

The symmetry of a single column is passed on when concatenating columns, this
is, when going from $\Phi_1$ to $\Phi_N$, Fig.~\ref{Fig:PEPS}d-f, in
analogy to the arguments given before. In order for this to lead to a
coupling between the two edge modes in the limit of an infinite cylinder,
as observed in the examples with chiral edge modes, it is additionally
required that $|\alpha_L|=|\alpha_R|$. Otherwise the symmetry becomes
localized at a single boundary.  This can be understood by exchanging
horizontal and vertical directions, leading to $e^{i k_x^0} = i \alpha_R /
\alpha_L$ (and $k_x^0 = 0, \pi$, too).  As we have seen in the last
Section, a coupling between the left and the right edge for $\chi = 1$ can
only emerge, if $k_y^0 = 0, \pi$ (and analogously $k_x^0 = 0, \pi$). Thus,
we have to require $\alpha_D / \alpha_U = \pm i$, $\alpha_R / \alpha_L =
\pm i$ for a symmetry leading to a chiral edge state. Since there can only
be one such symmetry for $\chi = 1$ (otherwise the virtual and physical
system decouple), we conclude that there can be a maximally entangled
 Majorana mode only for $k_y^0 = 0$ or $k_y^0 = \pi$, 
but not for both of them (and similarly for $k_x$). 
We thus find that $d_1$ must be of the form
\begin{equation}
d_1 = \alpha_L(c_L \pm i c_R) + \alpha_U (c_U \pm i c_D)\ . \label{d_max_ent}
\end{equation}
($\alpha_L, \alpha_U \neq 0$)
in order to be stable under concatenation.

Let us finally show that in order to have a non-trivial Chern number,
there is an additional constraint on $\alpha_L$ and $\alpha_U$, namely
that
\be
\label{eq:pm1-pmi}
\mathrm{arg}(\tfrac{\alpha_L}{\alpha_U}) \not\in \{0,\pi, \pm \tfrac\pi2\}\ .
\ee
This can be directly verified by explicitly constructing $D$ (given $\hat
d_1$, the only remaining freedom is the eigenvalue of the non-pure mode),
where one finds that the diagonal (off-diagonal) elements of $K$
[cf.~Eq.~\eqref{eq:CM-singlesite-virtonly-def}] vanish exactly if
$\mathrm{arg}(\tfrac{\alpha_L}{\alpha_U}) = 0,\pi$ 
[$\mathrm{arg}(\tfrac{\alpha_L}{\alpha_U}) = \pm\tfrac\pi2$]. As we have
seen in the last section, this in turn is equivalent to a trivial
(completely flat) edge spectrum, and thus to a trivial Chern
number. 

In summary, we find that we have a non-trivial Chern number whenever we
have exactly one symmetry $d_1$ which satisfies Eqs.~\eqref{d_max_ent} and
\eqref{eq:pm1-pmi}.

\subsection{Necessity of an on-site symmetry}\label{sec:necessity}

Let us now show the converse statement of the previous subsection: We
will show that for $\chi=1$, a maximally entangled Majorana pair between
the left and right boundary of a cylinder at
$k_y^0=0,\pi$, which is equivalent to the presence of a chiral edge mode,
implies the existence of a local symmetry of the form Eq.~\eqref{d_max_ent}.

Following the results in Sec.~\ref{sec:full-chi1}, the presence of a
maximally entangled Majorana pair on the boundary of 
a cylinder of arbitrary length is
equivalent to the presence of the symmetry on a single column (i.e., a
cylinder of length $N=1$), that is,
\begin{equation}
\hat D_1 (k_y^0) = \left(\begin{array}{cc}
0&\pm 1 \\
\mp 1&0
\end{array}\right)\ .
\label{eq:D1_fixpoint}
\end{equation}
According to Eq. \eqref{eq:hat-D1-fromHVK}, we also have
\begin{equation}
\hat D_1 (k_y^0) = H + K\left(V-\left(\begin{array}{cc}
0&\pm 1\\
\mp 1&0
\end{array}\right) \right)^{-1} K^\top
\label{eq:D1_realmap}
\end{equation}
where the upper sign is for $k_y^0 = 0$ and the lower for $k_y^0
= \pi$ (and is unrelated to the sign in Eq.~\eqref{eq:D1_fixpoint}). We
choose in both cases the upper sign; the other cases can be treated
analogously. Then, Eq.~\eqref{eq:D1_realmap} tells us that $\langle
\omega_v| \Psi_1 \rangle$ (with $\langle \omega_v|$ corresponding to the
projection on $\tfrac12(1 + i c_D c_U)$) is in a maximally entangled state of
the two horizontal Majorana modes. This maximally entangled state fulfills
\be
\langle \omega_h, \omega_v | \Psi_1 \rangle = 0 \label{eq:proj-zero}
\ee
with $\langle \omega_h |$ corresponding to the projection on $\tfrac12(1 +
i c_R c_L)$.

We now parameterize the reduced density matrix of the virtual system $\rho_\mr{vir}$ in the basis $\{|\Omega_\mr{vir} \rangle, |\omega_h\rangle, |\omega_v\rangle, |\omega_h, \omega_v\rangle\}$ ($|\Omega_\mr{vir} \rangle$ denoting the projection on the vacuum of the virtual particles and their subsequent discard). According to Eq.~\eqref{eq:proj-zero} its matrix representation is
\begin{equation}
\rho_\mr{vir} = \frac{1}{4} \left(\begin{array}{cccc}
\rho_{00}&0&0&0\\
0&\rho_{hh}&\rho_{hv}&0\\
0&\rho_{hv}^*&\rho_{vv}&0\\
0&0&0&0
\end{array}\right). \label{eq:Rho-proj}
\end{equation}
From it, we can calculate the elements of $D$ via $D_{p,q} =
\frac{i}{2} \tr\left(\rho_\mr{vir} [c_p, c_q]\right)$ with
$p, q = L, R, U, D$, cf.~Eq.~\eqref{eq:CM-rho}, and obtain
\be
D = \left(\begin{smallmatrix}
0&\rho_{00}-\rho_{hh}+\rho_{vv}&2\Im(\rho_{hv})&-2\Re(\rho_{hv})\\
-\rho_{00}+\rho_{hh}-\rho_{vv}&0&2\Re(\rho_{hv})&2\Im(\rho_{hv})\\
-2\Im(\rho_{hv})&-2\Re(\rho_{hv})&0&\rho_{00}+\rho_{hh} - \rho_{vv}\\
2\Re(\rho_{hv})&-2\Im(\rho_{hv})&-\rho_{00}-\rho_{hh}+\rho_{vv}&0
\end{smallmatrix}\right).  \label{eq:D-projected}
\ee
The fact that $\rho_\mr{vir}$ describes a Gaussian state is used by
inserting this into Eq.~\eqref{eq:D1_realmap}, which gives
\be
\rho_{00} = 1 - \sqrt{(\rho_{hh} - \rho_{vv})^2 + 4 |\rho_{hv}|^2}.
\ee
Given this restriction, one can check that $D$ in
Eq.~\eqref{eq:D-projected} has an eigenvalue $-i$ with the corresponding
symmetry $d_1 = \alpha_L(c_L - i c_R) + \alpha_U (c_U - i c_D)$ fulfilling
Eq.~\eqref{eq:P-annih-d}, where $\alpha_L = 2 \rho_{hv}^*$, $\alpha_U =
-\rho_{hh} + \rho_{vv} - \sqrt{(\rho_{hh} - \rho_{vv})^2 + 4
|\rho_{hv}|^2}$. After considering all possible sign cases in
Eqs.~\eqref{eq:D1_fixpoint}, \eqref{eq:D1_realmap}, one arrives at Eq.~\eqref{d_max_ent}.
We thus find that for a GFPEPS with $\chi=1$, a (unique) symmetry of this
form with $\mathrm{arg}(\tfrac{\alpha_L}{\alpha_U})\notin \{0,\pi,\pm \tfrac\pi2\}$ and $\alpha_L, \alpha_U \neq 0$
is both necessary and sufficient to have a divergence in the boundary
spectrum, and thus a Chern number $C = \pm 1$.  The states simultaneously
fulfilling Eq.~\eqref{d_max_ent} and
$\mathrm{arg}(\tfrac{\alpha_L}{\alpha_U})\in \{0,\pi,\pm \tfrac\pi2\}$, on
the other hand,  are the transition points between
GFPEPS with Chern number $C = -1$ and $C = +1$.

\subsection{Symmetries in the considered examples}\label{sec:symm-examples}

We will now study the symmetries in the examples given in
Sec.~\ref{sec:examples} and relate them to chiral edge modes in the light
of the results of the previous subsections.

\subsubsection{Chern insulator with Chern number $C = -1$}

For the Chern insulator introduced in Sec.~\ref{sec:Chern}, we consider
only one copy of the two superconductors constituting the Chern insulator.
Doing so is trivial on the virtual level, since the matrix $D$ is
block-diagonal with two identical blocks. Each of those blocks has an
eigenvalue $-i$ corresponding to the symmetry
\be
d_1 = -e^{-i\frac{\pi}{4}} c_L - e^{i\frac{\pi}{4}} c_R - i c_U + c_D \label{eq:symm-Chern}
\ee
for any $\eta \in (0,1)$.
Thus, the state $\Psi_1$ possesses two symmetries, $d_1^{(1,2)}$, with
$d_1^{(1)}\ket{\Psi_1}=d_1^{(2)}\ket{\Psi_1}=0$.  Both of them are the form
\eqref{eq:symm-Chern}, with one containing only the first Majorana operator in
left, right, down, up direction and the other only the second Majorana
operator.

\subsubsection{GFPEPS with Chern number $C = 2$\label{sec:sym-chern2}}

Let us now consider the topological superconductor with Chern number $C =
2$ introduced in Sec.~\ref{sec:Chern2}; in the following, all equalities
are to be understood up to numerical accuracy.  By diagonalizing the
CM $D$ of the virtual system, one obtains two
linearly independent eigenvectors with eigenvalue $-i$,
 i.e., there exist two operators
$d_1^{(0,1)}$ such that $(x_0d_1^{(0)}+x_1 d_1^{(1)})\ket{\Psi_1}=0$ for
all $x_i$. In order to find a basis $d_1^{(\pm)}$ of operators which
reveals the symmetries of the model, we start from the state
$\Phi_1$ on one column, which has zero modes at momenta
$k_y=\pm1$. We first focus on the symmetry at $k_y = k_y^0=1$, where
we find that horizontal modes of $\Phi_1$ at momentum $k_y^0$ are
annihilated by an operator $\sum_{\kappa=1}^2 \alpha^{(+)}_{L,\kappa}
[\hat c_{L,\kappa}(k_y^0) - i e^{i k_x^0} \hat c_{R,\kappa}(k_y^0)]$
with $k_x^0=-2.58$ (see Ref.~\onlinecite{note-values}
for the values of the $\alpha$'s).
This suggests to try to construct a $d^{(+)}_1$ which contains the above
operator: it turns out that $x_0d_1^{(0)}+x_1 d_1^{(1)}$ indeed contains
an operator of this form, which at the same time acts on the vertical
modes as $\sum_\kappa\alpha^{(+)}_{U,\kappa}(c_{U,\kappa} - i e^{i k_y^0}
c_{D,\kappa})$. We proceed identically for $k_y = -k_y^0=-1$, and obtain a
pair of (non-orthogonal) symmetries
\begin{align}
d_1^{(\pm)} = \sum_{\kappa=1}^2 [ &\alpha^{(\pm)}_{L,\kappa} (c_{L,\kappa}
- i e^{\pm i k_x^0} c_{R,\kappa}) \notag \\
+ &\alpha^{(\pm)}_{U,\kappa} ( c_{U,\kappa} - i e^{\pm i k_y^0} c_{D,\kappa})]. \label{annihilate_pm}
\end{align}
We thus find that also for this model, the existence of divergences in the
entanglement spectrum and thus of chiral edge modes is closely related to
local symmetries in $\Psi_1$ with the corresponding momenta. Note that
since all coefficients $\alpha$ are different, the only way to grow this
symmetries following the procedure of Sec.~\ref{sec:sufficiency-symmetry}
is to concatenate either exclusively $d_1^{(+)}$ or exclusively
$d_1^{(-)}$, which therefore gives rise to maximally entangled Majorana
pairs between the two boundaries with definite momenta $\pm k_y^0$ and $\pm k_x^0$,
respectively.

\subsubsection{Generic GFPEPS with Chern number $C = 0$}

Let us now consider the non-chiral family of states discussed in
Sec.~\ref{sec:C0-nontrivial}. As it has only one physical mode, there must
be a symmetry $d_1$ such that $d_1\ket{\Psi_1}=0$.  It can be calculated
to be
\be d_1 = - \frac{2 i \sqrt{f(\mu)}}{2-\mu} c_L + i c_R
-\frac{2 \sqrt{f(\mu)}}{2-\mu} c_U + c_D\ .  \ee
As it is not of the form Eq.~\eqref{d_max_ent} required for chiral edge
states, the Chern number of the family is zero.

\subsubsection{GFPEPS with flat entanglement spectrum and $C = 0$}

Let us finally consider the example of Sec.~\ref{sec:ex:nochern-flatspec},
which has a flat entanglement spectrum. It has a symmetry
$d_1\ket{\Psi_1}=0$ with $d_1 = c_L + c_R - i c_U - i c_D$, i.e., with
momentum $k_x^0=k_y^0=\tfrac\pi2$.  Since it is not at momentum $0$ or $\pi$,
there cannot be entangled Majorana modes between the left and the right
edge of a long cylinder.  However, as the amplitudes are equal, the
symmetry is stable under concatenation, and must therefore still be
present in an infinite cylinder.  The explanation is that in the limit
$N\rightarrow\infty$, a second symmetry at $k_y^0=\tfrac\pi2$ arises, such that on 
each edge the two modes at $k_y = \pm \tfrac\pi2$ can pair up locally.

\subsection{Symmetry and ground space}\label{sec:symmetry-GS}

The GFPEPS models discussed in this paper appear as ground states of two
types of Hamiltonians: On the one hand, there is is the flat band
Hamiltonian $\mathcal H_\mr{fb}$, Eq.~\eqref{eq:flat}, which by
construction has the GFPEPS $\Phi$ as its unique ground state. On the
other hand, we can construct the local parent Hamiltonian $\mc H_\mr{ff}$,
Eq.~\eqref{eq:HPEPS}, which is gapless for the chiral examples considered,
i.e., for any finite 
system size, it is exactly doubly degenerate with energy splittings to higher 
energies that are the inverse of a polynomial in the system size.
In the following, we will
show how this ground space can be parametrized by using the virtual
symmetry $d_1$ of the local state $\Psi_1$. This is in close analogy to
the case of conventional PEPS with topological order, where the ground
space can be parametrized by putting loops of symmetry operators on the
virtual bonds in horizontal and vertical direction around the torus on
which the GFPEPS is defined.

In the following, we will consider the example of
Sec.~\ref{sec:Intro-Examples} and show how to parametrize its doubly 
degenerate ground space
in terms of strings of symmetry operators. For simplicity, we will set
$\lambda = 1/2$. Let us start by recalling Eq.~(\ref{eq:symmetries}),
which defines operators $u$, $w$, and $d_1$ such that $u\ket{\Psi_1} =
w\ket{\Psi_1} = d_1\ket{\Psi_1}=0$, where
$u=\tfrac{1}{\sqrt2}(a^\dagger-b)$ and $w=\tfrac{1}{\sqrt2}(a+b^\dagger)$,
with $a$ the physical mode, and $b=\tfrac{1}{\sqrt{2}}(h+v)$,
$d_1=\tfrac{1}{\sqrt2}(-h+v)$, with $h=\exp(i\tfrac\pi4)(c_L-i c_R)/2$ and
$v=(c_U-ic_D)/2$, cf.~Eqs.~\eqref{eq:def-hv} and \eqref{eq:d}.

Let us now consider a lattice of size $N_h \times N_v$, and concatenate all
the $\Psi_1$ in this region by projecting onto $\langle \omega_{jn}|$ and
$\langle \omega_{jn}'|$
on all the horizontal and vertical links,
respectively, but without closing either of the boundaries, resulting in a
state $\Psi_{N_h \times N_v}$. Following the arguments given in
Sec.~\ref{sec:sufficiency-symmetry}, projecting onto the maximally
entangled states concatenates the symmetry operators $u$, $w$ and $d_1$,
which gives rise to three symmetries for the square region,
\[
\tilde u \ket{\Psi_{N_h \times N_v}} = \tilde w \ket{\Psi_{N_v\times
N_h}} = \tilde d_1 \ket{\Psi_{N_h \times N_v}}=0\ ,
\]
where
\begin{subequations}
\label{eq:k0-syms}
\begin{align}
\tilde u & =
\tfrac{1}{\sqrt{2}}(\tilde a^\dg -\tilde b)\ , \\
\label{eq:k0-syms-2}
\tilde w & =
\tfrac{1}{\sqrt{2}}(\tilde a + \tilde b^\dagger)\ , \\
\tilde d_1 & =\tfrac{1}{\sqrt{2}}(-\tilde h+\tilde v)\ ,
\label{eq:k0-syms-3}
\end{align}
\end{subequations}
where again
\begin{align}
\label{eq:k0-b-def}
\tilde b &= \tfrac{1}{\sqrt{2}} (\tilde h + \tilde v), \\
\nonumber
\tilde h= \tfrac{1}{2} e^{i \tfrac{\pi}{4}}(\tilde c_L-i \tilde c_R) &\mbox{\ and\ }
\tilde v= \tfrac{1}{2}(\tilde c_U-i \tilde c_D)\ ,
\end{align}
Here, $\tilde c_p = \sum_y c_{y,p}$, $\tilde c_q = \sum_x
c_{x,q}$ ($p = L, R$, $q = U, D$) 
are the zero-momentum
(center of mass) mode of the virtual Majorana modes on the corresponding
boundary, and $\tilde a = \sum_j a_j$ is the center-of-mass mode of the
physical fermion (with site index $j$).

In order to close the boundary, we first transform the entangled states
across the boundary into the Fourier basis (since they are translational
invariant, they are of the same form in $k$-space), and project onto all
entangled states at the boundary except those with momentum $k_x=0$ and
$k_y=0$.  This leaves us with the zero-momentum part of the state
$\Psi_{N_h \times N_v}$, which we denote by $\tilde \Psi_{N_h \times N_v}$,
where we disregard additional physical modes which are unentangled to the
boundary degrees of freedom. This state is exactly characterized by the
three symmetries of Eq.~(\ref{eq:k0-syms}), and thus
\begin{align*}
\tilde \Pi_{N_h \times N_v} & =
    \ket{\tilde\Psi_{N_h \times N_v}} \bra{\tilde \Psi_{N_h \times N_v}} =
    \tilde d_1 \tilde d_1^\dagger \tilde u \tilde u^\dg \tilde w \tilde w^\dagger\\
    & = \tfrac{1}{4} \tilde d_1 \tilde d_1^\dagger (\tilde a^\dg -\tilde b)(\tilde a-\tilde
b^\dagger)(\tilde a+\tilde b^\dagger)(\tilde a^\dg +\tilde b)\\
    & = \tfrac{1}{2} \tilde d_1 \tilde d_1^\dagger (-\tilde a^\dg \tilde b^\dagger +
	 \tilde a^\dg \tilde a \tilde b^\dagger \tilde b +
	 \tilde b \tilde b^\dagger \tilde a \tilde a^\dg +
	 \tilde a \tilde b)\ .
\end{align*}
Following Eq.~(\ref{eq:omegas}), the projection onto the remaining zero
momentum bonds is $\tilde \omega_h=\tfrac12(1 + i \tilde c_R \tilde c_L)$
and $\tilde \omega_v=\tfrac12(1 + i \tilde c_D \tilde c_U)$. Hence,
$\tilde h^\dagger \ket{\tilde\omega_h} = \tilde v^\dagger
\ket{\tilde\omega_v} = 0$,  and thus, using Eqs.~(\ref{eq:k0-syms-3}) and
(\ref{eq:k0-b-def}), also $\tilde d_1^\dagger
\ket{\tilde\omega_h,\tilde\omega_v} = \tilde b^\dagger\, \ket{\tilde
\omega_h, \tilde \omega_v}  = 0$, i.e.,
\begin{equation}
\label{eq:k0-bond-fockrep}
\ket{\tilde\omega_h,\tilde \omega_v} = \tilde d_1^\dagger \tilde b^\dagger
\ket{\Omega_\mathrm{vir}}
\end{equation}
(with $\ket{\Omega_\mathrm{vir}}$ the vacuum of $\tilde d_1$ and $\tilde
b$, or equivalently of $\tilde h$ and $\tilde v$; the phase can be
absorbed in $\ket{\Omega_\mathrm{vir}}$).

Let us now see what happens when we close the remaining $(k_x, k_y) = (0,0)$ boundary. 
Since $\tilde \Pi_{N_h \times N_v}$ is proportional to $\tilde d_1\tilde
d_1^\dagger$, we find
that
\[
\langle \tilde \omega_h, \tilde \omega_v| \, \tilde \Pi_{N_h \times N_v}
\, | \tilde \omega_h, \tilde \omega_v\rangle = 0
\]
-- the success probability for constructing the GFPEPS by projecting onto
entangled states is zero!  Indeed, this comes as no surprise, since the
success probability of any such projection is related to
$\sqrt{\det(D+\omega^{-1})}$ in Eq.~(\ref{eq:proj-ME})~\cite{Bra05}, which
in turn is the square root of the spectral function of the parent
Hamiltonian as constructed in Ref.~\onlinecite{Wah13} (which generally, and in
particular for the example considered, is equal to $\mc H_\mr{ff}$):
Having a gapless parent Hamiltonian requires the GFPEPS to vanish when
performing the projections. This raises the question of how to obtain a
proper PEPS description of the ground state subspace.

Fortunately, this problem can be overcome exactly by using the virtual
symmetry of $\Psi_1$. To this end, let us place a string of symmetry
operators
\[
\tilde c_L = \tfrac{1}{\sqrt{2}} e^{-i \tfrac{\pi}{4}} (-\tilde d_1+\tilde b)+\tfrac{1}{\sqrt{2}} e^{i \frac{\pi}{4}} (-\tilde d_1^\dagger+\tilde b^\dagger)
\]
at the left edge before closing the boundary, i.e., we replace the state
$| \tilde \omega_h, \tilde \omega_v\rangle=\tilde d_1^\dagger \tilde
b^\dagger \ket{\Omega_\mathrm{vir}}$ by
\[
\ket{\tilde \omega_L}=\tilde c_L | \tilde \omega_h, \tilde \omega_v\rangle =
\tfrac{-1}{\sqrt{2}} e^{-i \tfrac{\pi}{4}} (\tilde b^\dagger+\tilde
d_1^\dagger) \ket{\Omega_\mathrm{vir}}\ .
\]
Using that $\tilde \Pi_{N_h \times N_v}$ is proportional to $\tilde
d_1\tilde d_1^\dagger$, this immediately yields
\begin{equation}
\bra{\tilde \omega_L} \tilde \Pi_{N_h \times N_v}\ket{\tilde \omega_L} =
\tfrac12\bra{\Omega_\mathrm{vir}} \,\tilde b \,
 \tilde \Pi_{N_h \times N_v} \tilde b^\dagger \ket{\Omega_\mathrm{vir}} =
\tfrac{1}{4} \tilde a^\dg \tilde a\ ,
\label{eq:sym-gs:gs1}
\end{equation}
i.e., this way we obtain a GFPEPS for one of the ground states of the
parent Hamiltonian $\mc H_\mr{ff}$, namely the one with the gapless
center-of-mass mode occupied.  It is easy to see that we obtain the same
result when we insert a horizontal string instead, e.g., $\tilde c_U$. In
terms of the notation introduced in Sec.~\ref{sec:symm-top}, $\ket{\Phi} =
0$ and $\ket{\Phi_{{\mc C}_h}} \propto \ket{\Phi_{\mc C_v}}$. Note that
the string operators $\tilde c_L$ and $\tilde c_U$ can be deformed
without changing the state: This can be seen by consecutively using
Eq.~\eqref{eq:dPsi0} to deform the string as
\begin{align}
 \ket{\Phi_{{\mc C}_h}} &= \langle \omega_{\partial \mc R, \partial \bar{\mc R}} | \tilde c_L | \Psi_{\mc R}, \Psi_{\bar{\mc R}}\rangle = \langle \bar \omega | \tilde c_L | \bar \Psi\rangle \notag \\
 &= \langle \bar \omega | \left(\sum_{y =2}^{N_v} c_{y,L} + i c_{1,R} + e^{-i\tfrac{\pi}{4}} c_{1,U} - e^{i \tfrac{\pi}{4}} c_{1,D}\right)|\bar \Psi\rangle  \notag\\
 &= \ket{\Phi_{\mc C'_h}}
\end{align}
etc., where we defined $|\bar \Psi\rangle$ as the state of all virtual and
physical particles before any projection is applied, and $\langle \bar
\omega |$ denotes the projection on all virtual modes.

Let us now finally see what happens if we insert both a horizontal and a
vertical string: Then, we must replace
$\ket{\tilde \omega_h, \tilde \omega_v}$ by
\[
\ket{\tilde \omega_{UL}}=
\tilde c_U \tilde c_L | \tilde \omega_h, \tilde \omega_v \rangle =
-e^{-i\pi/4} |\Omega_\mr{vir}\rangle,
\]
and we find
\begin{equation}
\label{eq:sym-gs:gs2}
\bra{\tilde\omega_{UL}} \tilde\Pi_{N_h \times N_v}\ket{\tilde\omega_{UL}} =
    \bra{\Omega_\mr{vir}} \,
\tilde \Pi_{N_h \times N_v}\ket{\Omega_\mr{vir}} = \tfrac12 \tilde a \tilde a^\dg\ ,
\end{equation}
which is the second ground state, where the gapless center-of-mass mode is
in the vacuum.

Note that the second ground state can equivalently be obtained using that
$\tilde a \ket{\tilde \Psi_{N_h \times N_v}} = -\tilde b^\dg \ket{\tilde
\Psi_{N_h \times N_v}}$ [Eq.~\eqref{eq:k0-syms-2}], which exactly cancels the
$\tilde b$ in Eq.~(\ref{eq:sym-gs:gs1}), and thus yields
Eq.~\eqref{eq:sym-gs:gs2}.

In summary, we find that it is possible to parametrize the two-dimensional ground
state subspace of the model using the string operators given by the virtual symmetry
of $\Psi_1$: One of the ground states is obtained by inserting a single
string (either horizontally or vertically), while the other ground state is
obtained by inserting \emph{both} a horizontal and a vertical string.


\section{Conclusions and Outlook}

In this paper, we have established a framework for boundary and edge
theories for Gaussian fermionic Projected Entangled Pair States (GFPEPS),
and applied it to the study of chiral fermionic PEPS, and in particular
their underlying symmetry structure.

We have introduced two different kinds of Hamiltonians, the boundary
Hamiltonian $\mc H_N^\mr{b}$ and the edge Hamiltonian $\mc H_N^\mr{e}$.
The former reproduces the entanglement spectrum of the reduced density
matrix of a region as a thermal state $\exp(-\mc H_N^\mr{b})$, while the
latter contains the low energy physics of the truncated flat band
Hamiltonian $\mc H_\mr{fb}$. We have shown that in the context of GFPEPS,
both of these Hamiltonians act on the auxiliary degrees of freedom at the
boundary, which naturally imposes a one-dimensional structure, and that
they are related in a simple way.  As the physical edge modes
corresponding to $\mathcal H_N^\mr{e}$ are localized at the same edge of a
cylinder, the number of chiral edge modes and thus the Chern number of a
GFPEPS can be read off the virtual boundary and edge Hamiltonian.  We have
also provided constructive methods for analytically and numerically
determining $\mc H_N^\mr{b}$ and $\mc H_N^\mr{e}$ for general GFPEPS, and
in particular on infinite cylinders and tori.

We have subsequently provided a full analysis of the edge and boundary
Hamiltonian for the case of GFPEPS with one Majorana mode per bond,
$\chi=1$. We have put particular emphasis on the case of GFPEPS with
chiral edge modes, where we have shown that the presence of chiral edge
modes is equivalent to a maximally entangled state 
between the virtual Majorana modes at 
the two boundaries of a cylinder, which leads to a
divergence in the entanglement spectrum at the corresponding momentum.
Subsequently, we have related this global virtual symmetry in the GFPEPS
to a local virtual symmetry in the PEPS tensor $\Psi_1$. Identifying such
symmetries has proven extremely powerful in the case of non-chiral
topological models, where it has allowed for a comprehensive understanding
of ground state degeneracy, topological entropy, excitations, and more
from a simple local symmetry and the strings formed by it. We have shown
that the virtual symmetry of chiral GFPEPS is similarly powerful, as it
explains the origin of chiral edge modes, the topological correction to
the R\'enyi entropy, and it allows to parametrize the ground state space of
the gapless parent Hamiltonian using strings formed by the symmetry. It is
an interesting question to understand further implications of the
symmetry, such as the excitations obtained from open strings, or the role
played by symmetries for fermionic PEPS with higher bond dimension $\chi$.
Our numerical results indeed suggest that the same type of symmetries
underlies chiral edge modes for $\chi>1$.

Understanding the local symmetries underlying chiral topological order is
of particular interest when going to interacting models, since these local
symmetries will still give rise to maximally entangled Majorana 
modes between
distant edges even for interacting models; keeping the symmetry structure
of the local PEPS tensor untouched thus seems to be a crucial ingredient
when adding interactions. This can in particular be achieved by taking
several copies of a chiral GFPEPS and coupling the copies on the physical
level without changing the auxiliary modes, for instance by a Gutzwiller
projection (cf. Ref.~\onlinecite{Dub13}), similar to the way in which fractional
Chern insulators are constructed; we are currently pursuing
research in this direction.

\subsection*{Acknowledgements}
We thank the Perimeter Institute for Theoretical Physics (Waterloo) and
the Simons Institute for the Theory of Computing (Berkeley), where parts
of this work were carried out, for their hospitality.  TBW acknowledges
helpful discussions with T.~Shi and financial support by QCCC Elitenetzwerk Bayern. JIC and NS thank X.-L.~Qi,
F.~Verstraete, and M.~Zaletel for discussions. JIC acknowledges support by the Miller Institute in Berkeley, and NS acknowledges support by
the Alexander von Humboldt foundation.
Part of this work has been supported by the EU integrated project SIQS.

\onecolumngrid

\vspace*{0.4cm}
\begin{center}
\rule{14cm}{0.03cm}
\end{center}
\vspace*{0.4cm}

\twocolumngrid

\appendix

\section{\label{app:corr}
Decay of correlations in real space}

In this part of the Appendix we show that the correlations of the
GFPEPS defined via Eq.~\eqref{eq:Psi1ex} and therefore also the hoppings
of the corresponding flat band Hamiltonian $\mc H_\mr{fb}$ in
Eq.~\eqref{eq:H-Gout} decay like the inverse of the distance cubed. More
precisely, we will show that the $\hat d_j(k)$ in Eq.~\eqref{eq:G-param}
($j = x,y$) decay at least as fast as $\tfrac{\log(|r|)}{|r|^3}$, but not
faster than $\tfrac{1}{|r|^3}$ in real space (by analogous arguments it
can be shown that $\hat d_z(k)$ corresponds to a faster decay than the
inverse distance cubed). Crudely speaking, the reason for this decay is
that the $\hat d_j(k)$ have a non-analytical point at $k = (0,0)$, where
they are continuous, but not continuously differentiable. 

An important fact which we will need in the proof is the following
relation between the decay of Fourier coefficients and the smoothness of
the corresponding Fourier series, stated for the relevant case of two
dimenions: Given that the Fourier coefficients decay faster than
$|r|^{-(2+d)}$ (i.e., they are upper bounded by a constant 
times $|r|^{-(2+d+\delta)}$ for some
$\delta>0$), it follows that the Fourier series is $d$ times continuously
differentiable (continuous if $d=0$);  see, e.g., Proposition 3.2.12 in
Ref.~\onlinecite{grafakos}.

Let us start by considering the behavior of $\hat d(k)$ around the
non-analytical point $k = (0,0)$. For simplicity, we again restrict
ourselves to $\lambda = 1/2$, but the arguments for other $\lambda$ are
the same. We expand the
numerators and denominators in Eqs.~\eqref{eq:dx} and \eqref{eq:dy} to
second order and those in Eq.~\eqref{eq:dz} to fourth order around $k =
(0,0)$ to obtain 
\begin{align}
\hat d_x(k) &= \frac{-2 k_x k_y^2}{k_x^2 + k_y^2} + \mc{O} (k^2)  \label{eq:dx-approx},\\
\hat d_y(k) &= \frac{2 k_x^2 k_y}{k_x^2 + k_y^2} + \mc{O} (k^2)  \label{eq:dy-approx},\\
\hat d_z(k) &= -1 + \frac{2 k_x^2 k_y^2}{k_x^2 + k_y^2} + \mc{O} (k^3).  \label{eq:dz-approx}
\end{align}
This shows that the $\hat d_{x,y}(k)$ are continuous, but not continuously
differentiable at $k = (0,0)$, whereas $\hat d_z(k)$ is both (and only its
second derivative is non-continuous). This implies that the $\hat d_{x,y}$
cannot asymptotically decay faster than $\tfrac{1}{|r|^3}$ in real space,
since otherwise their Fourier transform would be continuously
differentiable.  This demonstrates the claimed lower bound bound on the
decay of the correlations. 

The upper bound is obtained by formally carrying out the Fourier transform
and bounding the terms obtained after
partial integration: To simplify notation we suppress the index $x$ or $y$
in $\hat d_{x,y}(k)$, respectively (the result applies to both of them and
also to the overall hopping amplitude of the Hamiltonian). Let us assume
that the site coordinates fulfill $|x| \geq |y|$ ($x \neq 0$); in the
opposite case the line of reasoning is the same. We integrate its Fourier
transform twice with respect to $k_x$ by parts, ($r = (x,y)$)
\begin{align}
d_{r} &= \int_\mr{BZ} \hat d(k) e^{-i k \cdot r} \mr{d} k_x \mr{d} k_y \notag \\
&= \left(-\frac{1}{-i x}\right)^2 \int_{-\pi}^\pi \mr{d} k_y \int_{-\pi}^\pi \frac{\partial^2 \hat d(k)}{\partial k_x^2} e^{-i k \cdot r} \mr{d} k_x, \label{eq:double-int}
\end{align}
where BZ denotes the first Brillouin zone, that is, $(-\pi,\pi] \times
(-\pi,\pi]$. Let us first show that the last double integral is defined,
although its integrand might diverge at $k = (0,0)$: For that, we will
demonstrate the bounds 
\begin{equation} 
\left| \frac{\partial^2 \hat
d(k)}{\partial k_x^2}\right| < \frac{c}{|k|}, \ \left| \frac{\partial^3
\hat d(k)}{\partial k_x^3}\right| < \frac{c'}{|k|^2} \label{eq:bounds}
\end{equation}
with $c, c' > 0$. In order to show the first bound, we realize that $\frac{\partial^2 \hat d(k)}{\partial k_x^2} \sqrt{k_x^2 + k_y^2}$ and $\frac{\partial^3 \hat d(k)}{\partial k_x^3} (k_x^2 + k_y^2)$
cannot diverge anywhere but at $k = (0,0)$. We expand them for $\hat d(k) = \hat d_x(k)$ around this point by setting $k = (|k| \cos(\phi), |k| \sin(\phi))$ and obtain
\begin{align}
\frac{\partial^2 \hat d(k)}{\partial k_x^2} \sqrt{k_x^2 + k_y^2} &\xrightarrow[|k| \rightarrow 0]{}  \frac{-4(\cos(3 \phi) \sin^2(\phi)) + \mc{O}(|k|)}{1 + \mc{O}(|k|)}, \label{eq:2nd-limit} \\
\frac{\partial^3 \hat d(k)}{\partial k_x^3} (k_x^2 + k_y^2) &\xrightarrow[|k| \rightarrow 0]{} \frac{12 \cos(4 \phi) \sin^2(\phi) + \mc{O}(|k|)}{1 + \mc{O}(|k|)}. \label{eq:3rd-limit}
\end{align}
Therefore, the limit $|k| \rightarrow 0$ exists for all
$\phi$ and is uniformly bounded, and as a result, the expressions on the
left hand side of Eqs.~\eqref{eq:2nd-limit} and~\eqref{eq:3rd-limit} are
bounded for any $k \in \mr{BZ}$. The same thing is encountered for $\hat
d(k) = \hat d_y(k)$.  Since the left hand sides of Eqs.
\eqref{eq:2nd-limit} and
\eqref{eq:3rd-limit} do not diverge for any $k$ and are defined for a
finite region (the first Brillouin zone), the bounds \eqref{eq:bounds} are
correct. The first bound implies that the double integral
\eqref{eq:double-int} is defined (and finite).

It will be convenient to split the integral \eqref{eq:double-int} into two parts, one with range over the full circle $C_\epsilon$ of radius $\epsilon$ centered at $k = (0,0)$ and the rest. The first part is bounded in absolute value by $\int_{C_\epsilon} \frac{c}{|k|}\mr{d}^2 k = 2 \pi c \epsilon$. Thus, employing another partial integration
\begin{align}
&|d_r| <  \frac{2 \pi c \epsilon}{x^2} + \frac{1}{x^2} \left| \int_{\mr{BZ} \setminus C_\epsilon}  \frac{\partial^2 \hat d(k)}{\partial k_x^2} e^{-i k \cdot r} \mr{d}^2 k \right| \notag \\
&= \frac{2 \pi c \epsilon}{x^2} + \frac{1}{x^2} \left| \left(\frac{1}{-i x}\right) \left(  \int_{-\pi}^\pi \mr{d} k_y \left[\frac{\partial^2 \hat d(k)}{\partial k_x^2} e^{-i k \cdot r}  \right]^{\sqrt{\epsilon^2-k_y^2}}_{-\sqrt{\epsilon^2 - k_y^2}} \times \right. \right. \notag \\
&\left. \left. \times \theta(\epsilon^2 - k_y^2) - \int_{\mr{BZ} \setminus C_\epsilon} \frac{\partial^3 \hat d(k)}{\partial k_x^3}e^{-i k \cdot r}  \mr{d}^2 k \right) \right|.
\end{align}
We use the bounds on the second and third derivative of $\hat d(k)$,
\begin{align}
|d_r| &< \frac{2 \pi c \epsilon}{x^2} + \frac{1}{|x|^3} \left(2 \pi c  + \left| \int_{\mr{BZ} \setminus C_\epsilon} \frac{\partial^3 \hat d(k)}{\partial k_x^3} e^{-i k \cdot r}  \mr{d}^2 k \right|\right) \\
&< \frac{2 \pi c \epsilon}{x^2} + \frac{1}{|x|^3} \left(2 \pi c + \int_{\mr{BZ} \setminus C_\epsilon} \frac{c'}{|k|^2} \mr{d}^2 k \right) \notag \\
&< \frac{2 \pi c \epsilon}{x^2} + \frac{1}{|x|^3} \left(2 \pi c  + 2 \pi c' (\ln(\sqrt{2} \pi) - \ln(\epsilon))\right).
\end{align}  
We now set $\epsilon = \tfrac{1}{|x|}$ to obtain
\be
|d_r| < \frac{2 \pi (2 c  + c' \ln(\sqrt{2} \pi |x|))}{|x|^3}.
\ee
After realizing that $|x| \geq \tfrac{|r|}{\sqrt{2}}$, this leads to
\be
|d_r| < \frac{a + b \ln(|r|)}{|r|^3}
\ee
($a, b > 0$). The decay of $\hat d_z$ in real space is faster, since its derivatives start diverging at a higher order. Hence, the hoppings decay at least as fast as $\tfrac{\ln(|r|)}{|r|^3}$ and, therefore, for large $|r|$ as the inverse distance cubed.

\section{\label{app:euler-maclaurin}
Momentum polarization and topological entanglement entropy}

In this Appendix, we derive analytical expressions for two quantities which
probe topological order based on the entanglement spectrum, namely
the momentum polarization and the topological entropy, for the case of
non-interacting fermions, i.e., Gaussian states.  First, we will prove
that the universal contribution to the momentum
polarization~\cite{mom-pol} is exactly determined by the number of
divergences in the entanglement spectrum ($\hat H^\mathrm{b}_N(k_y)$ in
the case of GFPEPS);  and second, we will prove that there is no additive
topological correction to the von Neumann entropy $S_\mathrm{vN}$ of the
entanglement spectrum.  Let us stress that both of these
arguments rely only on few properties of the entanglement spectrum and the
corresponding boundary Hamiltonian, and are thus not restricted to the
case of GFPEPS.

Both these proofs
are based on the Euler-Maclaurin formulas, which for our purposes say the
following: Given a function $f:[0,2\pi]\rightarrow \mathbb C$ which is $3$
times continuously differentiable, it holds that
\begin{widetext}
\begin{align}
\sum_{k=0}^N f\left(\frac{2\pi k}{N}\right)  - \frac{f(0)+f(2\pi)}{2} &=
\frac{N}{2\pi}\int\limits_{0}^{2\pi}f(x)\,\mathrm{d}x +
\frac{2\pi\,(f'(2\pi)-f'(0))}{12\,N} + \mc O(1/N^3),
\\
\label{eq:euler-maclaurin-2}
\sum_{k=1}^N f\left(\frac{\pi(2k-1)}{N}\right) & =
\frac{N}{2\pi}\int\limits_{0}^{2\pi}f(x)\,\mathrm{d}x -
\frac{2\pi\,(f'(2\pi)-f'(0))}{24\,N} + \mc O(1/N^3)\ .
\end{align}
\end{widetext}

Let us now first discuss how to compute the momentum polarization; for
clarity, we will focus on two copies of the superconductor defined in
Sec.~\ref{sec:Intro-Examples}, but the arguments can be readily
adapted. For a state $\ket\varphi$ on a long cylinder which is
partitioned into two cylinders $A$ and $B$, the momentum
polarization~\cite{mom-pol} is $\mu(N_v)=\bra\varphi T_A\ket\varphi$, where
$T_A$ translates part $A$ of the system around the cylinder axis, and
$N_v$ is the circumference of the cylinder; and it
is expected to scale as $\exp[-\alpha N_v + \tfrac{2\pi
i}{N_v}(h_a-\tfrac{c}{24})]$, where $c$ is the chiral central charge and
$h_a$ the topological spin, and $\alpha\in\mathbb C$ is non-universal. It
is immediate to see that this definition is equivalent to evaluating
$\mu(N_v)=\sum_\ell \lambda_\ell e^{i k_\ell}$, where $\lambda_\ell$ is the
entanglement spectrum of $A$, i.e., $\ket\varphi = \sum_\ell \sqrt{|\lambda_\ell|}
\ket{\varphi^A_\ell}\ket{\varphi^B_\ell}$, and $k_\ell$ is the momentum of
$\ket{\varphi^A_\ell}$.  In PEPS, the entanglement spectrum corresponds to
a state on the boundary degrees of freedom, and therefore this expression
can be evaluated directly at the boundary.  Concretely, in the case of two
states with one fermion per bond (i.e., $\chi=2$), such as two copies of
the superconductor of Sec.~\ref{sec:Intro-Examples}, the entanglement
spectrum corresponds to the thermal state of the non-interacting
Hamiltonian $H^\mr{b}_{N}$, so that the momentum polarization is given by
\begin{equation}
    \label{eq:mompol-sum}
\log(\mu(N_v)) = \sum_k
\underbrace{
\log\frac{e^{-\omega_k}+e^{ik+\omega_k}}{e^{-\omega_k}+e^{\omega_k}}}_{=:f(k)}\ ,
\end{equation}
where $\omega_k$ is the energy of the boundary mode with momentum $k\equiv
k_y$, as shown in Fig.~\ref{Gamma_R1}.  To evaluate the sum
(\ref{eq:mompol-sum}), we use the Euler-Maclaurin formulas, where $f(k)$
is defined via the summand in (\ref{eq:mompol-sum}) on
the open interval $(0,2\pi)$, and continuously extended to $[0,2\pi]$. In
order to ensure continuity of $f$, we follow the different branches of the
logarithm (i.e., we add $2\pi i$ as appropriate). Moreover, for examples
with a gapless mode at $k=\pi$ (such as the examples of
Sec.~\ref{sec:Intro-Examples}) $f(k)$ diverges,
which can be fixed by replacing $e^{ik}$ by $e^{2ik}$ above (and
subsequently correcting for the
factor of $2$ obtained in the scaling). For the examples considered, the functions
$f$ obtained this way are indeed $3$ times continuously differentiable.  Which of the two
Euler-Maclaurin equations we use depends on whether the sum in
(\ref{eq:mompol-sum}) runs over $k=2\pi n/N_v$ or $k=2\pi(n+\tfrac12)/N_v$ ($n = 0, \ldots, N_v-1$),
which is connected to the choice of boundary conditions.  We will focus on
the case $k=2\pi(n+\tfrac12)/N_v$, but let us note that the difference in
the relevant subleading terms is merely a factor of $-2$ in the $1/N_v$
term (which in the examples relates to a non-zero topological spin $h_a$)
and a trivial additive term proportional to $f(2\pi)-f(0)$
which relates to the treatment of the branches of the logarithm.

With this choice of $k$, using
(\ref{eq:euler-maclaurin-2}) we find that
\[
\log\,\mu(N_v) = \alpha N_v - \frac{2\pi i}{N_v}\tau + \mc O(1/N_v^3)\ ,
\]
where $\alpha=\tfrac{1}{2\pi}\int f(x)\,\mathrm{d}x$ is non-universal, and
$\tau = \frac{1}{24i} (f'(2\pi)-f'(0))$.  It is now easy to check that
for $k_0=0,2\pi$,
\[
f'(k_0) = \lim_{k\rightarrow k_0}
\left[\frac{i\,e^{2
\omega_k}}{1+e^{2\omega_k}} + \mc O(k-k_0)\right]
\]
and thus a divergence in the entanglement spectrum at $k_0=0$, such as for
the example of Sec.~\ref{sec:Intro-Examples},  implies that $f'(2\pi)-f'(0)=\pm
i$.  We thus find that $\tau$ is universal, with its value only depending
on the presence of a divergence in the entanglement spectrum, but not on
the exact form of $\omega_k$. In particular, with $\tau=c/24$, we find a
chiral central charge of $c=1$ for two copies of the superconductor, which
amounts to $c=1/2$ for a single copy of the topological superconductor.
Note that the Euler-MacLaurin formulas can be easily adapted to deal with
more discontinuities and with different values of $k$, by expanding $f(k)$ 
in terms of Bernoulli polynomials; thus, the outlined approach allows
for the analytical calculation of the momentum polarization for general
free fermionic systems with several boundary modes and arbitrary
fluxes through the torus.

Let us conclude by discussing the scaling of the topological entropy,
which is given by $S_{\mathrm{vN}}(N_v) = \sum_k g(k)$, $g(k)=-p_k\,\log p_k -
(1-p_k)\log(1-p_k)$, $p_k=e^{-\omega_k}/(e^{-\omega_k}+e^{\omega_k})$ (in
particular, $g(k)\rightarrow0$ for $k\rightarrow0,2\pi$). For the cases
discussed in the paper, $g'(k)$ is continuous and periodic, but its second
derivative diverges; thus, the error term in the Euler-Maclaurin formula
can be of order $o(1/N_v)$. Yet, this is sufficient as we are only
interested in \emph{constant} corrections to the entanglement entropy, and
one immediately finds that both for periodic and anti-periodic boundary
conditions, $S_{\mathrm{vN}}(N_v)=a N_v + o(1/N_v)$, with a non-universal
$a=\tfrac{1}{2\pi}\int_0^{2\pi}g(k)\,\mathrm{d}k$, and no constant
topological correction.

\section{Polynomial decay of the boundary Hamiltonian hoppings}\label{app:poly-hopping}

In this part of the Appendix, we prove that the hopping amplitudes
$|[H_\infty^{R}]_{1,1+y}|$ of the boundary Hamiltonian of the example of
Sec.~\ref{sec:Intro-Examples}, shown in Fig.~\ref{fig:Ham}, decay as
$\ln(y)/y$.

We start by calculating the single-particle entanglement spectrum on the right boundary:
For that we employ Eq.~\eqref{eq:hat-D1-fromHVK} to calculate $\hat D_1(k_y)$ for the topological superconductor defined by Eq.~\eqref{eq:Psi1ex} and from it $\hat \Sigma^R_\infty(k_y)$  via Eq.~\eqref{eq:sigma_es_chi1} as a function of $\lambda$. The result is
\be
\hat \Sigma^R_\infty(k_y) = i \frac{2 \lambda^2 \sin(k_y)}{\sqrt{\frac{g^2(k_y)}{|1-\lambda - e^{i k_y}|^4} + 4 \lambda^4 \sin^2(k_y)}}
\ee
with $g(k_y)$ some second order polynomial in $\cos(k_y)$. For $\lambda
\neq 0$ this function is analytic as long as $k_y$ is not an integer
multiple of $\pi$. One can check that $g(\pi) \neq 0$ for any $\lambda \in
(0,1)$, so $\hat \Sigma^R_\infty(\pi) = 0$ and the only possible
non-analytical point is $k_y = 0$. As shown in Sec.~\ref{sec:full-chi1},
these are the only $k_y$-points where $|\hat
\Sigma^R_\infty(k_y)| = 1$ is possible and where hence the spectrum of the
boundary Hamiltonian can diverge: One can check from the explicit function
$g(k_y)$ that $g(\delta k_y) = g(-\delta k_y) = g_0 \, \delta k_y^{2} (1 +
\mathcal{O}(\delta k_y^2))$ for $\lambda \in (0,1)$ (where $g_0$ depends on
$\lambda$). Therefore,
\begin{align}
\hat \Sigma^R_\infty(\delta k_y) &= i \frac{2 \lambda^2 \delta k_y}{\sqrt{\frac{g_0^2 \delta k_y^{4}(1 + \mathcal{O}(\delta k_y^2))}{(2-\lambda)^4} + 4 \lambda^4 \delta k_y^2}} \notag \\
&= i \left(1 - \frac{g_0^2}{8 \lambda^4 (2-\lambda)^4} \delta k_y^{2}\left(1 + \mathcal{O}(\delta k_y^2)\right)\right) \mr{sgn}(\delta k_y)
\end{align}
Owing to Eq.~\eqref{eq:HN} for $N \rightarrow \infty$, the single-particle spectrum is given by
\be
- i \hat H_\infty^R(k_y) = \ln \left( \frac{1 - i \hat \Sigma^R_\infty (k_y)}{1 + i \hat \Sigma^R_\infty(k_y)}\right).
\ee
Henceforth, we can expand
\begin{align}
- i \hat H_\infty^R(\delta k_y) &= \big[\ln\left(\frac{16 \lambda^4 (2-\lambda)^4}{g_0^2}\right) - 2\ln(\delta k_y) \notag \\
&- \ln(1 + \mathcal{O}(\delta k_y^2)) \big] \mr{sgn}(\delta k_y),
\end{align}
and we see that the non-analycity is only due to the term $2\ln(\delta k_y)$, the other ones being analytical around $k_y = 0$. The Fourier coefficients of an analytical function defined on $(-\pi,\pi]$ decay exponentially. Thus, the algebraic decay of $|[H_\infty^{R}]_{1,1+y}|$ is due to the diverging term we singled out,
\begin{align}
|[H_\infty^{R}]_{1,1+y}| &\xrightarrow[y \rightarrow \infty]{} \frac{4}{\sqrt{N_v}} \int_0^\pi \sin(k_y y) \ln(k_y) \d k_y \notag \\
&\xrightarrow[y \rightarrow \infty]{} \frac{4}{\sqrt{N_v}} \left[\frac{\ln(y)}{y} + \mc{O}\left(\frac{1}{y}\right)\right]
\end{align}
with the prefactor of the ${1}/{y}$ contribution being depending on whether $y$ is even or odd but constant otherwise.


\begin{thebibliography}{99}
\bibitem{Wen90} X.-G. Wen, Int. J. Mod. Phys. B \textbf{4}, 239 (1990).


\bibitem{Wen89} X.-G. Wen, Phys. Rev. B. \textbf{40}, 7387 (1989).

\bibitem{Kit03} A. Kitaev, Ann. Phys. \textbf{303}, 2 (2003), quant-ph/9707021.

\bibitem{Lev05} M. A. Levin and X.-G. Wen, Phys. Rev. B \textbf{71}, 045110
(2005), cond-mat/0404617.






\bibitem{Hal88} F. D. M. Haldane, Phys. Rev. Lett. \textbf{61}, 2015 (1988).

\bibitem{Kit08} A. Kitaev, the Proceedings of the L. D. Landau Memorial
Conference \textquotedblleft Advances in Theoretical
Physics\textquotedblright (2008), arXiv:0901.2686.

\bibitem{Sch08} A. P. Schnyder, S. Ryu, A. Furusaki, and A. W. W. Ludwig,
Phys. Rev. B \textbf{78}, 195125 (2008), arXiv:0803.2786.




\bibitem{Pol09} F. Pollmann, A. M. Turner, E. Berg, and M. Oshikawa, Phys.
Rev. B \textbf{81}, 064439 (2010), arXiv:0910.1811.

\bibitem{Che11} X. Chen, Z.-C. Gu, and X.-G. Wen, Phys. Rev. B \textbf{83},
035107 (2011), arXiv:1008.3745.

\bibitem{Sch11} N. Schuch, D. Perez-Garcia, and I. Cirac, Phys. Rev. B
\textbf{84}, 165139 (2011), arXiv:1010.3732.

\bibitem{Scholl11} U. Schollw\"o{ck},
\newblock Ann. Phys. (NY) \textbf{326}, 96 (2011), arXiv:1008.3477.

\bibitem{Hal08} H. Li and F. D. M. Haldane, Phys. Rev. Lett. \textbf{101},
010504 (2008), arXiv:0805.0332.

\bibitem{Ver04} F. Verstraete and J. I. Cirac, cond-mat/0407066; F.
Verstraete and J. I. Cirac, Phys. Rev. A \textbf{70}, 060302 (2004), quant-ph/0311130.


\bibitem{And73} P. W. Anderson, Mater. Res. Bull. \textbf{8}, 153 (1973).


\bibitem{Ver06} F. Verstraete, M. M. Wolf, D. Perez-Garcia, and J. I. Cirac,
Phys. Rev. Lett. \textbf{96}, 220601 (2006), quant-ph/0601075.

\bibitem{Sch10} N. Schuch, J. I. Cirac, and D. Perez-Garcia, Ann. Phys.
\textbf{325}, 2153 (2010), arXiv:1001.3807.

\bibitem{Bue08} O. Buerschaper, M. Aguado, and G. Vidal, Phys. Rev. B
\textbf{79}, 085119 (2009), arXiv:0809.2393.

\bibitem{Gu08} Z.-C. Gu, M. Levin, B. Swingle, and X.-G. Wen, Phys. Rev. B
\textbf{79}, 085118 (2009), arXiv:0809.2821.

\bibitem{Bue13}
O.~Buerschaper,
\newblock arXiv:1307.7763.

\bibitem{Sch13} N. Schuch, D. Poilblanc, J. I. Cirac, and D. Perez-Garcia,
Phys. Rev. Lett. \textbf{111}, 090501 (2013), arXiv:1210.5601.

\bibitem{Cir11} J. I. Cirac, D. Poilblanc, N. Schuch, and F. Verstraete,
Phys. Rev. B \textbf{83}, 245134 (2011), arXiv:1103.3427.

\bibitem{Shuo}
S. Yang, L. Lehman, D. Poilblanc, K. Van Acoleyen, F. Verstraete, J. I. Cirac, and N. Schuch,
\newblock Phys. \ Rev. \ Lett. {\bf 112}, 036402 (2014), arXiv:1309.4596.

\bibitem{Wah13} T. B. Wahl, H.-H. Tu, N. Schuch, and J. I. Cirac, Phys. Rev.
Lett. \textbf{111}, 236805 (2013), arXiv:1308.0316.

\bibitem{Dub13} J. Dubail and N. Read, arXiv:1307.7726.

\bibitem{kraus:fPEPS}
C.~V. Kraus, N.~Schuch, F.~Verstraete, and J.~I. Cirac,
\newblock Phys.\ Rev.\ A {\bf 81}, 052338 (2010), arXiv:0904.4667.

\bibitem{Cor10}
P. Corboz, R. Orus, B. Bauer, and G. Vidal,
\newblock Phys. Rev. B {\bf 81}, 165104 (2010), arXiv:0912.0646.

\bibitem{mom-pol}
H.-H. Tu, Y. Zhang, and X.-L. Qi,
Phys. Rev. B {\bf 88}, 195412 (2013), arXiv:1212.6951.

\bibitem{note-Majorana}
The name Majorana mode is not
really appropriate, as two of them define a single (fermionic) mode.
Nevertheless, we will use such a nomenclature here.

\bibitem{note:ME}
The motivation
for the name ``maximally entangled Majorana modes" comes from the following
construction (see Ref.~\onlinecite{kraus:fPEPS}): One can add an extra Majorana
operator per bond such that each bond is represented by a full fermionic
mode. In that case, the state generated by $\omega_{j,n}$ out of the vacuum
is an entangled state between the two fermionic modes along that bond
(for the formal definition of entangled fermionic modes, see Ref. \onlinecite{Ban07}). 
The extra Majorana operators do not play any role as long as antiperiodic
boundary conditions are imposed on them, and can thus be ignored so that
one ends up with the states as we are using them throughout this article.

\bibitem{Ban07}
M.-C. Banuls, J. I. Cirac, and M. M. Wolf, 
\newblock Phys. Rev. A {\bf 76}, 022311 (2007), arXiv:0705.1103.

\bibitem{Qi10}
X.-L. Qi, and S.-C. Zhang,
\newblock Rev. Mod. Phys. {\bf 83}, 1057 (2011), arXiv:1008.2026.

\bibitem{Has10}
M. Z. Hasan, and C. L. Kane,
\newblock Rev. Mod. Phys. {\bf 82}, 3045 (2011), arXiv:1002.3895.

\bibitem{Qi06}
X.-L. Qi, Y.-S. Wu, and S.-C. Zhang, 
\newblock Phys. Rev. B {\bf 74}, 085308 (2006), arXiv:cond-mat/0505308.


\bibitem{grafakos}
L.~Grafakos, Classical Fourier Analysis. Springer, 2008.

\bibitem{cirac:peps-boundaries}
J.~I. Cirac, D.~Poilblanc, N.~Schuch, and F.~Verstraete,
\newblock Phys. Rev. B {\bf 83}, 245134 (2011), arXiv:1103.3427.

\bibitem{schuch:topo-top}
N.~Schuch, D.~Poilblanc, J.~I. Cirac, and D.~Perez-Garcia,
\newblock Phys. Rev. Lett. {\bf 111}, 090501 (2013), arXiv:1210.5601.

\bibitem{fidkowski:freeferm-bulk-boundary}
L.~Fidkowski,
\newblock Phys.\ Rev.\ Lett. {\bf 104}, 130502 (2010), arXiv:0909.2654.

\bibitem{Fla09}
S. T. Flammia, A. Hamma, T. L. Hughes, and X.-G. Wen,
\newblock Phys. \ Rev. \ Lett. {\bf 103}, 261601 (2009), arXiv:0909.3305. 

\bibitem{Zha12}
Y. Zhang, T. Grover, A. Turner, M. Oshikawa, and A. Vishwanath,
\newblock Phys. Rev. B {\bf 85}, 235151 (2012), arXiv:1111.2342.

\bibitem{Cin13}
L. Cincio, and G. Vidal,
\newblock Phys. Rev. Lett. {\bf 110}, 067208 (2013), arXiv:1208.2623.

\bibitem{Zal13}
M. P. Zaletel, R. S. K. Mong, and F. Pollmann,
\newblock Phys. Rev. Lett. {\bf 110}, 236801 (2013), arXiv:1211.3733.

\bibitem{note-Zaletel}
We thank M. Zaletel for
suggesting to compute the momentum polarization.

\bibitem{Bra05}
S. Bravyi,
\newblock Quantum Inf. Comput. {\bf 5}, 216 (2005), arXiv:quant-ph/0404180.

\bibitem{Hat93}
Y. Hatsugai,
Phys. Rev. Lett. {\bf 71}, 3697 (1993).

\bibitem{kitaev:honeycomb-model}
A.~Kitaev,
\newblock Ann. Phys. {\bf 321}, 2 (2006), cond-mat/0506438.

\bibitem{Dem84}
S. Demko, W. F. Moss, and P. W. Smith,
Math. of Comp. {\bf 43}, 491 (1984).

\bibitem{note-differ}
The two families only differ by a change of
sign in the ``right'' and ``down'' column, which effectively
shifts the momentum by $(\pi,\pi)$.

\bibitem{note-values}
The values of the coefficients in
Sec.~\ref{sec:sym-chern2} are
\begin{align*}
\alpha_{L,1}^{(+)} &= -0.267 + 0.422i, & \alpha_{L,2}^{(+)} &= 0.488 + 0.105i, \\
\alpha_{U,1}^{(+)} &= 0.076 - 0.251i, & \alpha_{U,2}^{(+)} &= 0.147 + 0.259i, \\
\alpha_{L,1}^{(-)} &= 0.607, & \alpha_{L,2}^{(-)} &= -0.358 - 0.052i, \\
\alpha_{U,1}^{(-)} &= -0.058 + 0.412i, & \alpha_{U,2}^{(-)} &= 0.196 + 0.406i.
\end{align*}


\end{thebibliography}
\end{document}